\documentclass[fleqn,11pt]{wlscirep}
\usepackage[utf8]{inputenc}
\usepackage[T1]{fontenc}
\usepackage{bm}

\usepackage{amsmath,mathtools}
\usepackage{comment}

\usepackage{graphicx}

\def\beq{\begin{equation}}
	\def\eeq{\end{equation}}
 \def\vq{{\bf q}}
\def\vk{{\bf k}}

\def\vR{{\bf R}}

\def\psl{\boldsymbol{\ell}}

\newcommand{\Gm}{\Gamma}

\title{Hidden orders
in spin-orbit entangled
correlated insulators}

\author[1,2]{Leonid V. Pourovskii}
\author[1,2, 3]{Dario Fiore Mosca}
\author[3]{Lorenzo Celiberti}
\author[4]{Sergii Khmelevskyi}
\author[5]{Arun Paramekanti}
\author[4,6,*]{Cesare Franchini}
\affil[1]{CPHT, CNRS, \'Ecole polytechnique, Institut Polytechnique de Paris, 91120 Palaiseau, France}
\affil[2]{Coll\`ege de France, Université PSL, 11 place Marcelin Berthelot, 75005 Paris, France}
\affil[3]{University of Vienna, Faculty of Physics and Center for Computational Materials Science, Vienna, Austria}
\affil[4]{Vienna Scientific Cluster Research Center,  Vienna University of Technology, Vienna, Austria}
\affil[5]{Department of Physics, University of Toronto, Toronto, Canada}
\affil[6]{Department of Physics and Astronomy "Augusto Righi", Alma Mater Studiorum - Universit\`a di Bologna, Bologna, 40127 Italy}

\affil[*]{e-mail: cesare.franchini@univie.ac.at}

\begin{abstract}
In many materials, ordered phases and their order parameters are 
easily 
characterized by standard experimental methods.
"Hidden order" refers to a phase transition in which an ordered state emerges without such an easily detectable order parameter, despite clear thermodynamic evidence of the transition.
The underlying mechanisms for these unconventional states of matter stem from spin-orbit coupling, which intertwines inter-site exchange, classical electromagnetic interactions, and electron-lattice effects. This physics
 is elusive to experimental probes and beyond traditional theories of insulating magnetism,  requiring sophisticated methodologies for its exploration.
In this Review, we survey exotic hidden-order phases in correlated insulators, particularly focusing on the latest progress in material-specific theories and numerical approaches. The relevant degrees of freedom in these phases are local high-rank multipole moments of magnetic and charge density that emerge from spin-orbit entangled correlated shells of heavy $d$ and $f$ electron ions and interact on the lattice via various mechanisms. We discuss approaches to modelling hidden orders in realistic systems via direct ab initio calculations or by constructing low-energy many-body effective Hamiltonian. We also describe how these new theoretical tools have helped to uncover driving mechanisms for recently discovered multipolar phases in double perovskites of heavy transition metals, and how they have proved instrumental in disentangling the role of various interactions in “traditional” $f$-electron multipolar materials like actinide dioxides. In both cases, material-specific theories have played a key role in interpreting and predicting experimental signatures of hidden orders.
\end{abstract}
\begin{document}

\flushbottom
\maketitle

\thispagestyle{empty}

\section*{[H1] Introduction}

Spin-orbit coupling (SOC) and electronic correlation are two fundamental interactions that substantially influence the properties of many-body electronic systems. SOC in solids is a relativistic interaction that entangles the electron spin with its orbital angular momentum in a 
crystal field environment. SOC 
is strongly enhanced 
for elements with a large atomic number~\cite{landau1965,khomskii2014} and is primarily responsible for 
effects such as Rashba coupling~\cite{Rashba1960,Bihlmayer2022}, the spin Hall effect~\cite{hirsch1999,Maciejko2011}, and topological spin textures such as skyrmions~\cite{fert2017}. Electronic correlation is the repulsive interaction between electrons, which is particularly intense in strongly correlated materials with localized electronic states, typically partially filled 3$d$ orbitals, and leads to phenomena such as Mott insulating behavior and quantum criticality.~\cite{Mott1937,Mott1949,Imada1998,Morosan2012}. 

\begin{figure}[h!]
\centering
\includegraphics[width=\linewidth]{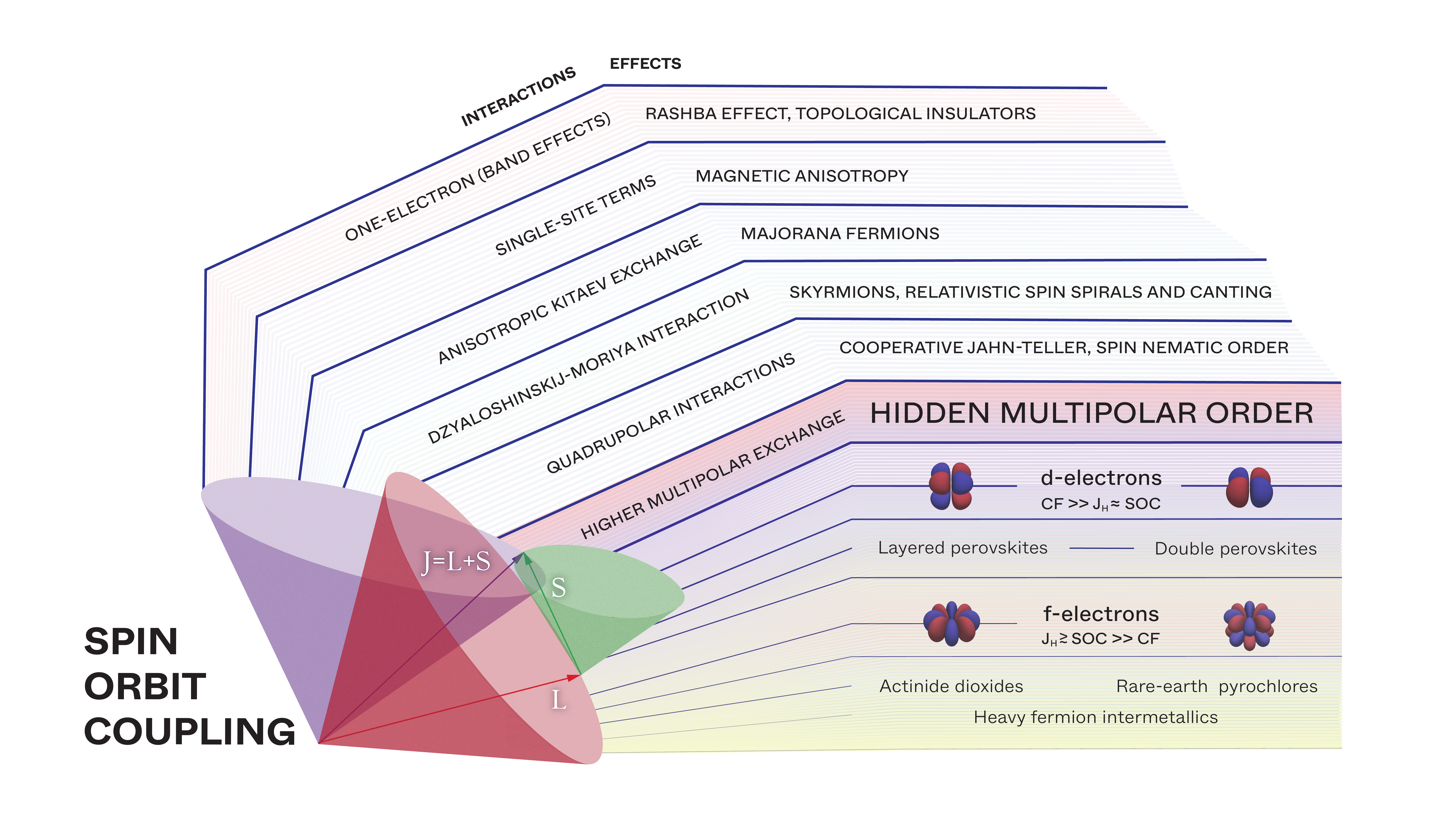}
\caption{\textbf{Spin-orbit entangled phenomena.} Interplay  
of spin-orbit coupling with band and correlation effects induces 
unusual interactions (left side), which, in turn, generate 
unconventional phenomena (right side), including hidden multipolar orders. Characteristic energy scales and key families of materials hosting hidden orders are listed for $d$- and $f$-electron systems. Representative multipolar order parameters are shown for both classes of materials:quadrupolar ($K=2$) and octupolar ($K=3$) moments for $d$ systems and hexadecapolar ($K=4$) and triakontadipolar ($K=5$) moments for $f$ systems.}
\label{fig:intro}
\end{figure}

The synergistic interplay of large SOC and strong electronic correlation can give rise to new phenomena distinct from conventional electronic and magnetic states,
\cite{Witczak2014,Khaliullin2005,
 Rau2016, Schaffer2016,Takayama2021,Khomskii2021,Browne2021
} such as spin-orbit coupled Mott insulators~\cite{Jackeli2009}, insulator-to-metal transitions driven by the combined action of SOC and magnetism~\cite{Calder2012,Kim2016}, exotic magnetic states such as quantum spin liquids~\cite{Balents2010}, valence-bond glasses~\cite{deVries2010}, high-rank multipolar orders~\cite{Sasaki1970}, skyrmions~\cite{Heinze2011,Sitte2018}, non-collinear Dzyaloshinskii-Moriya  interactions~\cite{Dziailoshinskii1957,Moriya1960,Morrish1994}, non-trivial topological Mott systems~\cite{Pesin2010} and Rashba effects~\cite{Bihlmayer2022}. These effects are typically observed in systems containing heavy 5$d$ transition metals or heavy 4$f$-5$f$ elements, which constitute an extensively studied class of materials and still represent a frontier in condensed matter research.~\cite{Celiberti2024,Fioremosca2024,Hart2024,Khmelevskyi2024,Soh2024,Yahne2024}. A compendium of the most relevant phases is provided in \textbf{Table~\ref{tab:socU}} and pictorially represented in \textbf{Fig.~\ref{fig:intro}} together with the one-body and many-body interactions that are at their origin.

A common characteristic of spin-orbit entangled correlated insulators is the formation of unconventional spin-orbital phases, often developing in geometrically frustrated lattices. In conventional magnetic materials, a phase transition to a well-defined pattern of local magnetic moments is typically associated with an order parameter detectable experimentally by both thermodynamic and spectroscopic probes such as magnetic susceptibility, specific heat, and neutron scattering measurements. In contrast, in a correlated spin-orbit coupled background, the strong entanglement of spin and orbital moments 
can lead to spontaneous symmetry breaking driven by a complex order parameter that, though clearly observed in thermodynamic probes, is challenging to characterize by standard magnetic measurements and neutron spectroscopy. The resulting order is typically referred to as `hidden', a term introduced in the early 2000s for the unknown ordered state observed in the heavy-fermion metal URu$_2$Si$_2$~\cite{Shah2000, Amitsuka2002, Chandra2002}.
Since then, hidden-order phases (\textbf{Box 1}) have been reported for different types of correlated insulators with localized and partially filled $f$ and $d$ spin-orbital manifolds.~\cite{Cricchio2009,Jackeli2009b, Chen2010, Gardner2010, Mydosh2011,Tsirlin2012,Kim2012,Sugiyama2014,Huang2014,Zhao2016,Cameron2016, Li2016, Lu2017,Li2017,Kim2017, LiuC2018, Ishikawa2019, Gaudet2019, Shen2019, Rau2019,Hirai2020,Aeppli2020,Zvereva2020, Maharaj2020, Sibille2020,Fioremosca2021, Pourovskii2021,Pourovskii2021f,Smith2022,Chen2023,Voleti2023,Verbeedk2023}

In conventional magnetic phase transitions, the order parameter is associated with distinct ordering patterns of the electronic spin dipole moments in a crystal. These emergent orders are typically induced by a Heisenberg-type effective Hamiltonian describing spin-spin interactions, and they can be 
investigated by ab initio electronic structure theory using density functional theory (DFT) and dynamical mean-field theory (DMFT) techniques. This well-established approach facilitates a quantitative comparison of computed intersite dipolar interactions with experimental data, allowing for a material-specific analysis.~\cite{Szilva2023} In spin-orbit entangled correlated materials, the interplay of spin and unquenched orbital degrees of freedom may lead to the activation of higher-rank multipole interactions, which can be responsible for the onset of hidden orders ~\cite{Santini2009}. Compared to dipolar couplings, evaluating multipolar intersite interactions is more complex due to the necessity of accounting for tensorial operators in the extended multipolar Hamiltonian. 
Equally challenging is the extraction of high-rank coupling from experimental measurements, which are often "blind" to multipolar hidden orders.~\cite{Bultmark2009,Pourovskii2016,Pi2014,Pi2014b,FioreMosca2022,Schaufelberger2023}
Nevertheless, pioneering experiments have provided insights in deciphering the microscopic nature of hidden orders using different techniques: 
X-ray magnetic circular dichroism (XMCD) has been employed to inspect 
the ferroic order of the magnetic octupole in Mn$_3$Sn~\cite{Kimata2021}, and a variety of tools including neutron and X-ray diffraction (XRD), muon spin resonance ($\mu$SR), inelastic neutron scattering, resonant and non-resonant elastic X-ray scattering (REXS) and nuclear magnetic resonance (NMR) were used to elucidate the octupolar order in Os-based double perovskites~\cite{Maharaj2020,Soh2024,Lu2017} and UPd$_3$~\cite{McMorrow2001} as well as the rank-5 (trikontadipolar, the name referring to 32 lobes of this multipole) order in NpO$_2$ \cite{Santini2006,Magnani2008}.

Unraveling and understanding hidden orders is challenging, demanding advanced theories and computational methodologies. These efforts can uncover novel quantum states of matter, advancing the boundaries of materials research. 

In this Review, we provide an overview of recent \emph{ab initio} quantitative approaches for studying the complex hidden multipolar orders in correlated insulators. We 
detail how these methods can offer essential insights into material-specific properties, which are crucial for a microscopic understanding of non-conventional order parameters.
We begin by reviewing the fundamental theoretical aspects of multiplet physics and the formalism necessary to construct low-energy microscopic Hamiltonians. Next, we present computational approaches to derive these Hamiltonians from first principles, focusing specifically on multipolar DFT and DMFT frameworks.
We then demonstrate how these theories have clarified the role of multipolar physics in two main classes of materials: "traditional" $f$-electron Mott insulators~\cite{Santini2009} and 5$d$ double perovskites~\cite{VASALA2015,Chen2024}.  Very complex low-energy Hamiltonians involving couplings between high-rank multipolar moments have been fully derived  from first principles for prototypical "hidden-order" $f$-electron systems\cite{Pourovskii2021f,Iwahara2022,Khmelevskyi2024}. The double perovskite family
has recently garnered significant attention as a rich playground for discovering novel high-rank multipolar quantum phases and unraveling the coupling of hidden orders with phonons~\cite{Ning2023, Hart2024} and vibronic effects~\cite{Bersuker2006,Iwahara2023,Fioremosca2024b,Soh2024} and the role of charged defects, polarons and doping~\cite{Celiberti2024}, providing knowledge that can be transferable to other quantum materials.  Although this Review focuses primarily on theoretical and computational aspects, we discuss direct comparisons with experimental observations of multipolar orders as a necessary validation for the proposed numerical protocols. 

We note that ideas and symmetry organization principles similar to those discussed in this Review also apply to materials with exotic magnetic orders such as toroidal order~\cite{Spaldin2008} or the recently proposed altermagnetic order, 
which can occur in systems with larger unit cells \cite{Smejkal2022,jungwirth2024}. In such materials, while the net magnetic dipole moment 
within the unit cell integrates to zero, one can nevertheless find emergent higher-rank multipoles constructed out of the local dipole moments.
These include
quadrupolar symmetry breaking, which is akin to nematic order, time-reversal breaking octupolar order, which may be viewed as a $d$-wave ferromagnet, 
or even multipolar variants of multiferroic materials.
\cite{Ederer2007,Hayami2019,Yatsuhiro2021,Bhowal2024a}

\begin{table}[h!]
\centering
\begin{tabular}{|l|l|l|}
\hline
\multicolumn{3}{|c|}{Phases of matter and exotic effects in spin-orbit entangled correlated materials}\\
\hline

Year of Report &  {Type of phase} & Materials\\\hline
  1958~\cite{DZYALOSHINSKY1958}, 1960~\cite{Moriya1960,Moriyab}  &   Dzyaloshinskii-Moriya interaction & Fe$_2$O$_3$, SrFeO$_3$~\cite{Takeda1972} Sr$_2$IrO$_4$~\cite{Liu2015},  polar and frustrated magnets~\cite{Tokura2021},  \\
   &  & multiferroics~\cite{Xu2019,Yang2023}, 2D magnets~\cite{Fert2023}, 2D Van der  Waals~\cite{Pan2020} \\ \hline
   &  & Reviews~\cite{HEIDE2009, Moskvin2019,Moskvin2019,Tokura2021,Yang2023,Fert2023,Kuepferling2023,CAMLEY2023} \\ \hline 
  1970~\cite{Sasaki1970},1996~\cite{BUYERS1996}     & Multipolar (hidden) order&  NpO$_2$, UO$_2$~\cite{Sasaki1970} (actinide dioxides~\cite{Santini2006, Santini2009}), CeB$_6$~\cite{ERKELENS1987,Shiina1997}, URu$_2$Si$_2$~\cite{BUYERS1996,Shah2000,Mydosh2011}, \\
  & & $R_3$Pd$_{20}$$X_6$ ($R$=Ce, Pr: $X$=Si, Ge)~\cite{Kitagawa1996}, Pr$X_3$ ($X$=Pb,Mg)~\cite{Tayama2001, Tanida2006}, \\ 
  & & Sr$_2$VO$_4$~\cite{Jackeli2009b,Sugiyama2014}, Ba$_2M$OsO$_6$~\cite{Chen2010, Maharaj2020,Pourovskii2021}, Sr$_2$RuO$_4$~\cite{Liu2019, Ning2023}, Sr$_2$IrO$_4$~\cite{Zhao2016} \\ 
  & & Rare-earth pyrochlores~\cite{Huang2014,Yahne2024,Rau2019}, further heavy fermions~\cite{Suzuki2018}, Mn$_3$Sn~\cite{Kimata2021}  \\ 
  &   &  { magnetoelectric oxides (Ca$_3$Ru$_2$O$_7$~\cite{Thole2018}, Cr$_2$O$_3$ \& Fe$_2$O$_3$ ~\cite{Verbeedk2023})}\\ \hline
  &  & Reviews~\cite{Kuramoto2008,Kuramoto2009,Santini2009,Rau2019,Aeppli2020, Takayama2021}\\ \hline
    1973~\cite{Anderson1973} &  Quantum spin liquid & Candidates~\cite{Savary2017}: Trianglular organics~\cite{Shimizu2003,Powell2011}, Kitaev materials~\cite{Jackeli2009,Plumb2014}, \\
  &   &  Rare-earth pyrochlores~\cite{Gardner2010}, Kagom\'e Herbertsmithite~\cite{Mendels2010}\\ \hline
 &  & Reviews~\cite{Powell2011,Savary2017,Winter2017,Takagi2019,Rau2019} \\ \hline
 2009~\cite{Jackeli2009}  &  Spin-orbit coupled Mott insulator & Sr$_2$IrO$_4$ (Layered)~\cite{Kim2008,Jackeli2009}, Na$_2$IrO$_3$ (Honeykomb)~\cite{Comin2012}, BaIrO$_3$~\cite{Ju2013,Franchini2014} \\ 
  &  & NaOsO$_3$~\cite{Calder2012,Kim2016}, other 4$d$ and 5$d$ oxides~\cite{Martins2017}\\ \hline
  &  & Reviews~\cite{Witczak2014,Rau2016,Martins2017,Takayama2021,Chen2024}\\ \hline
  2010~\cite{Dzero2010} & Non-trivial topological phases & SmB$_6$~\cite{Dzero2010}, SrIrO$_3$~\cite{Xiao2011}, Ce$_3$Bi$_4$Pd$_3$~\cite{Dzsaber2017}, Ce$_2$Au$_3$In$_5$~\cite{Chen2022} \\ \hline
  &  & Reviews~\cite{Pesin2010} \\ \hline
   2010~\cite{Caviglia2010,Rashba1960}   &Rashba effects in correlated systems & SrTiO$_3$~\cite{Caviglia2010}, KTaO$_3$~\cite{Wadehra2020}, SrNbO$_3$~\cite{Okuma2024}, RETMIr$_2$Si$_2$~\cite{Generalov2017,Generalov2018,Bihlmayer2022}\\ 
      &   & (RE=rare earth; TM=transition metal)\\ \hline
      &  & Reviews~\cite{Bercioux2015,Bihlmayer2022,Manchon2015} \\ \hline
\end{tabular}
\caption{\label{tab:socU} \textbf{Phases and behaviours originating from the coexistence of spin-orbit coupling and electronic correlation. Historical overview and exemplary materials. }}
\end{table}

\section*{[H1] Theories and methods}

This section provides an overview of theoretical approaches to multipolar interactions and orders in realistic materials. We start by briefly introducing the concepts of multipolar moments (and the corresponding operators) acting on a localized atomic shell in a correlated insulator. 
Various mechanisms providing coupling between multipoles at different sites are also briefly reviewed
together with model approaches designed to derive those couplings from tight-binding hopping parameters. 

We then focus on material-specific theoretical tools for modeling multipolar phases and extracting experimental signatures of multipolar moments. 
Ab initio total-energy DFT-based methods are extensively used to study conventional magnetic orders~\cite{DFTmag}. Generalizations of DFT functionals in conjunction with the +$U$ correction have been elaborated to treat multipolar orders.  
 An alternative route is to construct, from DFT or its combination with DMFT, a many-body effective Hamiltonian (MBEH) encoding low-energy multipolar physics. We  review newly developed "force-theorem" approaches to correlated insulators that  derive the MBEH from linear-response of the DFT (+U/+DMFT)  total energy functional, 
 as well as another DFT+DMFT-based approach that extracts the multipolar MBEH from the dynamical susceptibility.
A different class of methods  -- cluster MBEH derivation utilizing quantum-chemistry or exact diagonalization methods within a strong-coupling perturbative  perspective
 -- is considered afterwards. Theoretical approaches to electron-lattice coupling and the Jahn-Teller effect in multipolar systems are also briefly reviewed, as well as methods for solving the MBEH and for modeling experimental responses of multipolar orders.

\newpage

\newcounter{boxnum}
\setcounter{boxnum}{1} 

\noindent\fbox{
\parbox{\textwidth}{

\textbf{\large Box \theboxnum. Hidden order }

\vspace{2mm}

\begin{minipage}{0.6\textwidth}%
		\begin{itemize}
	\item {\it\bf Ordered phases and order parameters.} Second-order phase transitions in solids 
	involve the emergence of an ordered phase at a specific temperature, with a symmetry lower than that of the high-T disordered phase. This low-T ordered phase is characterized by a non-zero 
	of a certain physical observable, known as the \emph{order parameter} (OP), which gradually vanishes as the temperature approaches the transition point, as shown  in the schematics. 
	The order parameters  can be usually directly measured  due to their interactions with either external probes, e.~g., neutron diffraction can detect superstructural peaks due to an antiferromagnetic OP.

	\item{\it\bf Hidden-order phases.} In certain cases, experimental observations reveal clear thermodynamic signatures of a phase transition, such as the $\lambda$-type anomaly in specific heat $C_v$ shown in the schematics,  without any detectable spontaneous symmetry breaking. These scenarios pose long-standing challenges in solid-state physics and are referred to as \emph{hidden order} transitions. 
\end{itemize}	
 \end{minipage}%
\begin{minipage}{0.40\textwidth}
			\centering
		\includegraphics[width=0.7\textwidth]{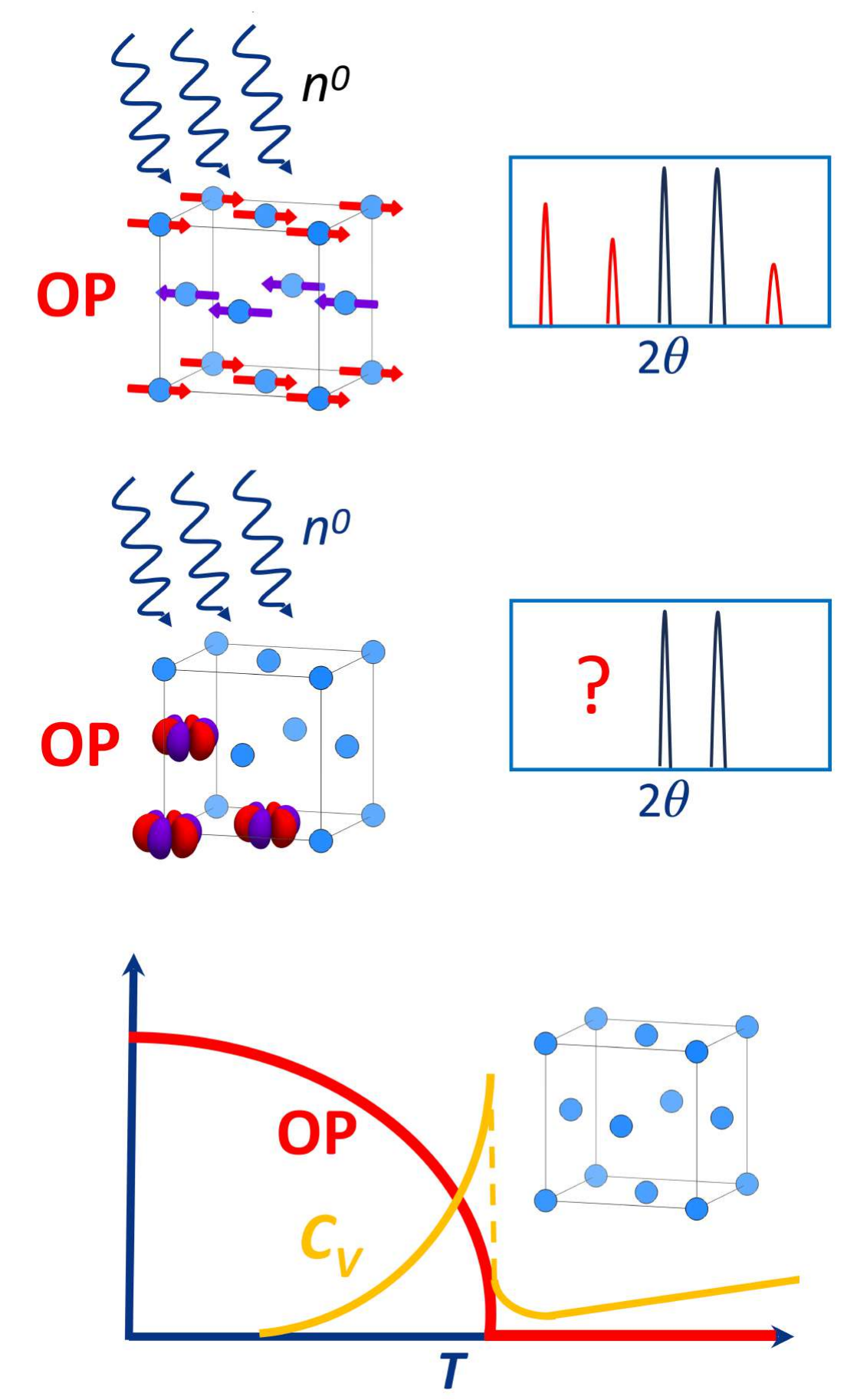}
\end{minipage}%
		
\vspace{-3mm}
			
\begin{itemize}
	
\item{\it\bf Multipolar orders.} In a common class of hidden orders, the OPs are high-rank local-moment degrees of freedom (multipole moments).
{
These systems are classified by the rank of their multipolar order parameters and are typically parity-even, with the exception of magnetic and electric toroidal moments, which are parity-odd (see Supplemental Material)}:

\vspace{-4mm}

\begin{itemize}
\item{\it Quadrupolar orders} The simplest case is the ordering of quadrupoles (rank 2), which are, in a quantum treatment, represented by symmetrized quadratic polynomials of  angular momenta. 
Since quadrupolar moments interact directly with the lattice and are a source of magnetic anisotropy, quadrupolar orders have been characterized in various materials by studying anisotropic susceptibility and magnetoelastic effects.
\vspace{-4mm}
\item{\it High-rank multipolar orders.}
Octupolar (rank 3) and higher-rank orders arise from electronic coupling between multipols rather than due to a coupling by the lattice. They are extremely challenging to detect. 
Promising techniques for identifying such hidden orders include methods sensitive to time-reversal symmetry and to the splitting of ionic energy levels: X-ray magnetic circular dichroism, neutron and X-ray diffraction, muon spin resonance, resonant and non-resonant elastic X-ray scattering, and nuclear magnetic resonance.
{{\it Magnetoelectric hidden order.} Magnetoelectric materials exhibit a net change in magnetization in response to an applied electric field or vice versa. The leading, linear contribution to the magnetoelectric response—requiring the simultaneous breaking of both spatial and time-reversal symmetries—is closely associated with the presence of magnetoelectric  multipoles. These are odd-parity, second-order multipoles arising from the magnetization density.}
\end{itemize}

\vspace{-4mm}

\item{\it\bf Multipolar systems.}  "Hidden-order" phases have been found in insulators, e.~g. actinide dioxides or double perovskites of heavy transition metals, {magnetoelectric materials, }as well as in heavy-fermion intermetallics, see Table~\ref{tab:socU} for selected references. 
Multipolar systems share key features that serve as preconditions for "hidden order":

\begin{minipage}{0.4\textwidth}
	\begin{itemize}
		\item{Strong electronic correlations}
		\item{High lattice symmetry}
		\item{Strong spin-orbit coupling}
   \end{itemize}
\end{minipage}%
\begin{minipage}{0.5\textwidth}
		\begin{itemize}
		 \item{Large spatial separation between magnetic ions}
		 \item{Geometrical frustration with respect to \mbox{conventional} antiferromagnetism}
	\end{itemize}
\end{minipage}%
\end{itemize}
}}

\setcounter{boxnum}{1} 

\noindent\fbox{
\parbox{\textwidth}{

\textbf{\large Box \theboxnum. Hidden order (continued)}
\begin{itemize}
\item{\it\bf Ab initio modelling.} A major difficulty lies in determining the true spatial symmetry of the multipolar order, even when associated local symmetry breaking is revealed by experiment. This challenge underscores the critical role of theoretical methods and models in exploring hidden-order phases in solid-state systems. In recent years, several \emph{ab initio} methodologies have been developed to investigate multipolar order in correlated systems and their underlying microscopic quantum interactions.%
\end{itemize}
}
}

\subsection*{[H2] The ground-state multiplet and its moments}  

Local multipolar degrees of freedom are hosted by strongly correlated, typically $d$ or $f$, shells in Mott insulators.  
The local electronic Hamiltonian for such a shell reads
\begin{equation}\label{eq:Hloc}	H_{\mathrm{loc}}=H_{\mathrm{CF}}+H_{\mathrm{SO}}+H_{\mathrm{C}},
\end{equation}
where the terms on the right-hand side are the crystal field, SOC and on-site Coulomb repulsion (C). 
 With charge fluctuations suppressed due to a large value of the on-site Coulomb repulsion parameter $U$, the correlated shell has a well-defined integer ground-state occupancy.  Other local energy scales -- Hund's rule coupling $J_H$,  crystal field splitting, SOC constant $\xi_{\mathrm{SO}}$ -- are,  as a rule, much larger than the relevant energy scale of intersite coupling. The highest known transition temperature for the onset of a purely "hidden" order without any observable conventional dipole moments, about 50~K, is probably observed in $d^2$ double perovskites of Os~\cite{Maharaj2020}. This energy scale is to be compared with the magnitude of $J_H$ (0.2-1~eV) and $\xi_{\mathrm{SO}}$ (several hundreds meV) in heavy transition metals and $f$-electron materials~\cite{Carnall1989,Moore2009,Takayama2021,Taylor2017,Paramekanti2018}. Hence, a well-separated ground-state multiplet (GSM) of the local Hamiltonian can be defined for a given shell; only those GSM states are thus relevant for low-temperature orders. 

The choice for the GSM depends on the problem at hand. In $f$-electron materials the crystal field  is much weaker than $\xi_{\mathrm{SO}}( \sim  J_H)$. However, the relevant crystal field splitting is still generally substantially larger than the magnitude of intersite coupling, and only the lowest crystal field level is relevant for the corresponding order. In this situation, it is useful to introduce the concept of orbital "pseudo-spin" operators, a fictitious angular momentum encoding the crystal field ground-state manifold~\cite{Santini2009}.

In contrast, in correlated insulators of heavy (4$d$ and 5$d$) transition metals, the crystal field term in equation (\ref{eq:Hloc}) is typically larger than $J_H$. The crystal field thus defines a relevant manifold  (such as $t_{2g}$ orbitals in cubic perovskites and double perovskites with $d$ shell occupancy $N_d \le 6$), which is in turn split by the action of spin-orbit and smaller crystal field terms~\cite{Takayama2021}. In the case of $t_{2g}$ systems, it is  convenient to introduce a pseudo angular-momentum operator $\ell=1$ 
whose matrix elements within this manifold are the same as those of the physical orbital momentum $l=2$ with the inverse sign~\cite{abragam_bleaney_book}. The SO term for the full $d$ shell,  $\xi_{\mathrm{SO}}\sum_i \mathbf{\hat{l}}_i\mathbf{\hat{s}}_i$, transforms into $-\xi_{\mathrm{SO}}\sum_i \hat{\psl}_i\mathbf{\hat{s}}_i$ acting within the $t_{2g}$ space. 
The flip of the SOC sign results in the inversion of the atomic third Hund's rule, with the shell total angular momentum $J=L+S$ or $L- S$ for the less than or more than half-filled shell, respectively~\cite{Takayama2021}.

The GSM with occupancy $N$ and effective $J$ is fully determined by the corresponding $N$-electron density matrix $\rho$; thus, the value of any operator $A$ acting within the GSM is given by $\mathrm{Tr}\left[\rho A \right]$. Apart from its usual matrix representation, the density matrix operator can be equally well defined by its multipole expansion\cite{Santini2009,Blum_DM} 
\begin{equation}\label{eq:mult_exp}
	\rho=\langle O_{K}^Q \rangle O_{K}^Q,
\end{equation}
where $O_{K}^Q$ with $K \le 2J$ are normalized, $\mathrm{Tr}\left[O_{K}^Q O_{K'}^{Q'} \right]=\delta_{KK'}\delta_{QQ'}$, Hermitian spherical tensor operators of rank $K$ and projection $Q$ constructed from the standard multipolar spherical tensors\cite{Sakurai1993Modern} akin to spherical harmonics\cite{Santini2009}. 
Alternatively, the degrees of freedom within the GSM space can also be defined as polynomials of angular momentum operators (Stevens operators) or as products of pseudo-spin-1/2 operators. An expansion similar to that in equation (\ref{eq:mult_exp}) can also be introduced for the one-electron density matrix of a given correlated shell using one-electron multipoles $\hat{D}_k^q$.

The states belonging to a single $J$-multiplet have the same parity, leading to only parity-even -- charge (even $K$) or magnetic (odd $K$) -- multipoles contributing to the expansion in equation (\ref{eq:mult_exp}). Two additional kinds of (parity-odd) multipoles  -- magnetic and electrical toroidal moments -- need to be included in the multipolar expansion of a general electronic system \cite{Dubovik1990}, that is, in the case of either an atomic cluster or hybridized orbitals on a single atomic site. These multipoles play a crucial role in the magnetoelectrical response~\cite{Spaldin2008}. One-electron operators for such multipoles have been recently derived~\cite{Spaldin2013,Hayami2018} and classified by their symmetry \cite{Watanabe2018,Hayami2018_1}; the corresponding toroidal order parameters have been proposed in a number of correlated $d$ and $f$-electron systems~\cite{Watanabe2018}. Recently, magnetic multipoles have been utilized in the cluster-multipole expansion, enabling the decomposition of arbitrary magnetic configurations into magnetic multipole components. This approach facilitates the systematic and efficient generation of magnetic configurations, which can be screened in high-throughput studies to identify the magnetic ground-state phase.~\cite{Suzuki2019,Huebsch2021}

The experimental determination of  the nature of  
multipolar order parameters
in spin-orbit entangled materials presents 
a substantial
challenge. Quadrupolar moments exhibit strong coupling to phonons with the same symmetry, and can be studied using polarization-resolved
Raman spectroscopy, which can probe the symmetry of crystal field multiplets as well as the relevant phonon modes as a function of temperature.~\cite{Kung2015,Ye2019a,Ye2019b,Volkov2021} Quadrupolar ordering also leads to weak local lattice displacements, which  
may be measured using high-resolution synchrotron X-ray diffraction, thus allowing one to infer the nature of quadrupolar ordering.~\cite{Hirai2020}
Higher-order multipolar symmetry breaking, such as octupolar ordering, could be probed using a judicious combination of magnetic field and strain.~\cite{Patri2019,Ye2023} Crudely, we may view octupolar order as a product of dipolar and quadrupolar orders, so that applying a magnetic
field can pin the dipolar order, allowing one to detect the emergent quadrupolar signature via lattice response. Conversely, the application of
a symmetry-tailored strain field can reveal the octupolar order via an emergent dipolar signal. Polarized neutron scattering,~\cite{Lovesey2020,Urru2023}, resonant X-ray scattering~\cite{Santini2009,Soh2024},
unconventional magnetoelectric responses,~\cite{Urru2022} second-harmonic generation,~\cite{Harter2017,vanderLaan2021}
and nuclear magnetic resonance studies \cite{Taniguchi2016,Taniguchi2019,Lu2017} 
may also be used to infer partial symmetry information for higher multipolar orders, but their 
complete characterization remains a complex task.

We detail various definitions of multipolar operators and their connection to the classical multipolar expansion of the charge and magnetic densities in the Supplemental Material.  Intersite coupling mechanisms between multipolar moments are outlined in \textbf{Box~2}.

\newpage

\setcounter{boxnum}{2} 

\noindent\fbox{
\parbox{\textwidth}{

\textbf{\large Box \theboxnum. Intersite interactions between multipole moments} \\

The local multipolar moments introduced in the text interact on the lattice through various mechanisms, including purely electronic interactions—such as exchange and classical electromagnetic interactions—as well as electron-lattice coupling. These couplings were initially uncovered and intensively investigated in more conventional magnetic and orbital-ordered 3$d$ systems~\cite{Anderson1950,Goodenough1955,Kanamori1959,Gehring1975,kugel_khomskii,Khomskii_book_2014}. In the context of  high-rank multipole moments in spin-orbit insulators, the relative importance of those coupling terms may change as we discuss below while introducing most important coupling mechanisms.  

\begin{itemize}
	\item  {\it Intersite exchange interactions (IEI)}. In  strongly-correlated systems, direct exchange due to correlated orbitals on neighboring sites overlapping   is usually small, and kinetic exchange due to correlated hopping dominates. In complex systems like double perovskite,
    a significant contribution due to hopping through non-magnetic cations is revealed by ab initio analysis~\cite{Pourovskii2021}. In $f$-electron materials, the CF splitting is weak and one starts with deriving symmetry-allowed coupling between one-electron hopping terms and the full $J$-multiplet~\cite{Hirst1978,Fulde1985}. The resulting formulae are quite cumbersome and semi-empirical superexchange is often used together with the assumption that dipolar superexchange dominates over multipolar one.~\cite{RareEarthMag_book}. The latter assumption is not  of general validity as shown by recent direct ab initio calculations of superexchange  in rare-earth oxides and nitrides~\cite{Iwahara2022,Khmelevskyi2024}.

Whatever the exchange mechanism is, the low-energy Hamiltonian  for a chosen GSM 
reads
\beq\label{eq:HIEI}
H_{\mathrm{IEI}}=\sum_{\langle ij\rangle}\sum V_{KK'}^{QQ'}(\vR_{ij})O_K^Q(i)O_{K'}^{Q'}(j),
\eeq
in terms of the multipolar operators $O_K^Q$  (\ref{eq:mult_exp}) and the intersite exchange interactions (IEI) $V_{KK'}^{QQ'}(\vR_{ij})$ for the bond $\langle ij\rangle$ linking two corresponding correlated sites. The second sum in (\ref{eq:HIEI}) is over the  repeated $K$ and $Q$ indices. The IEI constant $\hat{V}$ for a given bond is thus a $(2J_i+1)^2\times (2J_j+1)^2$ matrix. The bond point-group and time-reversal symmetries 
nullify some matrix elements\cite{Santini2009} in  $\hat{V}$, though non-zero couplings are still numerous for systems with large effective $J$~\cite{Pourovskii2021,Pourovskii2021f,Iwahara2022,Khmelevskyi2024}.
	
	\item {\it Classical electromagnetic coupling}. The classical dipole-dipole ($K=K'=1$) magnetic coupling $\sum I_{11}^{QQ'}(\vR_{ij})g_i O_1^Q(i) g_j O_1^{Q'}(j)$ is well known to be important in rare-earth magnets with large on-site dipole moments and gyromagnetic ratios $g$ as well as weak exchange couplings, where it may provide a leading contribution to the two-site magnetic anisotropy \cite{RareEarthMag_book}. Since classical electronmagnetic interactions decay as $\sim r^{-K-K'-1}$, the higher-rank  magnetic (odd $K$)  coupling are expected to be small. In contrast, electrostatic interaction between quadrupoles ($K$=2)  has been suggested to play an important role in some systems, in particular,  due to a large spacial extend of 5$d$ orbitals, in multipolar 5$d$ double perovskites \cite{Chen2010,Chen2011}. Its magnitude remains, however, poorly constrained by ab initio calculations or experiments \cite{Svoboda2021}.

	\item {\it Electron-lattice (EL) coupling}. In accordance with the Jahn-Teller (JT) theorem \cite{Jahn_Teller1937}, the on-site JT term $g_{\mathrm{JT}}O_K^Q(i) q_Q(i)$ that linearly couples electric (even $K$) operators with a local lattice distortion mode $q_Q$ of the same symmetry must be present in the Hamiltonian whenever the GSM has a non-Kramer's (non-time-reversal related) degeneracy\cite{bersuker_vibronic_1989}.  The local distortions on different sites are then coupled elastically by the lattice  resulting in an effective lattice-mediated  intersite interaction between electric multipoles \cite{Gehring1975,Khomskii_book_2014}. In practice, this cooperative JT effect in strongly-correlated systems always involve quadrupoles ($K$=2); relevant hexadecapoles ($K$=4) of the same symmetry are expected to give a small contribution. The shape of quadrupoles is affected by SO entanglement resulting in qualitative changes in the JT potential energy surface between weak to strong SO limit\cite{Streltsov2020,Streltsov2022}. In high-symmetry (cubic) multipolar systems, the EL coupling is generally important and can give a leading contribution to intersite interactions~\cite{Fioremosca2021, Iwahara2023,Fioremosca2024}. In lower-symmetry systems with the GSM that is formally a singlet or a Kramer's doublet, a pseudo-JT effect\cite{Bersuker2013,Liu2019} may arise due to inter-multiplet mixing induced by $q_Q$.
	
\end{itemize}
}}

\setcounter{boxnum}{2} 

\noindent\fbox{
\parbox{\textwidth}{

\textbf{\large Box \theboxnum. Intersite interactions between multipole moments (continued)} \\

Overall, the low-energy many-body effective  Hamiltonian (MBEH) thus reads 
\beq\label{eq:MBEH}
H=H_{\mathrm{IEI}}+H_{\mathrm{EM}}+H_{\mathrm{EL}}+\sum_i H_{1s}(i),
\eeq
where the classical electromagnetic term $H_{\mathrm{EM}}$ has the same form as the IEI term (\ref{eq:HIEI}).  The electron-lattice term $\hat{H}_{\mathrm{EL}}$ can be written differently depending whether ionic degrees of freedom are kept explicitly (thus JT dynamics is included) or integrated out.  In the latter case the EL term also takes the same bi-linear form (\ref{eq:HIEI}) with $K,K'=2$. Finally, $\hat{H}_{1s}$ contains all the remaining single-site terms that are active within the GSM like its CF splitting. 
}
}

\subsection*{[H2] Model superexchange Hamiltonians}

\begin{figure}[b!]
\centering
\includegraphics[width=\linewidth]{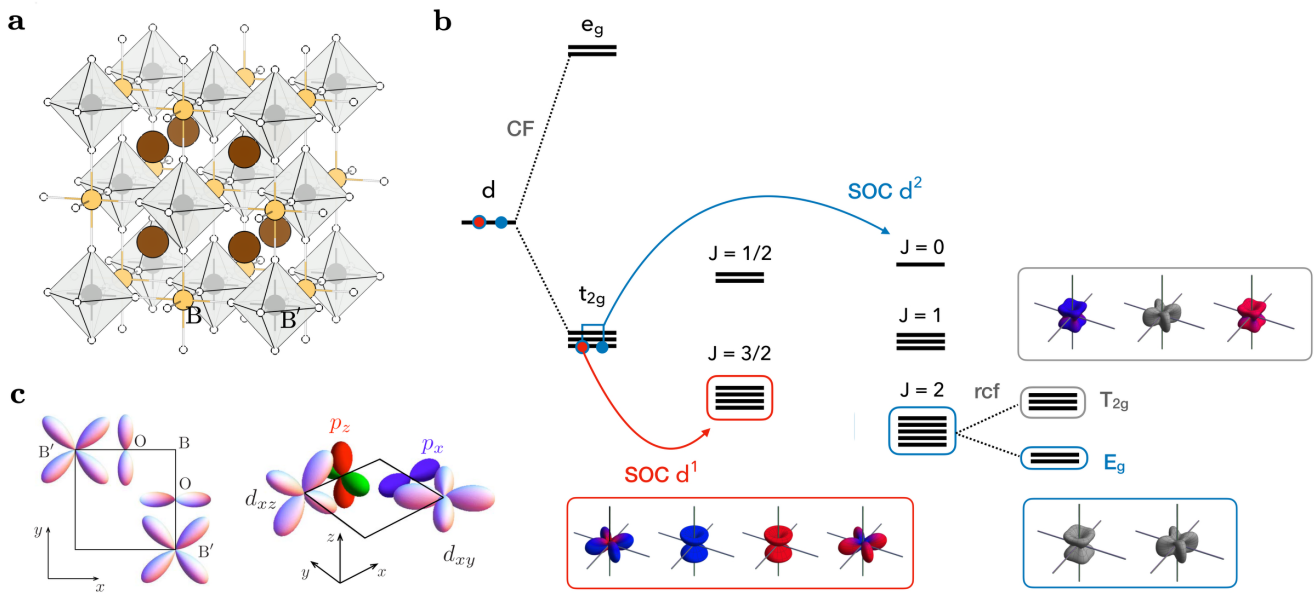}
\caption{\textbf{Levels structure, ground state multiplets and hopping paths in $d^1$ and $d^2$ spin-orbit double perovskites}. (a) The conventional unit cell of $A_2BB'O_6$ cubic double perovskites. (b) The level  splitting on the  $d^1$ and $d^2$ shells of
 magnetic $B'$ sites due to the octahedral crystal field and spin-orbit coupling. Electron density distribution for the ground-state multiplet orbitals is shown after Ref.~\cite{Iwahara2024}, the spin-up/spin-down densities are indicated by red-blue color. (c) Hopping paths included in the superexchange model of Chen~{\it et al.}~\cite{Chen2010} (after Ref.~\cite{Chen2010}).} 
\label{fig:levels}
\end{figure}

Qualitative features of the multipolar orders, in particular in $d$-electron materials, have been intensively studied by the traditional method of deriving intersite exchange interactions from tight-binding Hamiltonians.  This framework, initiated by the classical work of Philip Anderson~\cite{Anderson1950},  revealed the importance of   superexchange  through  the $p$-orbitals of surrounding ligands. The tight-binding (Goodenough-Kanamori)~\cite{Goodenough1955,Kanamori1959} rules  relating the transition metal-ligand-transition metal bond geometry with the sign of the resulting superexchange between transition metal spins have been intensively used to analyze coupling in 
correlated magnetic insulators. The orbital (Kugel-Khomskii) superexchange~\cite{kugel_khomskii} needs to be simultaneously  included. These approaches have been generalized to the case of strong SOC entangling orbital and spin inter-site exchange ~\cite{Jackeli2009,Chen2010,Chen2011,Takayama2021}.

 In particular, this approach was employed to elucidate low-temperature orders in spin-orbit double perovskites (\textbf{Fig.2a}) ~\cite{Chen2010,Chen2011}.   First, the spin-orbit GSM is established by focusing on the  $t_{2g}$ manifold, separated from the excited $e_g$ one by a large octahedral crystal field (\textbf{Fig.~\ref{fig:levels}b}). In the absence of SOC,  the superexchange coupling within this manifold comprises  separate $t_{2g}$-orbital and Heisenberg spin   terms.  Following well-established tight-binding rules for $dp$ hopping~\cite{Goodenough1955,Kanamori1959,Khomskii_book_2014}, the leading hopping term is assumed to be due to $\pi$-hopping between $t_{2g}$ and an appropriate  ligand-$p$ orbital lying in the same plane $\alpha$ (such as an $xy$ electron along the $y$ direction hopping to the $p_x$ orbital, \textbf{Fig.~\ref{fig:levels}c})~\cite{Chen2010}.  From the "diagonal" process between $t_{2g}$ orbitals of the same kind one obtains strong antiferroic spin and ferroic orbital superexchange
 $$
 H^{\alpha}_{\mathrm{dr}}=J_{\mathrm{SE}}\sum_{\langle ij\rangle \in \alpha}\left(S_{i\alpha}S_{j\alpha}-\frac{1}{4}n_{i\alpha}n_{j\alpha}\right),
 $$
 where $n_{i\alpha}$ is the occupancy operator for the corresponding site and orbital, $J_{\mathrm{SE}}$ is the superexchange coupling constant.
 Hopping between two orthogonal $t_{2g}$ orbitals along two orthogonal bonds (\textbf{Fig.~\ref{fig:levels}c}) leads, due to Hund's rule on the ligand-$p$ shell, to a weaker superexchange  of the opposite sign with respect to the "direct" term.

 In the limit of strong SOC, well justified for heavy 5$d$ transition metal ions, this spin-orbital Hamiltonian can be projected onto the spin-orbit GSM. In the case of $d^1$ systems, the GSM is the $J_{\mathrm{eff}}=3/2$ quadruplet shown in \textbf{Fig.~\ref{fig:levels}b}, which, in accordance with equation (\ref{eq:mult_exp}), can host dipoles ($K=1$), quadrupoles ($K=2$), and octupoles ($K=3$). The time-odd spin operators are projected onto superpositions of time-odd $J=3/2$ dipoles and octupoles, while $n_{i\alpha}$ become time-even $J=3/2$   quadrupoles of $e_g$ symmetry. Hence, the $t_{2g}$ charge and magnetic degrees of freedom project onto the corresponding degrees of freedom of the spin-orbit GSM; moments of opposite symmetry under time reversal do not couple, as expected.

 A similar analysis was carried out for $d^2$ double perovskites~\cite{Chen2011}, with the projection onto the $J_{\mathrm{eff}}=2$ GSM of $d^2$ ions. In this case, however, it is crucial~\cite{Maharaj2020} to take into account the ``remnant'' crystal field splitting of the $J_{\mathrm{eff}}=2$ manifold by the cubic crystal field (\textbf{Fig.~\ref{fig:levels}b}). The lower $E_g$ manifold does not carry any dipole $K=1$ moments. This can be shown either by a group-theory analysis or by explicitly projecting the dipole-moment operators onto the $E_g$ space, where they become zero. The $E_g$ doublet is thus described by three moments -- two quadrupoles and an octupole -- that project onto $x$, $z$, and $y$ pseudo-spin-1/2 operators of the $E_g$ space. Hence, by using such symmetry analysis one can immediately show that in the limit of dominating remnant cubic crystal field only multipolar phases can be realized in $d^2$ double perovskites.

The approach used in Refs.~\cite{Chen2010,Chen2011} has been intensively employed to explain experimental trends in spin-orbit double perovskites. Despite its success, its predictions on the relative magnitude of various multipolar intersite exchange interactions (IEIs) are based on a rather simplified tight-binding picture and not always supported by recent ab initio calculations. For example, the treatment used in Ref.~\cite{Chen2010} predicts no superexchange coupling between $J=3/2$ $t_{2g}$ quadrupoles in $d^1$ double perovskites, while this coupling is found to be the leading term in the $d^1$ double perovskite Ba$_2$MgReO$_6$   by first principles methods (see \textbf{Fig.~\ref{fig_v_mult}} and the section on materials). Hence, we now turn to ab initio methodologies for multipolar ordering.

\vspace{4mm}

\subsection*{[H2] Total energy DFT-based methods}

Total-energy DFT-based approaches do not attempt to construct the low-energy MBEH (equation \ref{eq:MBEH}) explicitly. Instead, they treat all electronic degrees of freedom on equal footing and rely on the well-known precision and robustness of DFT~\cite{Lejaeghere2016} to capture the subtle contribution of hidden orders in the total energy. In this framework, an effective MBEH can be constructed {\it a posteriori} from the computed dependence of the total energy on order parameters. Strong electronic correlations are not properly described by standard DFT with local or semi-local treatment of exchange and correlations; hence, on-site Coulomb interactions need to be  treated explicitly, by adding the Hubbard $U$ term for the correlated shell or employing more advanced corrections such as hybrid functionals~\cite{Franchini2005, Franchini2011} and the self-interaction correction methods~\cite{Archer2011}.\\

{\it [H3] DFT+U.} The simplest and well-tested approach of this type, DFT+U~\cite{Anisimov1991}, consists of treating a shell with a local Hubbard interaction within the Hartree-Fock approximation, producing a static orbital-dependent contribution to the DFT one-electron potential and the corresponding correction to the total energy. 
Once DFT+U iterations are converged, the multipolar moments can be calculated from the one-electron density matrix $\rho^{1el}$ using one-electron multipolar operators $\hat{D}_k^q$~\cite{Santini2009,Bultmark2009,Kusunose2008,Lovesey2015,Hayami2018} (which are defined in the Supplemental Material). 

This framework has been intensively applied to multipolar orders in $f$-electron materials~\cite{Suzuki2018}. 
 Its main difficulty stems from the dependence of DFT+U on the initial guess for $\rho^{1el}$, since the DFT+U total energy exhibits multiple local minima in the space of possible order parameters~\cite{Shick2001,Larson2007,Gerald2008}. Hence, to obtain a correct multipolar order one generally needs to preset it by a correct initial guess for  $\rho^{1el}$ on all sites in the magnetic unit cell\cite{Dorado2009,Allen2014,Suzuki2018}. Moreover, since DFT+U always converges to a certain total-energy minimum, it is not possible to estimate the total energy dependence on a continuous change in the order parameters. The straightforward DFT+U also has difficulty treating the higher-temperature orders that only partially lift the GSM degeneracy (such as quarupolar orders, which preserve Kramer's degeneracy), because the Hartree-Fock approach representing the state by a single Slater determinant cannot describe multiplet physics. 

New DFT+U-based approaches that overcome or mitigate these limitations have recently been proposed. In particular, constrained versions of DFT+U\cite{Liu2015,Ma2015} have been devised to evaluate the total energy upon continuous evolution of the order parameters (dipole moments) in their phase space. To fix the moment of a given site $i$ along the predefined direction $\mathbf{M}^0_i$ an energy penalty
\beq\label{eq:constr_M}
E=E_0+\sum_i \gamma\left[\mathbf{M}_i-\mathbf{M}_i^0(\mathbf{M}_i^0\cdot\mathbf{M}_i)\right]^2,
\eeq
where $\gamma$ is the penalty energy parameter, is added to the DFT+U total energy. The corresponding "penalty field" then appears in the one-electron  Kohn-Sham potential acting to rotate the moment $\mathbf{M}_i$ towards the chosen direction. While the "penalty" energy contribution is unphysical, it becomes negligibly small once $\gamma$ is sufficiently large to force the moment to align along   $\mathbf{M}^0_i$. This approach has been first applied to calculate the MBEH of the $d^5$  $J_{\mathrm{eff}}=1/2$ iridate Sr$_2$IrO$_4$~\cite{Liu2015} and, subsequently, to estimate multipolar intersite interactions in  UO$_2$~\cite{Dudarev2019} and in the $J_{\mathrm{eff}}=3/2$ double perovskite Ba$_2$NaOsO$_6$~\cite{Fioremosca2021}.  
 In Sr$_2$IrO$_4$ and Ba$_2$NaOsO$_6$ the  magnetic order is  planar; the magnitude of the total-energy contribution from IEIs between various multipoles (equation \ref{eq:HIEI}) 
 is parameterized as a function of the ordered moments' canting angle $\phi$ in the $xy$ plane. 
 This method does not enable full separation of all bilinear contributions in equation (\ref{eq:HIEI}), since multipoles of different $K$ with the same $Q$ exhibit the same $\phi$-dependence. However, the method can identify multipolar Dzyaloshinskii-Moriya IEIs that arise due to inversion-symmetry breaking by lattice distortions 
  \cite{Fioremosca2021}. 
  
  Another version of DFT+U~\cite{Schaufelberger2023} in the spirit of standard constrained DFT~\cite{Dederichs1984}  adds the term 
  $\sum_{i kq} s^i_{kq}(\hat{w}_{kq}^i-\tilde{w}_{kq}^i)$
    to the DFT(+U) functional for a  chosen subset of sites $i$ and one-electron multipole operators\cite{Bultmark2009} 
      $\hat{w}_{kq}^i$ with target expectation values $\tilde{w}_{kq}^i$ . The Lagrange multiplier $s^i_{kq}$ is in fact a staggered multipole field  inducing corresponding  on-site  moments. Its   contribution to the total energy is then subtracted to obtain the physical total energy for the induced multipolar order.
 
  The constrained DFT+U methods \cite{Liu2015,Ma2015,Schaufelberger2023} described above cannot treat purely electric (such as quadrupolar)  orders, as they relay on magnetic polarization to obtain a non-degenerate ground state, which can be represented by a single Slater determinant, on correlated shells. 
 A recent development~\cite{Tehrani2021} addresses this limitation through an approach similar to the DFT+U treatment for non-spin orbit correlated paramagnetic systems with local moments~\cite{Yoon2019,Varignon2019}. Namely, atomic multiplet physics is mimicked  by a static magnetic disorder within a large supercell. The random orientation of the moments is enforced by the penalty term in equation (\ref{eq:constr_M}).  While time-reversal symmetry is broken locally at each site, it is globally preserved due to the random orientations of the moments, enabling the evaluation of quadrupole moments in a paramagnetic environment.  \\

{\it [H3] DFT+DMFT}.

 A more advanced treatment of electronic correlations is provided by combining DFT with DMFT~\cite{Georges1996,Anisimov1997_1,Lichtenstein_LDApp,Kotliar2006,Pavarini2011}. In this framework, a single correlated shell is embedded into a self-consistent electronic bath representing the rest of system; this "quantum impurity problem" is then solved beyond the static Hartree-Fock approximation, with local correlations fully encoded by a dynamical single-site potential: DMFT  self-energy $\Sigma_i(\omega)$. Solving the quantum impurity problem becomes increasingly more complex and time-consuming with increasing degeneracy of the correlated shell, reducing its symmetry and lowering the temperature, which corresponds exactly to the regime where multipolar orders in spin-orbit insulators arise. Direct DMFT calculations of multipolar ordered phases are thus  very rare~\cite{Haule2009}. The quadrupolar phase of the d$^1$ double perovskite Ba$_2$ReMgO$_6$ has been   recently studied by DFT+DMFT~\cite{Merkel2024}. In this case, also especially hard for DFT+U, direct DFT+DMFT calculations of the ordered phase were prohibitively computationally expensive. Hence, the relative stability of competing orders was estimated by applying suitable small staggered quadrupolar  fields in the symmetry-unbroken high-temperature phase; the transition temperatures were then extracted from Curie-Weiss fits of corresponding static susceptibilities. 

 \subsection*{[H2] Force-theorem methods}
 
 Force-theorem approaches are based on a general DFT "force theorem"~\cite{Mackintosh1980} that states that the change of DFT total energy upon a small perturbation is given by the corresponding change in the one-electron band energy only. The magnetic force theorem (MFT)~\cite{Liechtenstein1984,Liechtenstein1987} then evaluates  the change in DFT grand thermodynamic potential $\Omega$
  in an ordered magnetic state upon an infinitesimal tilting $\delta \Theta$  of two moments at the sites $i$ and $j$ away from the ground state direction.  
The resulting expression for intersite coupling in an effective classical Heisenberg model involves one-electron propagators  
between sites $i$ and $j$ as well as the on-site DFT exchange  field on the two sites.
The MFT formula 
can be straightforwardly  generalized to DFT+DMFT~\cite{Katsnelson2000} with 
the  DFT exchange  field substituted by the difference between DMFT spin-up and spin-down self-energies. MFT approaches of different flavors~\cite{Liechtenstein1987,Bruno2003,Kvashnin2015} have been widely used to investigate magnetic systems~\cite{Szilva2023} and its formalism generalized to treat Dzyaloshinskii-Moriya IEIs~\cite{Solovyev1996}. Its generalization to multipolar IEIs is, however, not straightforward. First, an ordered state for a given set of multipolar moments needs to be obtained before the MFT can be applied, which is a daunting task for correlated insulators. Second, multipolar operators do not have any obvious classical counterparts, and their effective Hamiltonian should naturally   be quantum.

The MFT was generalized to multipolar orders within DFT+U~\cite{Pi2014,Pi2014b}. This approach consists in starting with an ordered state and calculating the change in the DFT+U band energy upon simultaneous flipping of two multipole moments $\langle O_K^Q(i)\rangle$ and $\langle O_{K'}^{Q'}(j)\rangle$ on neighboring sites $i$ and $j$ while freezing the converged DFT+U Kohn-Sham potential. 
 The method can in principle extract  the IEIs for all the moments that are consistent with the single Slater-determinant ground state. As other DFT+U based techniques, this method has difficulties with time-even orders only partially lifting the GSM degeneracy, such as the purely quadrupolar order in the case of a half-integer $J_{\mathrm{eff}}$.\\

{\it [H3] Force theorem in Hubbard-I}. A reasonable starting point for treating correlated insulators can often  be obtained within a quasi-atomic Hubbard-I (HI)~\cite{Hubbard1963}  approximation. This DFT+HI approach~\cite{Lichtenstein_LDApp} is a substantial simplification over the general DFT+DMFT framework, because the hopping between the effective bath and quantum impurity is neglected. The DMFT impurity problem then reduces to diagonalizing a local Hamiltonian for a single correlated shell subjected to external (crystal or exchange) fields
, which contribute to the one-electron level positions
of the local Hamiltonian. The atomic Green's function and self-energy are then evaluated.
DFT+HI contains all single-shell multiplet effects and produces a rather good description of the electronic spectrum in local-moment paramagnets~\cite{Lebeque2005,Pourovskii2009,Shick2009,Locht2016}. The straightforward DFT+HI  is not suitable for ordered phases, because it neglects kinetic exchange, which  is contained in electronic hopping from the bath.

The "kinetic energy" term  
of the DFT+DMFT grand thermodynamic potential $\Omega$~\cite{Kotliar2006,Georges2004}, however, still contains all electron hopping terms through Kohn-Sham Hamiltonian $H_{\mathrm{KS}}$ even when the HI approximation is used for the  DMFT self-energy. 
 The force theorem in Hubbard-I (FT-HI)~\cite{Pourovskii2016} utilizes this fact to extract IEIs from the paramagnetic  (symmetry unbroken)  DFT+HI electronic structure. Namely, one considers fluctuations corresponding to the simultaneous appearance of small moments within the GSM of two sites in an otherwise paramagnetic environment and, similar to the MFT, evaluates the response of $\Omega$.   
 Introducing trace-conserving  density matrix fluctuations $\rho_{M_1M_2}$ within the GSM and neglecting, in the force-theorem spirit, their impact on the Kohn-Sham potential, the following formula was derived~\cite{Pourovskii2016}  for the matrix elements of the quantum MBEH (equation \ref{eq:HIEI}):
\beq\label{eq:V}
\langle M_1^iM_3^j|\hat{H}_{\mathrm{IEI}}| M_2^iM_4^j\rangle=\mathrm{Tr} \left[ G_{ij}\frac{\delta\Sigma_{j}}{\delta \rho_{M_3M_4}^{j}} G_{ji}\frac{\delta\Sigma_{i}}{\delta \rho_{M_1M_2}^{i}}\right],
\eeq
where $i$ and $j$ are site labels. The  FT-HI  structure  is 
very similar to the  MFT one, but uses the paramagnetic intersite propagators $G_{ij}$ together 
with the on-site "vertices", which are derivatives of the HI self-energy over fluctuations calculated with analytical formulas \cite{Pourovskii2016}. Once all MBEH matrix elements for a given bond $ij$ are computed, they are transformed to the form of interaction between multipoles (equation~\ref{eq:HIEI}) by using the orthonormality properties of multipolar operators $O_K^Q$.

The FT-HI method can thus calculate the IEI matrix (equation~\ref{eq:HIEI}) for all multipoles starting from a single self-consistent DFT+HI calculation for a paramagnetic phase. Its applicability is limited  to Mott insulating  phases, where the FT-HI formula (equation~\ref{eq:V}) reproduces the lowest-order in  hopping/U contribution to superexchange~\cite{Pourovskii2016}. The mapping to a multipolar MBEH also requires a specific choice of the phases for GSM states $|M\rangle$, that is, a proper pseudo-spin basis should be constructed \cite{Iwahara2018b}. 

\begin{figure}[t!]
\centering
\includegraphics[width=\linewidth]{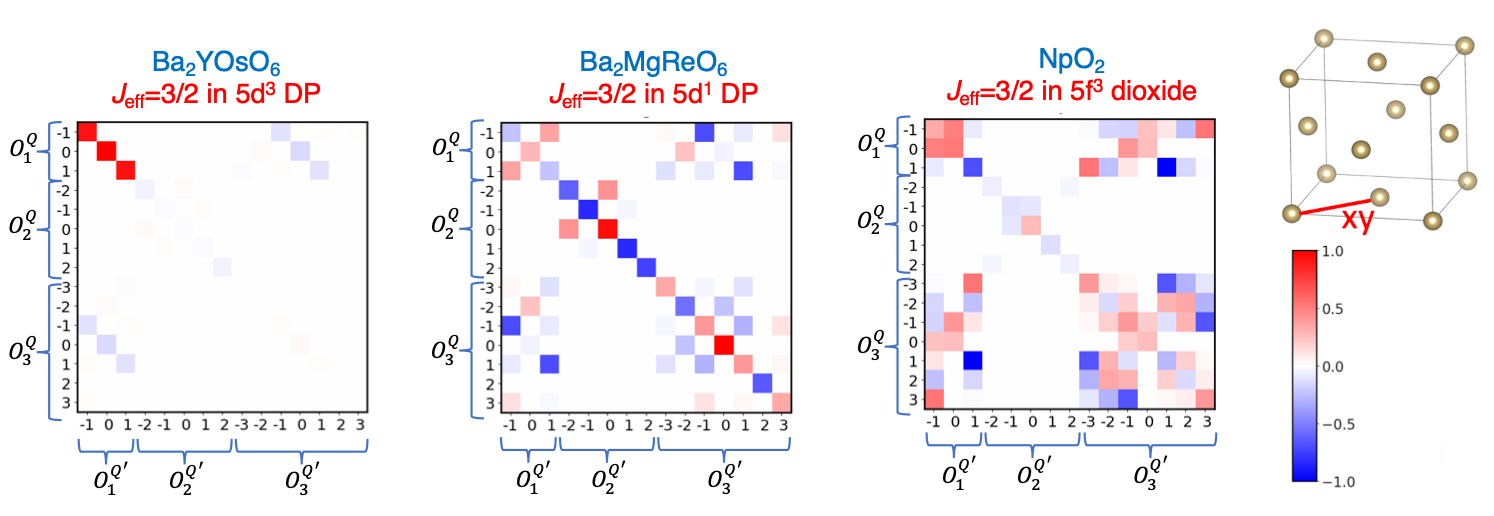}
\caption{\textbf{
Intersite exchange interaction (IEI) matrices for 
three $J_{\mathrm{eff}}$=3/2 SO insulators 
calculated by the force theorem in Hubbard-I (FT-HI) method.} \textbf{Left.} The  d$^3$ double perovskite (DP) Ba$_2$YOsO$_6$. In this case there is no spin-orbit entanglement within the $J_{\mathrm{eff}}$=3/2 ground state (GS), and the dominating IEI are  antiferroic dipole-dipole Heisenberg terms as expected for a pure spin-3/2 system. Perturbative admixture of excited multiplets by SO gives rise to weaker multipolar IEI, in particular, to IEI between dipoles ($K$=1) and octupoles ($K$=3) that induce a large gap in the magnetic excitation spectrum of  Ba$_2$YOsO$_6$ \cite{Pourovskii2023}. \textbf{Middle.} The d$^1$ DP Ba$_2$MgReO$_6$ exhibits a spin-orbit entangled $J_{\mathrm{eff}}$=3/2 GS. The states within this multiplet feature a strongly anisotropic charge distribution i.~e. large quadrupole (QP) moments. The QP-QP ($K$=2) IEI thus dominate in this system, leading to a competition between ferroic $t_{2g}$ and antiferroic $e_g$ electronic QP orders \cite{Soh2024,Fioremosca2024b}. \textbf{Right.} The $f^3$ dioxide NpO$_2$. Proliferation of non-zero  IEI in this case stems from complex shapes of $f-$electron orbitals leading to many non-zero contributions to superexchange. The time-odd (dipole-octupole) IEI are seen to dominate. These degrees of freedom within Np $J_{\mathrm{eff}}$=3/2 are in fact downfolded high-rank magnetic moments of the physical $J=$9/2 atomic GS of Np $f^3$; the dipole-octupolar order in the $J_{\mathrm{eff}}$ space of NpO$_2$ thus corresponds to a rank-5 (trikontadipole) order of physical multipoles\cite{Santini2006,Pourovskii2021f}.
All IEI matrices are plotted for the $\vR=[1/2,1/2,0]\equiv xy$ nearest-neighbor bond of the face-centered cubic (fcc) lattice shown in the top right corner.}
\label{fig_v_mult}
\end{figure}

The method has been extensively applied to multipolar $d$ and $f$ electron compounds. The IEI matrices calculated by FT-HI for several $J_{\mathrm{eff}}$=3/2 systems with magnetic fcc sublattice are displayed in \textbf{Fig.~\ref{fig_v_mult}}; a clear connection between the nature of the ground state multiplet and the structure of IEIs can be observed.  All IEI matrices in \textbf{Fig.~\ref{fig_v_mult}} exhibit no coupling between charge quadrupoles ($K=$2) and magnetic dipoles/octupoles, because the IEI Hamiltonian calculated in the paramagnetic state must be invariant under time reversal. Zero matrix elements within various $KK'$ blocks can be understood from a group theory analysis~\cite{Santini2009}, as demonstrated, for example, for a $d^1$ double perovskite~\cite{Fioremosca2024b}. \textbf{Fig.~\ref{fig_v_mult} }shows purely electronic IEIs; the  electron-lattice coupling can also be mapped into an effective interaction between quadrupoles~\cite{Gehring1975,kugel_khomskii,Polinger2009}o which is particularly important in the case of $d^1$ double perovskites, as discussed in the section on materials. Since electron-lattice quadrupole-quadrupole terms obey the same symmetry as electronic IEI ones, the electron-lattice quadrupole-quadrupole blocks feature the same non-zero matrix elements.  

Various other applications of the FT-HI are discussed in the section on materials.

\subsection*{[H2] Susceptibility-based approaches}

Second-order phase transitions can be detected by the divergence of the general static susceptibility $\chi_{K}^{Q}(\vq)=\langle O_K^Q(\vq)O_K^Q(-\vq)\rangle$ for a given primary order parameter $\langle O_K^Q\rangle$ at the ordering $\vq$-vector in the Brillouin zone. 

In the DMFT framework, the general static susceptibility $\chi(\vq) = \sum_{n,n'} X(i\omega_n,i\omega_{n'},\vq)$ is given by the two-electron Green's function, which is a function of two fermionic Matsubara frequencies and four $(m,\sigma)$ indices labeling the correlated-shell orbitals. This Green's function is obtained by solving a Bethe-Salpeter equation\cite{Jarell1992,Georges1996} :
\beq\label{eq:BS}
\hat{X}^{-1}(\vq)=\hat{X}_{\mathrm{imp}}^{-1}-\hat{X}_{\mathrm{imp},0}^{-1}+\hat{X}_0^{-1}(\vq),
\eeq
where the subscripts "$\mathrm{imp}$" and "0"  label the impurity and bare (single bubble) susceptibilities, respectively, and the hats label matrices in frequency indices. The bare susceptibilities are diagonal in the $nn'$ indices, while the two-particle Green's function $\hat{X}_{\mathrm{imp}}$ is a large non-diagonal matrix
with the rank rapidly increasing with the number of Matsubara points included and correlated shell degeneracy. Though the impurity susceptibilities can be efficiently calculated by some many-body methods~\cite{Gull2011}, solving the Bethe-Salpeter equation is numerically very heavy, especially for large shell degeneracies and in the low-temperature regime relevant to multipolar orders.

To tackle this problem, a single-value decomposition of the impurity susceptibility was performed~\cite{Otsuki2019}, keeping only the leading eigenvalue $\hat{X}_{\mathrm{imp}}(i\omega_n,i\omega_{n'}) \propto\phi_0(i\omega_n)\phi_0^*(i\omega_{n'})$ and showing that this approximation is valid for leading fluctuation channels in one and two-band Hubbard models in the strong-coupling limit. The susceptibility is then rewritten in a 
random-phase approximation (RPA)-like form, $\chi^{-1}(\vq) =[\chi_{\mathrm{imp}}^{-1}-I(\vq)]$, where the IEI  reads
 $I(\vq)=\mathrm{Tr}_n |\phi_0(i\omega_n)|^2 \left(\hat{X}_{\mathrm{imp},0}^{-1}(i\omega_n)-\hat{X}_0^{-1}(\vq,\omega_n)\right)$.
The method was applied~\cite{Otsuki2024}, in conjunction with DFT+HI, to multipolar interactions in CeB$_6$.

\subsection*{[H2] Cluster methods}\label{sec:cluster_methods}

Cluster methods consider a many-electron problem on finite-size clusters of correlated sites with its parameters (hopping between the sites and their local Hamiltonian) derived in an ab initio way for a continuous system. Subsequently, within a strong-coupling perspective suitable for correlated insulators, the single-site terms are treated exactly, while the hopping within the cluster is typically included perturbatively to obtain the IEI between the cluster's sites. To derive the kinetic exchange IEI, a pair of correlated sites is typically considered.

A generalization of this traditional approach to superexchange~\cite{Anderson1959,Anderson1963} to the case of multipolar interactions~\cite{Mironov2003,Iwahara2015,Iwahara2022} includes all active orbitals on correlated sites into the cluster Hamiltonian that reads
$$
H_{\mathrm{cl}}=\sum_{il}H_{\mathrm{loc}}^{il}+H_{\mathrm{C}}^{\mathrm{int}}+H_{t}=H_0+V
$$
where local terms (Eq.~\ref{eq:Hloc}) are included for active orbitals $l$ on each correlated site $i$, and $H_{\mathrm{C}}^{\mathrm{int}}$ and $H_{t}$ are the intersite Coulomb interaction and hopping terms within the cluster, respectively. On the right-hand side, those terms are regrouped into those that contribute to the splitting within the GSM of correlated shells ($V$) and those that do not ($H_0$). Subsequently, by projecting $H_{\mathrm{cl}}$ to the manifold of GSM, 
$P_0=\bigotimes_i\left(\sum_M |iJM\rangle\langle iJM|\right)$, the MBEH is evaluated. It includes crystal field splitting within the GSM as well as various intersite terms. In particular, the kinetic exchange is evaluated within the second-order perturbation theory
$$
H_{\mathrm{KE}}=P_0H_t\left(\sum_{s \notin \mathrm{GSM}}P_s\frac{1}{E_0-H_0}P_s\right)H_t P_0,
$$
where the projection $P_s$ runs over all states outside of the GSM manifold, and $E_0$ is the GSM eigenenergy within $H_0$ (which is the same for all states within the GSM manifold). 
In addition to the kinetic exchange, other contributions like direct exchange are also included. 

Another approach~\cite{Voleti2021}, which is suitable for 
small clusters, performs an exact diagonalization  of $H_{\mathrm{cl}}$ followed by a unitary (Schrieffer-Wolff \cite{Schrieffer1966}) transformation $\mathcal{R}_{P\to P_0}$ from the low-energy subspace $P$ 
 of the dimension ($(2J_{\mathrm{eff}}+1)^2$) of a two-site cluster to  the GSM subspace $P_0$ of two disconnected sites defined as above.   The resulting Hamiltonian within the $P_0$ subspace then becomes $H_{\mathrm{eff}}=P_0 \mathcal{R}_{P \to P_0} H_{\mathrm{cl}} \mathcal{R}_{P \to P_0}^{\dagger}P_0$. This Hamiltonian $H_{\mathrm{eff}}$ contains on-site and IEI terms; the latter can be again transformed to the multipolar form (equation \ref{eq:HIEI}).
 Several similar methods based on exact diagonalization~\cite{Beom2012,Winter2016,Beom2019} also perform a mapping between non-interacting space $P_0$ and $P$ by expanding $P$ into the basis of pseudo-spin states $P_0$.

\subsection*{[H2] Approaches to electron-lattice interactions}

The electron-lattice term in (Eq.~\ref{eq:MBEH}) can provide a crucial contribution to the energetics of multipole orders~\cite{Iwahara2023,Soh2024,Fioremosca2024}. It arises due to the  Jahn-Teller coupling between an electronic quadrupole operator and local distortions of the same symmetry (both labeled by their irreducible representation $\Gm$ and projection $Q$).
The electron-lattice term consists of single-site and intersite contributions~\cite{Gehring1975}:

\def\hbar{{\mathchar'26\mkern-9mu h}}
\beq\label{eq:Hel}
H_{\mathrm{EL}}=H_{\mathrm{harm}}+H_{\mathrm{JT}}+H_{\mathrm{EC}}=
 \sum_{i\Gm Q} \left(\protect \hbar \omega_{\Gm}(\hat{n}_{\Gm Q}^i+1/2)- g_{\Gm} O^{\Gm Q}_2 (i) q_{\Gm Q} (i)\right) +\sum \Phi_{\Gm \Gm'}^{QQ'} (ij) q_{\Gm Q} (i) q_{\Gm' Q'} (j),
\eeq
where $\Phi_{\Gm \Gm'}^{QQ'}(ij)$ is the intersite elastic coupling, also termed the force constant matrix between irreducible local distortion modes on correlated sites $i$ and $j$, and the summation over all repeated indices is implied in the last term. Hence, to define the electron-lattice interaction one needs to determine all so-called vibronic parameters, that is, the local force constant matrix $\Phi_{\Gm \Gm'}^{QQ'}(ii)$ that determines the single-site oscillator strength $\omega_{\Gm}$,  the intersite one  $\Phi_{\Gm \Gm'}^{QQ'}(ij)$, and the  Jahn-Teller couplings $g_{\Gamma}$.

In the context of multipolar order in spin-orbit double perovskites, a framework for single-site terms in the electron-lattice Hamiltonian was developed ~\cite{Iwahara2018} based on evaluating the adiabatic potential energy surface, that is, the electronic energy as a function of fixed ionic coordinates $\{q_{\Gm Q}\}$, by ab initio DFT or quantum chemisty methods. This potential energy surface was calculated at a grid in the $\{q_{\Gm Q}\}$ space and then fitted with the expected form~\cite{bersuker_vibronic_1989} for $t_{2g}$ electrons on a transition metal ion coupling to all $e_g$ and $t_{2g}$ symmetry distortions of a ligand octahedron surrounding it. Going beyond the linear  Jahn-Teller coupling in equation (\ref{eq:Hel}),  higher-order  Jahn-Teller couplings were included.

Alternative and computationally less demanding approaches~\cite{Soh2024,Fioremosca2024} are based on DFT or DFT+HI calculations for continuous systems to extract the linear  Jahn-Teller coupling only. In particular, the DFT total energy as well as the quadrupoles of the correlated shell were calculated~\cite{Soh2024} 
upon the $\vk$-space distortion mode  amplitude  $q_{\vk \theta}$, where $\theta$ labels irreducible distortion modes at a given $\vk$ point. By writing the total energy as the sum of a bilinear  Jahn-Teller term and a quadratic elastic term~\cite{Peil2019,Georgescu2019},  
  Jahn-Teller couplings and corresponding elastic constants
are obtained from a fit of total energy vs $q_{\vk}^{\theta}$.

Jahn-Teller couplings in DFT+HI were calculated by a similar approach of applying small distortion modes~\cite{Fioremosca2024}. The splitting of the GSM levels that are degenerate in the undistorted structure is due to the  Jahn-Teller term $g_{\Gm}  q_{\Gm Q} O^{\Gm Q}_2$. Since the GSM eigenvalues and states are directly computed in the HI approach, the magnitude of the coupling constant can be simply extracted from a linear fit of the splitting versus distortion amplitude. The electronic quadrupole IEI cannot contribute to this splitting within the HI approximation, since the electron hopping inducing them is neglected while solving the impurity problem. The spurious self-interaction contribution to the  Jahn-Teller splitting is eliminated by charge-density averaging within GSM~\cite{Delange2017}.
The force constant matrix was evaluated 
by  standard DFT methods~\cite{Baroni2001} and then projected into the space of irreducible displacement modes $q_{\Gm Q}(i)$~\cite{Fioremosca2024}.

\subsection*{[H2] Approaches to electron-lattice interactions}

The electron-lattice term in (Eq.~\ref{eq:MBEH}) can provide a crucial contribution to the energetics of multipole orders~\cite{Iwahara2023,Soh2024,Fioremosca2024}. It arises due to the  Jahn-Teller coupling between an electronic quadrupole operator and local distortions of the same symmetry (both labeled by their irreducible representation $\Gm$ and projection $Q$).
The electron-lattice term consists of single-site and intersite contributions~\cite{Gehring1975}:

\def\hbar{{\mathchar'26\mkern-9mu h}}
\beq\label{eq:Hel}
H_{\mathrm{EL}}=H_{\mathrm{harm}}+H_{\mathrm{JT}}+H_{\mathrm{EC}}=
 \sum_{i\Gm Q} \left(\protect \hbar \omega_{\Gm}(\hat{n}_{\Gm Q}^i+1/2)- g_{\Gm} O^{\Gm Q}_2 (i) q_{\Gm Q} (i)\right) +\sum \Phi_{\Gm \Gm'}^{QQ'} (ij) q_{\Gm Q} (i) q_{\Gm' Q'} (j),
\eeq
where $\Phi_{\Gm \Gm'}^{QQ'}(ij)$ is the intersite elastic coupling, also termed the force constant matrix between irreducible local distortion modes on correlated sites $i$ and $j$, and the summation over all repeated indices is implied in the last term. Hence, to define the electron-lattice interaction one needs to determine all so-called vibronic parameters, that is, the local force constant matrix $\Phi_{\Gm \Gm'}^{QQ'}(ii)$ that determines the single-site oscillator strength $\omega_{\Gm}$,  the intersite one  $\Phi_{\Gm \Gm'}^{QQ'}(ij)$, and the  Jahn-Teller couplings $g_{\Gamma}$.

In the context of multipolar order in spin-orbit double perovskites, a framework for single-site terms in the electron-lattice Hamiltonian was developed ~\cite{Iwahara2018} based on evaluating the adiabatic potential energy surface, that is, the electronic energy as a function of fixed ionic coordinates $\{q_{\Gm Q}\}$, by ab initio DFT or quantum chemisty methods. This potential energy surface was calculated at a grid in the $\{q_{\Gm Q}\}$ space and then fitted with the expected form~\cite{bersuker_vibronic_1989} for $t_{2g}$ electrons on a transition metal ion coupling to all $e_g$ and $t_{2g}$ symmetry distortions of a ligand octahedron surrounding it. Going beyond the linear  Jahn-Teller coupling in equation (\ref{eq:Hel}),  higher-order  Jahn-Teller couplings were included.

Alternative and computationally less demanding approaches~\cite{Soh2024,Fioremosca2024} are based on DFT or DFT+HI calculations for continuous systems to extract the linear  Jahn-Teller coupling only. In particular, the DFT total energy as well as the quadrupoles of the correlated shell were calculated~\cite{Soh2024} 
upon the $\vk$-space distortion mode  amplitude  $q_{\vk \theta}$, where $\theta$ labels irreducible distortion modes at a given $\vk$ point. By writing the total energy as the sum of a bilinear  Jahn-Teller term and a quadratic elastic term~\cite{Peil2019,Georgescu2019},  
  Jahn-Teller couplings and corresponding elastic constants
are obtained from a fit of total energy vs $q_{\vk}^{\theta}$.

Jahn-Teller couplings in DFT+HI were calculated by a similar approach of applying small distortion modes~\cite{Fioremosca2024}. The splitting of the GSM levels that are degenerate in the undistorted structure is due to the  Jahn-Teller term $g_{\Gm}  q_{\Gm Q} O^{\Gm Q}_2$. Since the GSM eigenvalues and states are directly computed in the HI approach, the magnitude of the coupling constant can be simply extracted from a linear fit of the splitting versus distortion amplitude. The electronic quadrupole IEI cannot contribute to this splitting within the HI approximation, since the electron hopping inducing them is neglected while solving the impurity problem. The spurious self-interaction contribution to the  Jahn-Teller splitting is eliminated by charge-density averaging within GSM~\cite{Delange2017}.
The force constant matrix was evaluated 
by  standard DFT methods~\cite{Baroni2001} and then projected into the space of irreducible displacement modes $q_{\Gm Q}(i)$~\cite{Fioremosca2024}.

\begin{table}
\centering
\renewcommand{\arraystretch}{1.3}
\begin{tabular}{|l|c|c|}
\hline
\multicolumn{3}{|c|}{\textbf{Theories and methods to explore multipolar magnetism and hidden orders}}\\
\hline

\textbf{Properties} & \textbf{Theory}  & \textbf{Materials}\\\hline

\emph{IEI from the ordered phase} & Magnetic Force Theorem~\cite{Pi2014} & LaMnO$_3$~\cite{Solovyev1996}, UO$_2$~\cite{Pi2014, Pi2014b} \\ 
  & Constrained DFT + U~\cite{Liu2015} & UO$_2$~\cite{Dudarev2019}, Ba$_2$NaOsO$_6$~\cite{Fioremosca2021} \\ \hline

\emph{IEI from the paramagnetic phase} 
& Cluster perturbation theory~\cite{Mironov2003,Iwahara2015,Iwahara2022}  & UO$_2$~\cite{Mironov2003}, NdN~\cite{Iwahara2022} \\
& Exact diagonalization~\cite{Kim2012,Voleti2021} & 
  Ba$_2M$OsO$_6$ ($M$ = Ca, Mg, Zn)~\cite{Voleti2021} \\
 & & Ba$_2$NaOsO$_6$~\cite{Fioremosca2021} \\
& Force Theorem in Hubbard-I~\cite{Pourovskii2016}   & 
UO$_2$~\cite{Pourovskii2019}, NpO$_2$~\cite{Pourovskii2021f}, PrO$_2$~\cite{Khmelevskyi2024}, Ba$_2$MgReO$_6$~\cite{Fioremosca2024b}, \\ 
& &  Ba$_2$Y$M$O$_6$ ($M$ = Os, Ru)~\cite{Pourovskii2023} \\
& & Ba$_2M$OsO$_6$ ($M$ = Ca, Mg, Zn) \cite{Pourovskii2021} \\
& DMFT susceptibility & \\
 &in strong coupling limit\cite{Otsuki2019} & CeB$_6$~\cite{Otsuki2024} \\
 \hline

 \emph{Constrained multipoles}  & Polymorphous DFT+U~\cite{Trimarchi2018,Yoon2019} & Ba$_2$MgReO$_6$~\cite{Tehrani2021}\\
 & DFT + HI + Constrained DFT + U~\cite{FioreMosca2022, Schaufelberger2023} & Ba$_2M$OsO$_6$  ($M$ = Ca, Mg, Zn)~\cite{FioreMosca2022} \\
&  Constrained DFT + U & KCu$_{1-x}$Zn$_x$F$_3$~\cite{Schaufelberger2023}  \\

\hline

\emph{Multipolar susceptibilities} & Classical Monte Carlo~\cite{Voleti2021, Voleti2023}  & Ba$_2$Ca$_{1-\delta}$Na$_{\delta}$OsO$_6$~\cite{Voleti2023} \\ 
& DFT + DMFT~\cite{Soh2024} & Ba$_2$MgReO$_6$~\cite{Soh2024, Merkel2024}  \\
\hline

\emph{Electron-lattice interactions} & Cluster~\cite{Iwahara2018b, Soh2024} & d$^1$ Double perovskites~\cite{Iwahara2018b, Iwahara2023}, Ba$_2$MgReO$_6$~\cite{Soh2024} \\
& DFT, DFT + Hubbard-I~\cite{Fioremosca2024b} & Ba$_2$MgReO$_6$~\cite{Fioremosca2024b} \\ 
 \hline

\emph{Multipolar contributions to E$_H$ and E$_x$} & DFT+U~\cite{Bultmark2009} & Actinides~\cite{Bultmark2009} \\ \hline

\emph{Excitation spectra} & Mean field + RPA~\cite{RareEarthMag_book,Rotter2012} & \\ 
& +multipolar form-factors~\cite{Shiina2007,Pourovskii2021} &  Ba$_2M$OsO$_6$  ($M$ = Ca, Mg, Zn)~\cite{Pourovskii2021}, \\ 
& &  Ba$_2$Y$M$O$_6$ ($M$ = Os, Ru)~\cite{Pourovskii2023} \\ 
\hline

\end{tabular}
\caption{\label{tab:methods} \textbf{Theoretical approaches to multipolar magnetism and hidden orders in materials.} For each target property (left column) we list theoretical approaches developed for its calculation (middle column). Applications of each approach to specific materials are listed in the right column.}
\end{table}

\subsection*{[H2] Solving the many-body effective Hamiltonian and calculating properties}

The quantum ab initio MBEH (equation \ref{eq:MBEH}) for the system at a given volume that includes electronic and, possibly, electron-lattice terms, needs to be solved to obtain ordered phases as a function of temperature and external fields (such as an applied magnetic field). While there is a vast array of advanced methods to solve quantum  spin-1/2 models, especially in low dimensions~\cite{quantum_magnetism_2004,Sandvik2010,Schollwock2011}, the choice is much more limited for multipolar orders. The size of the Hilbert space rapidly increases for $J_{\mathrm{eff}}> 1/2$; moreover, the IEI Hamiltonian (equation \ref{eq:HIEI}) typically has rather low symmetry, unlike typical spin-$1/2$ models, where such symmetries (such as total $\langle S_z\rangle$ conservation) are employed to partition the Hilbert space into subblocks. Mapping the full multipolar MBEH into a classical model is generally also not possible.

Hence, multipolar MBEHs are typically solved using a generalized single-site mean-field approximation~\cite{RareEarthMag_book,Rotter2012}. 
The intesite terms in equation (\ref{eq:HIEI})  are mean-field decoupled
, that is, products of the fluctuation terms are neglected. This leads to the mean-field Hamiltonian 
$\sum_{\alpha KQ} \left( O_K^Q(\alpha) -1/2 \langle O_K^Q\rangle_{\alpha}\right)V^{\mathrm{MF}}_{KQ}(\alpha)$ , where each site $\alpha$ in the magnetic unit cell interacts with the Weiss field  $V^{\mathrm{MF}}_{KQ}(\alpha)$ produced by the rest of systems. This mean-field problem is solved self-consistently to obtain the set of order parameters $\{\langle O_K^Q\rangle\}$ and the free energy $F_{\mathrm{MF}}$ for a set of possible ordered structures identifying the one minimizing $F_{\mathrm{MF}}$ at a given temperature.  

The electron-lattice term (equation \ref{eq:Hel}) can be mapped into effective lattice-mediated quadrupole-quadrupole coupling~\cite{Gehring1975,kugel_khomskii,}, but this treatment neglects dynamical   Jahn-Teller effects. Alternatively, the mean-field approximation can be extended to electron-lattice terms~\cite{Englman1970} by an analogous mean-field decoupling of the intersite elastic term in eq.~\ref{eq:Hel}. 
A  vibronic single-site   basis that consists of the electronic wavefunction and local phonon modes 
is introduced \cite{bersuker_vibronic_1989}  
and the  Jahn-Teller single-site problem in electronic and elastic mean field is solved by exact diagonalization~\cite{Iwahara2018}.
When the  Jahn-Teller couplings $g$ are sufficiently weak or strong,  weak or strong-coupling perturbative approaches can be employed\cite{Iwahara2023,Fioremosca2024}. 

As in the case of conventional magnetic orders, the mean-field approximation leads to a systematic overestimation of transition temperatures. The overestimation factor depends on the system's effective dimensionality and lattice geometry. In particular, overestimation by about a factor of two was observed in a number of 3D  multipolar systems studied with a FT-HI Hamiltonian solved by mean-field~\cite{Pourovskii2019,Pourovskii2021f,Pourovskii2023,Khmelevskyi2024}. In those  cases, the magnetic sublattice is fcc and thus frustrated with respect to antiferromagnetism. Alternatively, a classical model can be introduced for a subset of multipoles, such as quadrupoles or octupoles, and solved by classical Monte Carlo~\cite{Khaliullin2021,Voleti2023}.

Once a stable order is identified,  single-site excitation within the GSM space can also be calculated from mean-field eigenstates using the Fermi golden rule, $\sum P_i |\langle f| \mathcal{V}| i \rangle|^2 \delta(\omega -E_{f}-E_{i})$, where $P_i$ is the thermal occupation probability for the initial mean-field state  $i$, $\mathcal{V}$ is an appropriate scattering operator for a given probe (such as Raman~\cite{Devereaux2007} or resonant inelastic X-ray scattering (RIXS)~\cite{Kim2017rixs}).
To evaluate  $\vq$-dependent excitations measured, for example, by inelastic electron scattering (INS) or RIXS, a generalized RPA can be employed~\cite{RareEarthMag_book,Rotter2012} to obtain generalized lattice susceptibility.
Subsequently, the fluctuation-dissipation theorem is used to obtain from it the relevant scattering cross-section. Appropriate
form-factors 
for scattering from multipoles can be obtained by expanding,  within the GSM, the  
corresponding scattering  operator 
in multipolar operators $ O_K^Q$~\cite{Lovesey_book,Shiina2007,Pourovskii2021}.

Finally, electronic excitations in a wide energy range can be affected by the ordering of the low-energy degrees of freedom that are included in the MBEH. These effects are outside of the GSM space and, hence, cannot be obtained from the MBEH alone. Instead,  a mean-field ordered state found by solving the MBEH can be utilized to provide a proper initial guess for all electron DFT+U or DFT+DMFT calculations. A scheme was implemented within DFT+U to initialize the one-electron density matrix $\rho^{1el}$ in accordance with a multipolar order derived from a FT-HI MBEH.
Once an appropriate $\rho^{1el}$ is constructed, it is used as an initial guess for DFT+U~\cite{FioreMosca2022}. This typically results into DFT+U converging to the local minimum in total energy corresponding to a given multipolar order; its electronic spectrum is then calculated.

A summary of representative results obtained using the approaches discussed in this section is provided in \textbf{Table~\ref{tab:methods}}.

\section*{[H1] Materials}

After reviewing theories and methods for studying multipolar hidden orders, we now turn to a survey of the emergence of multipolar phases in materials, focusing on two main materials classes: 5$d$ double perovskites and $f$-electron systems.

\subsection*{[H2] $d$-electrons: Double Perovskites}

The emergence of multipolar degrees of freedom relies on having atomic orbital degeneracy, electron interactions, 
and strong spin-orbital entanglement. These
conditions are met in oxides of heavy $4d$/$5d$ transition metals with partially filled $t_{2g}$ orbitals.
Due to the larger spread of the $4d$/$5d$ atomic wavefunctions, which tends to favor metallicity,
the route to access localized multipole moments is to explore 
crystals where the heavy $4d/5d$ transition metal ions are spaced farther apart than in typical perovskite materials. 
Ordered double perovskites with the chemical formula {A$_2$BB$^\prime$O$_6$ provide ideal platforms for such phenomena. 
Here, B is occupied by an electronically inert ion, and B$^\prime$ by magnetically active $4d$/$5d$ ions. In these compounds, the B and B$^\prime$ sites form two sublattices of a cubic lattice, resulting in a B$^\prime$--B$^\prime$ spacing that is $\sqrt{2}$ times greater than in single perovskites.
Moreover, the relatively large on-site Hubbard interaction $U$, in the range of 2–4 eV, combined with substantial SOC strength of a few hundred meV, further facilitates the development of multipolar phases. Consequently, these perovskite compounds are commonly classified as "multipolar Mott insulators."~\cite{Chen2010,Lu2017,Hirai2020,Paramekanti2020,Fioremosca2021,Pourovskii2021}
Indeed, various compounds in this class, featuring electron fillings such as $d^1$, $d^2$, and $d^3$, have been experimentally and theoretically shown to exhibit clear evidence of distinct multipolar degrees of freedom.

Early studies utilizing RIXS spectra in conjunction with single-site or cluster exact diagonalization successfully extracted the effective Hund's coupling and SOC in these systems. These investigations provided insights into the ground-state and excited-state multiplet structures, laying the groundwork for the identification of multipolar phases in this class of materials~\cite{Yuan2017, Taylor2017, Paramekanti2018, Nag2018}.

\begin{figure}
\centering
\includegraphics[width=\linewidth]{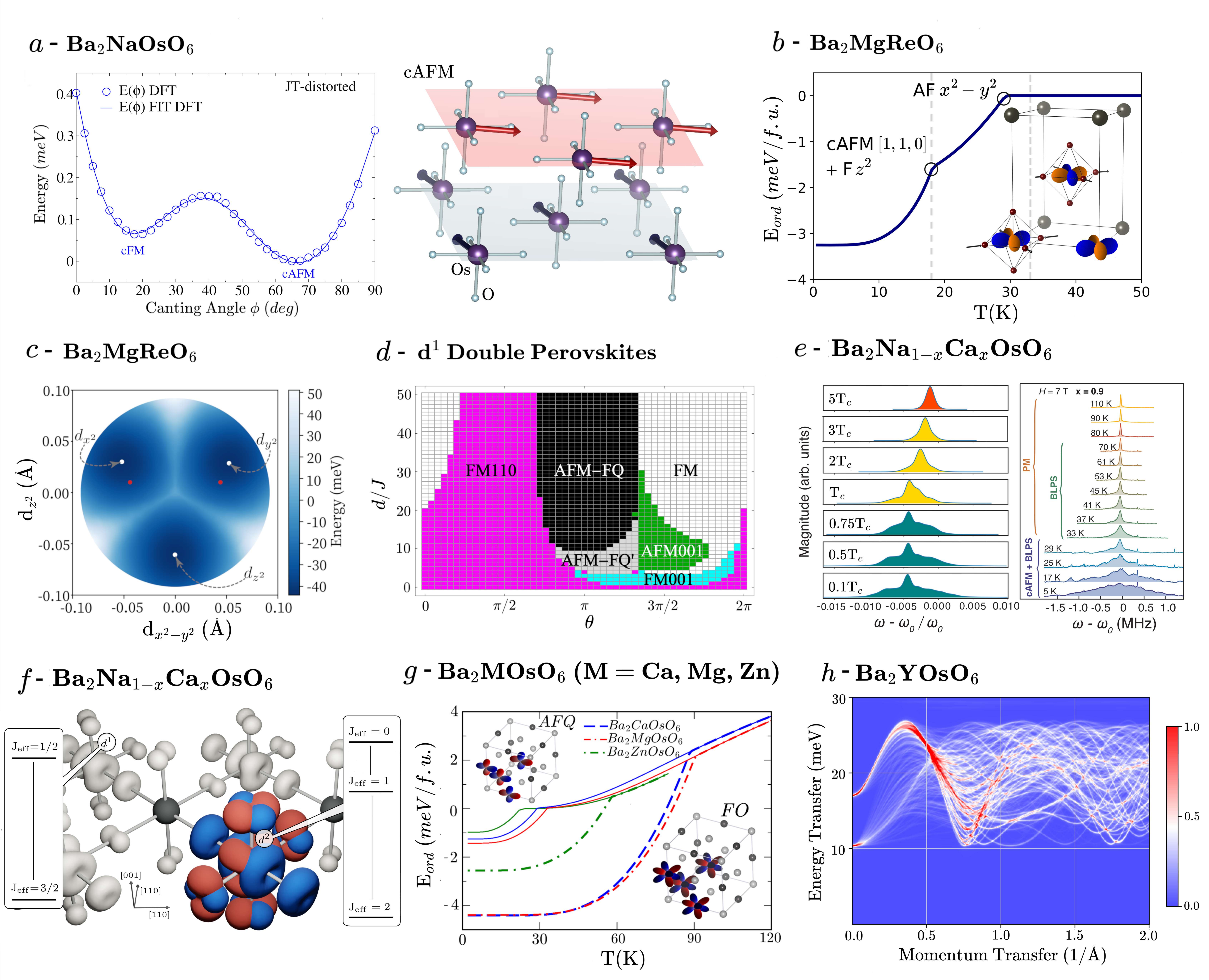}
\caption{\textbf{Multipolar properties of d-electron systems computed with different methods} \textbf{a} | Energy as a function of the canting angle for the 5d$^1$ double perovskite (DP) Ba$_2$NaOsO$_6$ in the Jahn-Teller (JT) distorted phase with constrained DFT + U~\cite{Fioremosca2021} (left panel);  graphical representation of the canted anti-ferromagnetic (cAFM) phase (right panel)~\cite{Fioremosca2021}. \textbf{b} | Mean field ordering energy vs temperature for the 5d$^1$ DP Ba$_2$MgReO$_6$ with low-energy effective Hamiltonian including both superexchange and vibronic interactions, as computed with force theorem in Hubbard-I (FT-HI) and  DFT(+HI) methods, respectively. The anti-ferro quadrupolar AF$x^2-y^2$ phase is shown in the inset~\cite{Fioremosca2024b}. \textbf{c} | The potential energy surface of the ReO$_6$ octahedron in the space of E$_g$ JT normal modes in 5d$^1$ DP Ba$_2$MgReO$_6$ calculated by quantum chemistry methods~\cite{Soh2024}. The result is compared with the experimental static octahedral distortions plotted as red dots.~\cite{Soh2024} \textbf{d} | Spin-vibronic phase diagram of 5d$^1$ DPs as a function of magnetic ($J$) and vibronic (d, $\theta$) parameters as obtained within a mean-field approximation~\cite{Iwahara2023}. \textbf{e} | Computed $^{23}$Na nuclear magnetic resonance (NMR) spectra displaying the shift and broadening of the NMR peak while lowering temperature  for lightly Na doped 5d$^2$ DP Ba$_2$CaOsO$_6$ (left panel)~\cite{Voleti2023}; experimental NMR spectra for comparison with Na concentration x = 0.9 (right panel)~\cite{Cong2023}.
\textbf{f} | The magnetic density of the 5d$^2$ polaron coexisting with the charge density of the 5d$^1$ sites the Ca doped 5d$^1$ DP Ba$_2$NaOsO$_6$ as computed in DFT+U~\cite{Celiberti2024}. \textbf{g} | Mean field ordering energy vs temperature for the 5d$^2$ DPs Ba$_2$MOsO$_6$ (M = Ca, Mg, Zn) obtained from a many-body effective Hamiltonian computed with DFT+HI and FT-HI. The ferro-octupolar/anti-ferro quadrupolar (FO/AFQ) are graphically represented in the insets~\cite{Pourovskii2021}  \textbf{h} |  The  spherically-averaged inelastic neutron scattering (INS) intensity for the  5d$^3$ DP Ba$_2$YOsO$_6$ computed for the ground-state 2\vq\ AFM  phase from FT-HI intersite exchange interactions within the random-phase approximation.~\cite{Pourovskii2023}.
}
\label{fig:d}
\end{figure}

\subsubsection*{[H3] $d^1$}
$d^1$ double perovskites are well-studied and include
Ba$_2$LiOsO$_6$, \cite{Stitzer2002,Erickson2007,Steele2011} Ba$_2$NaOsO$_6$, \cite{Stitzer2002,Erickson2007,Steele2011}  Ba$_2$YMoO$_6$, \cite{Carlo2011}
Ba$_2$CaReO$_6$, \cite{Yamamura2006,Ishikawa2021a}  Ba$_2$CaOsO$_6$, \cite{Yamamura2006}
Ba$_2$MgReO$_6$, \cite{Marjerrison2016,Hirai2019,Hirai2020} and Ba$_2$ZnReO$_6$. \cite{Marjerrison2016} While early work on many of these
nominally cubic double perovskites with a $J=3/2$ ground state multiplet reported a single low-temperature magnetic transition 
into weakly ordered antiferromagnetic or ferromagnetic phases with small ordered moments, more recent work has revealed that there is
a higher-temperature transition at which the crystal point group symmetry is subtly broken, splitting the 
$J=3/2$ quartet into two Kramers doublets. This weak crystal symmetry lowering occurs either through local distortions, as 
seen in Ba$_2$NaOsO$_6$ via NMR and nuclear quadrupole resonance experiments as well as DFT  calculations,\cite{Lu2017,LiuW2018,Willa2019,Kesavan2020,Barbosa2022, Fioremosca2021}, or through global symmetry breaking, as seen 
via high-resolution X-ray diffraction measurements of tiny lattice displacements in Ba$_2$MgReO$_6$.\cite{Hirai2019,Hirai2020,Arima2022,Soh2024} 
Such symmetry breaking has been proposed to arise from ordering 
of quadrupolar degrees of freedom $\vec Q =(J_x^2-J_y^2, 3 J_z^2-J^2)$ due to orbital charge repulsion.\cite{Chen2010,Svoboda2021} 
However, more recent work 
assigned a more central role
to the lattice degrees of freedom to explain the symmetry-breaking 
pattern seen in Ba$_2$MgReO$_6$, invoking a cooperative spin-orbit Jahn-Teller distortion of phonon modes coupled to these
quadrupolar moments.\cite{Fioremosca2024,Fioremosca2024b,Soh2024,Frontini2023,Zivkovic2024, Iwahara2024} 
In certain materials,
such as Ba$_2$YMoO$_6$,\cite{Aharen2010, deVries2010} which shows no sign of a magnetic transition down to the lowest temperature, dynamic Jahn-Teller fluctuations
might prevent magnetic ordering, leading to exotic spin liquid states, although a valence bond glass has also been proposed as a
possible explanation for these observations.

\textbf{Figure \ref{fig:d}} presents examples of predicted multipolar properties. Magnetically constrained non-collinear DFT successfully identifies the canted antiferromagnetic phase with a canting angle of $\phi = 66^\circ$ as the ground state of Jahn-Teller-distorted Ba$_2$NaOsO$_6$ (\textbf{Figure \ref{fig:d}a}), in excellent agreement with NMR measurements~\cite{Lu2017, Fioremosca2021}. The vibronic mechanisms underlying the antiferroic quadrupolar hidden order in Ba$_2$MgReO$_6$ have been elucidated using a low-temperature many-body effective Hamiltonian (equation \ref{eq:MBEH}) incorporating electronic and electron-lattice coupling mechanisms~\cite{Fioremosca2024b}. This Hamiltonian is derived from ab initio calculations by combining FT-HI superexchange interactions with DFT(+HI) elastic and Jahn-Teller couplings. This approach reproduces the two phase transitions observed in thermodynamics, XRD and REXS experiments~\cite{Marjerrison2016,Hirai2020,Soh2024},  which correspond to the breaking of cubic symmetry and the subsequent emergence of an antiferroic quadrupolar (AFQ) phase and a canted AFM phase (\textbf{Figure \ref{fig:d}b}).

The role of electron-lattice coupling in  Ba$_2$MgReO$_6$ and the regime of Jahn-Teller problem relevant for this compound are the subjects of ongoing debate. In particular, a weak-coupling Jahn-Teller regime and a subtle interplay of electron-lattice and superexchange interactions were predicted~\cite{Fioremosca2024b} to stabilize the observed AFQ order. In contrast, other work~\cite{Iwahara2023} assumed a strong Jahn-Teller regime, that is, the ReO$_6$ Jahn-Teller system tunneling between the minima of the Jahn-Teller potential energy surface. This potential energy surface with respect to local Jahn-Teller distortion computed by quantum-chemical
approaches~\cite{Soh2024} is shown in \textbf{Fig.~\ref{fig:d}c}; it reveals three degenerate minima, in qualitative consistency with the experimental distortion amplitudes~\cite{Hirai2020}. Using the strong Jahn-Teller-regime approximation~\cite{Iwahara2023} various quadrupole orders were obtained, as well as 
a global diagram of magnetically ordered states in $d^1$ double perovskites (\textbf{Fig.~\ref{fig:d}d)}.  
In this picture, 
the emergence of various ordered phases is mainly driven by the cooperative effects of spin-orbit coupling and electron-lattice interactions. Moreover, only $R\ln 2$ magnetic entropy seems to be recovered by heating Ba$_2$MgReO$_6$ up from zero to the AFQ transition point \cite{Pasztorova2023,Zivkovic2024}; this should be contrasted with the value of $R\ln 4$  expected upon recovering the $J=3/2$ degeneracy in the paramagnetic phase. None of the approaches discussed above can account for this missing entropy in Ba$_2$MgReO$_6$.

\subsubsection*{[H3] $d^2$}
For the $d^2$ double perovskites Ba$_2$CaOsO$_6$, \cite{Yamamura2006} 
Ba$_2$MgOsO$_6$,\cite{Marjerrison2016} and Ba$_2$ZnOsO$_6$,\cite{Marjerrison2016} 
thermodynamic measurements reveal a single magnetic phase transition, with $T_c \sim 30$-$50$K,
which was attributed to the onset of  an 
antiferromagnetic order, although muon spin resonance ($\mu$SR) oscillations found puzzling evidence for unusually weak internal fields. 
While earlier theoretical studies had suggested that the $5d^2$ configuration could form a robust $J=2$ moment,\cite{Chen2011,Svoboda2021} 
more recent theoretical work \cite{Paramekanti2020,Maharaj2020} argued that this five-fold multiplet gets split, even in a cubic environment, 
to yield a non-Kramers $E_g$ doublet ground state $\{ |0\rangle, |2\rangle+|-2\rangle \}$ and an excited $T_{2g}$ magnetic triplet.
This splitting arises from SOC-induced mixing of $t_{2g}$-$e_g$ orbitals, while it is absent in models projected from the outset onto the
$t_{2g}$ orbitals. This non-Kramers doublet acts as a pseudospin-1/2 degree of
freedom, with the pseudospin operators $(\tau_x,\tau_z)\sim (J_x^2-J_y^2, 3 J_z^2-J^2)$ encoding quadrupoles, while $\tau_y
\sim {\rm Sym}[J_x J_y J_z]$ encapsulates an octupolar moment (with `Sym' denoting symmetrization; these Stevens operators are the same, up to a normalization, as the operators $O_2^0$, $O_2^2$, and $O_3^{-2}$ introduced in the Supplemental Material). 
This scenario, together with
the assignment of the observed phase transition to ferro-ordering of $\tau_y$, was shown to capture the entropy, magnetic susceptibility, 
and weak $\mu$SR internal fields.\cite{Paramekanti2020,Maharaj2020,Voleti2020}
An ab initio 
electronic MBEH derived for these systems by DFT+HI and FT-HI approaches~ \cite{Pourovskii2021} exhibited dominant ferro-octupolar exchange, with subdominant
quadrupolar interactions; these results were supported by later calculations using a multiorbital tight-binding model with interactions
and SOC.\cite{Voleti2021,Churchill2022} 
\textbf{Figure \ref{fig:d}e} shows the temperature dependence of the total energy for the $d^2$ series Ba$_2$$R$OsO$_6$ ($R$ = Ca, Mg, Zn) obtained by mean-field solution of the MBEH~\cite{Pourovskii2021}, highlighting the greater stability of the ferromagnetic octupolar order compared to the competing antiferromagnetic quadrupolar phase.

The competition between quadrupolar
and octupolar order in these systems is highly susceptible to local and global strains that can break the crystalline point group 
symmetry.\cite{Voleti2021} Indeed, strong coupling of this non-Kramers doublet to phonons might favor local Jahn-Teller 
distortions and quadrupolar order, \cite{Khaliullin2021} potentially explaining the absence of an
octupolar ordering transition in Ba$_2$CdOsO$_6$.\cite{Marjerrison2016,Rayyan2023}
This competition is also revealed in
NMR experiments \cite{Cong2023} on 
Ba$_2$Na$_{1-x}$Ca$_x$OsO$_6$. 
Chemical substitution of Na$^+$ with Ca$^{2+
}$ on the $B$ site induces an effective electron transfer to the Na site~\cite{Kesavan2020}, accompanied by local structural modifications that alter the nature of the multipolar ground state~\cite{Cong2023, Celiberti2024}. Surprisingly, the system remains insulating throughout the entire doping range. The Mott gap is protected by the formation of Jahn-Teller polarons, which generate additional mid-gap SOC levels leading to the unusual coexistence of $J=3/2$ and $J=2$ states~\cite{Celiberti2024} (\textbf{Figure \ref{fig:d}f}).
In the $x \sim 1$ regime, dilute impurities
act as local probes of $d^2$ multipolar order
but also induce local strains that can locally pin quadrupoles and generate parasitic dipole moments tied to the octupolar ordering,
providing an explanation for the observed NMR lineshapes (\textbf{Figure \ref{fig:d}g}) \cite{Voleti2023}.

\subsubsection*{[H3] $d^3$}
Turning to the $d^3$ double perovskites, Ba$_2$YRuO$_6$ \cite{Aharen2009,Carlo2013,paddison2023} and Ba$_2$YOsO$_6$ \cite{Kermarrec2015,Taylor2017}, though the
half-filled $t_{2g}$ configuration could be expected to quench the orbital angular momentum and suppress SOC effects, the observation of 
spin gaps in inelastic neutron scattering (INS) \cite{Carlo2013,Kermarrec2015} and RIXS in Ba$_2$YOsO$_6$ \cite{Taylor2017} suggests that this scenario is incorrect. Indeed, ab initio IEI Hamiltonians (equation \ref{eq:HIEI}) derived by the FT-HI method for Ba$_2$YOsO$_6$ and Ba$_2$YRuO$_6$ exhibit, apart from expected AFM Heisenberg couplings, also substantial dipole-octupole IEIs\cite{Pourovskii2023}. These terms scale as the square of spin-orbit coupling. They lift the continuous symmetry of the Hamiltonian, opening an excitation gap, as shown\cite{Pourovskii2023} by direct calculations of INS spectra of Ba$_2$YOsO$_6$ (\textbf{Fig.~\ref{fig:d}h}) and Ba$_2$YRuO$_6$ from the mean-field ground state of the IEI Hamiltonian. The theoretical excitation gap and shape of the INS spectra are in good qualitative agreement with the measurements~\cite{Kermarrec2015,paddison2023}. 

The same dipole-octupole IEIs are also predicted\cite{Pourovskii2023} to stabilize a non-collinear double-\vq\ magnetic structure in both compounds. At the same time,  
non-coplanar triple-\vq\  spin textures in Ba$_2$YRuO$_6$ have been proposed~\cite{paddison2023} on the basis of INS measurements. The competition between various non-collinear orders stabilized by multipolar effects hence remains to be clarified.

\subsubsection*{[H3] Vacancy-ordered double perovskites}
We finally turn to the vacancy-ordered halide double perovskites A$_2$MX$_6$, a recently explored class of materials in the
same category as A$_2$BB$^\prime$O$_6$ but with oxygen replaced by ligand halides $X$ (=Cl, Br, I) and B site ions replaced by 
vacancies. The set of $d^1$ compounds Cs$_2$MX$_6$ (M=Ta,Nb;X=Cl),\cite{Ishikawa2019,Ishikawa2021b,Mansouri2023}
shows evidence of weak moments, $R\ln 4$ entropy, and ordering consistent with a $J=3/2$ ground state multiplet. Experiments on the
$d^2$ compounds $A_2$WCl$_6$ ($A$ = Cs, Rb, CH$_3$NH$_3$) \cite{Morgan2023} find evidence of strong deviations from the
classic Kotani formula for the magnetic susceptibility, which has been theoretically ascribed, via exact diagonalization calculations,
to the formation of a non-Kramers 
doublet and the resulting spin gap;\cite{Pradhan2024} this scenario has been confirmed 
through more detailed cluster calculations
.\cite{Li2024} The predicted ferro-octupolar order \cite{Pradhan2024} in A$_2$WCl$_6$ is 
yet to be experimentally confirmed. An important quantitative difference between these systems and the oxide double perovskites 
is their much narrower bandwidth, which results in weaker exchange interactions and lower transition temperatures for
symmetry breaking.
There is recent evidence from RIXS on the $d^5$ compound K$_2$IrCl$_6$ that strong coupling to 
phonons results in a clear vibronic character of the $J=3/2$ excited states.\cite{Warzanowski2024} This suggests that
coupling to the lattice may play an important role also in the vacancy-ordered halide double perovskites.

\subsection*{[H2] $f$-electron systems}

$f$-electron systems have been a traditional avenue to explore mutipolar phases 
 starting from the discovery~\cite{Osborn1953,Cox1967,Dunlap1968} of a hidden-order phase transition in NpO$_2$ in the 1960s, with multipolar orders identified and characterized in more than a dozen compounds \cite{Suzuki2018,Kuramoto2009,Santini2009}.

These systems have several distinctive features compared to their $d$-electron counterparts. The SOC is very strong, especially in the 5$f$-electronic shell of actinides. Together with a pronounced spatial localization of $f$-electron atomic states, this leads to the SOC dominating over crystal-field effects. The latter can be considered as a perturbation splitting the energy levels of the ionic ground-state $J$-multiplet~\cite{RareEarthMag_book}. Unlike in $d$-electron materials, the crystal-field potential does not substantially reduce the $f$-shell orbital moment; hence, physics of the multipolar interactions, including electron-lattice ones, emerges rather often.  
Another important consequence of the strong localization and large SOC 
is that  $f$-shells often keep their multiplet structure in solids and can be treated in the framework of standard crystal-field theory even in the metallic state.  In this case, a well-localized $f$-electrons subsystem coexists with $s$-$d$ metallic conduction bands. This situation occurs in elemental rare-earth metals and their inter-metallic compounds. Therefore, multipolar effects can play an essential role even in metallic systems. Multipolar interactions are suggested to be behind the iconic hidden-order phase of URu$_2$Si$_2$ \cite{Haule2009} and to mediate superconductivity in PrV$_2$Al$_{20}$ \cite{Tsujimoto2014,Sumita2020}. It is generally well established that metallic rare-earth-based compounds, such as DyB$_2$C$_2$ \cite{Yamauchi1999, Staub2005}, CeB$_6$ \cite{ERKELENS1987}, CeAg \cite{Morin1988}, TmCd \cite{Aleonard1979}, and TmZn \cite{Giraud1986}  harbor hidden quadrupolar orders. Since quadrupolar moments of $J$-multiplets directly couple to the lattice via magneto-elastic coupling, ferroic and more complex antiferroic quadrupolar orders in rare-earths and actinides materials \cite{Morin1990} can be identified and characterized 
 by lattice strains and local atomic distortions below the quadrupolar phase transition.  Early examples of the antiferroic quadrupole order detected in this way include measurements in TbGa$_3$  \cite{Morin1987} and PrPb$_3$  \cite{Morin1982}. Interestingly, spin-liquid phases associated with localized $f$-electrons multipolar degrees of freedom have also been reported for Gd$_3$Ga$_5$O$_{12}$ \cite{Paddison2015} and Ce$_2$Sn$_2$O$_7$ \cite{Sibille2020}.

\begin{figure}
\centering
\includegraphics[width=\linewidth]{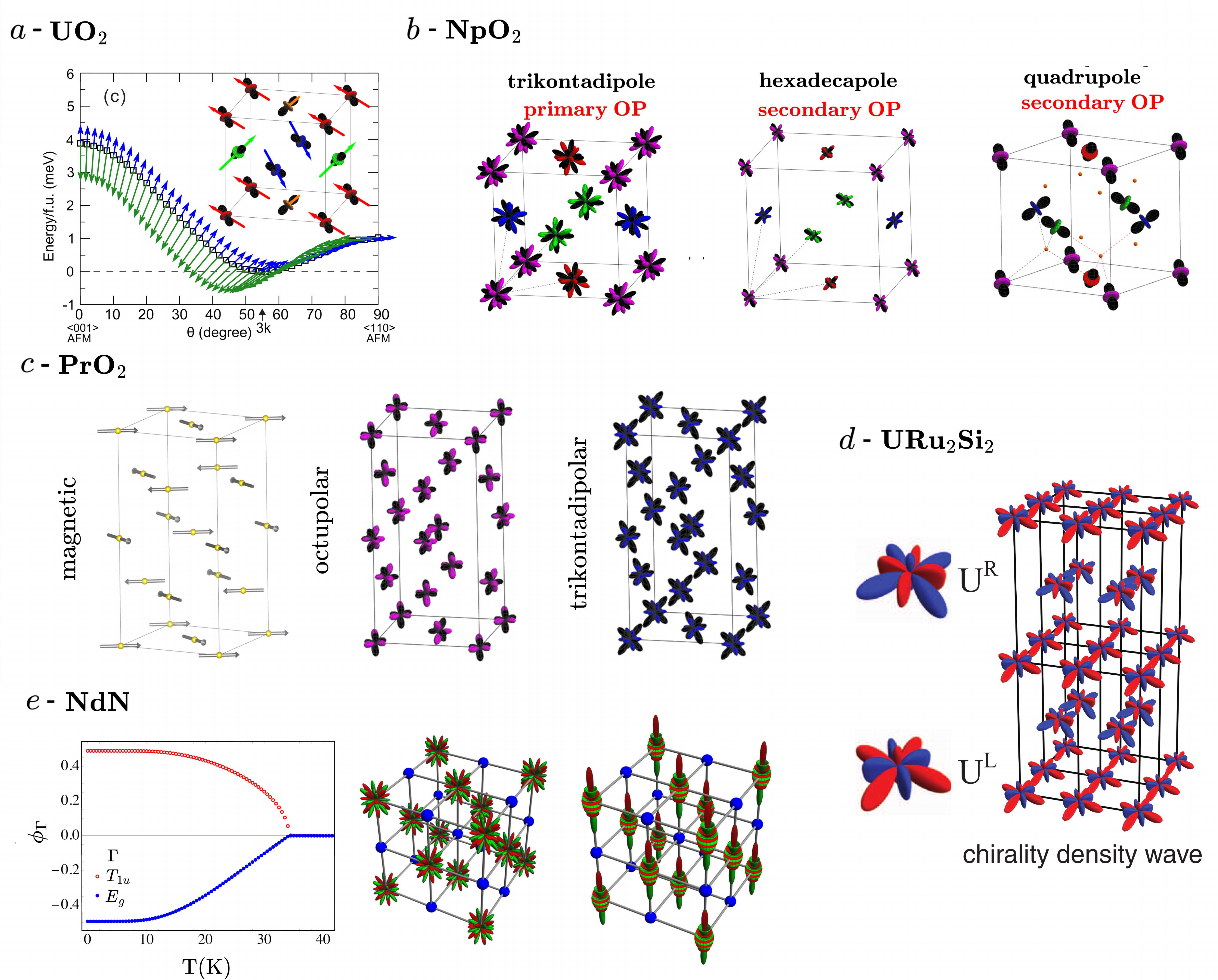}
\caption{\textbf{Multipolar orders of f-electron systems computed with different methods} \textbf{a}  | Total energy of UO$_2$ vs magnetic canting angle along the  1\vk\ antiferromagnetic  (AFM) $\langle 001\rangle  \to$ 3\vk\ AFM  $\to$ 1\vk\ AFM $\langle 110\rangle$ transformation pathway calculated by constained DFT+U\cite{Dudarev2019}. Inset shows 3\vk\ quadrupolar and magnetic orders predicted from intersite exchange (IEI) Hamiltonian derived with the force theorem in Hubbard-I (FT-HI) method\cite{Pourovskii2019}. \textbf{b}| Non-zero order parameters (OP) in the "hidden-order" phase of NpO$_2$ predicted from FT-HI IEI Hamiltonian\cite{Pourovskii2021f}. The primary OP is a rank-5 (trikontadipolar) antiferromagnetic (left); charge hexadecapole (middle) and quadrupole (right) are secondary OPs. The quadrupole OP is scaled up by a factor of 4 to be visible.
\textbf{c} | Dipole magnetic structure of Jahn-Teller-distorted PrO$_2$ (left), which is induced by "hidden" octupole (middle) and trikontadipole (right) OPs. These OPs have been derived from an effective Hamiltonian comprising   DFT+HI crystal-field splitting and FT-HI intersite exchange~\cite{Khmelevskyi2024}  \textbf{d}| Chiral hexadecapolar OP  proposed\cite{Kung2015} for the "hidden-order" phase of URu$_2$Si$_2$ on the basis of DFT+dynamical mean-field theory (DMFT) calculations \cite{Haule2009}. \textbf{e} |
The primary and secondary OP  vs $T$ as well as the largest rank-7 (middle) and rank-9 (right) contributions into the primary OP in NdN\cite{Iwahara2022}; the IEI Hamiltonian  in this work was calculated by cluster quantum chemistry methods. 
}
\label{fig:f}
\end{figure}

\subsubsection*{[H3] Actinide and praseodymium dioxides}

Actinide and praseodymium dioxides are prototypical cases of multipolar order and have been extensively investigated for decades by a combination of experiment, models and DFT calculations\cite{Santini2009,Suzuki2018}. We emphasize recent developments brought about by advanced ab initio methodologies.

Perhaps the most influential example of a magnetic system in which quadrupolar degrees of freedom play a fundamental role is UO$_2$ with rutile crystal structure \cite{Santini2009}.  The U ions feature a spherically symmetric ground-state crystal-field triplet, which is well separated from excited crystal field levels. They occupy a geometrically frustrated fcc sublattice. For any set of antiferro dipolar IEIs,  three different AFM structures -- single, double and triple-\vq\ -- have exactly the same energy and ordering temperature. Although UO$_2$ exhibits conventional dipolar magnetic ordering at low temperatures, the secondary quadrupolar order parameter is believed to be the main reason for the stabilization of the observed  \cite{Amoretti1989} complex non-collinear 3\vq\  AFM structure with respect to competing non-collinear 2\vq\  and collinear 1\vq\ AFM orders.  Several mechanisms have been proposed to explain the stabilization of the 3\vq\  magnetic structure by the secondary quadrupolar order parameter: via indirect quadrupolar exchange interactions mediated by phonons associated with oxygen vibrations \cite{Giannozzi1987}; purely electronic quadrupolar direct IEIs \cite{Mironov2003}; and due to static distortions of the oxygens in the magnetic ground state caused by interactions with quadrupolar degrees of freedom  \cite{Shiina2022}. The effects of quadrupole-quadrupole IEIs in UO$_2$ were investigated \cite{Pi2014} using the DFT+U mean-field theory.
Mapping magnetically constrained DFT total energies onto the model introduced in ref.~\cite{Mironov2003} provided an ab initio confirmation of the important role of quadrupole-quadrupole interactions in establishing the 3\vq\ phase.~\cite{Dudarev2019} The corresponding non-collinear magnetic energy surface is displayed in \textbf{Fig.~\ref{fig:f}a}. However, the relative stability of the  3\vq\ ground state with respect to the other competing AFM phases is largely overestimated, a known issue in DFT~\cite{Fioremosca2021}.
This scenario has been confirmed by FT-HI calculations of the full IEI Hamiltonian,\cite{Pourovskii2019} predicting quadrupole-quadrupole IEIs  that favor the 3\vq\ order of both magnetic moments and quadrupoles (inset in \textbf{Fig.~\ref{fig:f}a}).  However, the stabilization energy for this structure with respect to the predicted 2\vq\ one is extremely small (of the order of 0.01 meV). This casts doubt on the purely electronic quadrupole-quadrupole origin of the 3\vq-structure stabilization. The static distortions of the oxygen sublattice are also unlikely to stabilize the 3\vq\ structure, because the gain due to the magneto-elastic energy of the static Jan-Teller distortion is fully compensated by the elastic energy. Hence,  such distortions cannot affect the ordering temperature of the competing magnetic phases, as it was shown almost 50 years ago for the 1\vq\  structure \cite{Allen1968}.  Therefore, the principal mechanism to resolve the magnetic frustration in UO$_2$ still remains to be uncovered. Taking into account the importance of this material for nuclear technology, UO$_2$ remains of high interest as an archetypal system for studying the physics of the multipolar order \cite{Santini2009}.

High-rank multipolar IEIs can play a significant role also in the case of a conventional magnetic dipole order in an $f$-electron magnetic insulator.  
This situation occurs in PrO$_2$, where the experimentally observed \cite{Gardiner2004} complex non-colinear AFM order (\textbf{Fig.~\ref{fig:f}b})  cannot be obtained from any physically reasonable choice of dipolar IEI, \cite{Jensen2007} suggesting the importance of multipolar couplings\cite{Santini2009}.  The full IEI matrix between nearest-neighbor $J=5/2$ Pr$^{4+}$ ions calculated by the FT-HI approach\cite{Khmelevskyi2024} 
exhibits leading couplings between high-rank multipoles. The main contribution to the magnetic ordering stems from those high-rank multipoles, with the corresponding contribution due to the conventional dipolar de Gennes IEIs being about three orders of magnitude lower. Hence, the experimentally observed magnetic order is entirely due to non-dipolar IEIs and arises as a slave order to the complex antiferromagnetic ordering of the octupoles and trikontadipoles   \cite{Khmelevskyi2024} (\textbf{Fig.~\ref{fig:f}b}), to which it is entangled by the crystal-field potential. Recent calculations of multipolar IEIs \cite{Iwahara2022,Khmelevskyi2024} reveal their importance in the conventional magnetism of $f$-electron insulators. The standard paradigm\cite{RareEarthMag_book} that the observed magnetic order is due to crystal field (or magnetic anisotropy) and relativistic de Gennes dipolar terms 
can be thus violated in many cases  \cite{Khmelevskyi2024}.

The same ab initio  FT-HI approach has also been applied to the purely multipolar order in the oldest known hidden-order material, NpO$_2$ \cite{Pourovskii2021f}. The calculated multipolar IEIs predict a complex antiferromagnetic trikontadipolar order together with small secondary quadrupolar and hexadecapolar order parameters (\textbf{Fig.~\ref{fig:f}c}). The magnetic dipolar moments cancel each other exactly on each Np site, in agreement with experiment \cite{Osborne1953}. Moreover, the predicted mutual orientation of the quadrupolar moments on different Np sites is such that the magnetoelastic coupling with the Np $f$-electrons subsystem is fully compensated for all oxygen atoms. As a result, the cubic lattice structure of NpO$_2$ is preserved in the ordered phase. The predicted quadrupole structure also explains the lowering of the NpO$_2$ symmetry detected by NMR measurements~\cite{Tokunaga2005}. Indeed,  the quadrupole lobes of neighboring Np either all point towards O (for two oxygen sites along one of the cube diagonals) or all point away from it (for six other oxygen sites in the unit cell, \textbf{Fig.~\ref{fig:f}c}). Hence, the local environment for these two sets of oxygen sites becomes different in the ordered phase, as detected by NMR probes.  The structure of primary and secondary order parameters in NpO$_2$ was previously inferred from the results of a battery of experimental probes \cite{Paixao2002,Tokunaga2005,Santini2006,Magnani2008, Santini2006}; it has been fully reproduced using the ab initio FT-HI Hamiltonian. It is also shown~\cite{Pourovskii2021f} that the volume dependence of the multipolar IEIs leads to spontaneous volume-striction effects, similarly to the ordinary two-site exchange magnetostriction effect in magnetic materials  \cite{Morin1990}. This multipolar magnetostriction accounts for a tiny uniform volume contraction observed in NpO$_2$   below the hidden-order transition\cite{Amoretti1992}.

\subsubsection*{[H3] Rare-earth mononitrides}

The importance of multipolar IEIs was also demonstrated for  
rare-earth mononitrides $R$N, where $R$ is a rare-earth 3+ ion. They form a simple rock-salt structure and exhibit rather conventional ferromagnetic or antiferromagnetic orders \cite{Duan2007}. 

Nevertheless, kinetic IEIs obtained for $R_2$N$_2$ complexes~\cite{Iwahara2015} using a cluster strong-coupling approach exhibit multipolar terms. This approach was subsequently applied to derive the full IEI interaction matrix between the Nd$^{3+}$ $J=$9/2 ground state multiplets in NdN. High-rank multipolar IEI contributions up to the 7th and 9th rank were shown to provide an important contribution to the thermodynamics of the NdN ferromagnetic phase; this contribution is comparable to that of the usual magnetic dipolar IEI \cite{Iwahara2022}. The observed ferromagnetic phase in NdN is thus revealed to be a complex multipolar order (\textbf{Fig.~\ref{fig:f}d)} exhibiting substantial high-rank contributions to the order parameters together with the dipolar component.

\subsubsection*{[H3] Heavy-fermion intermetallics}

Ab initio theories of multipolar orders still find limited application in $f$-electron metals compared to $f$-electron insulators.  
DFT+U and DFT+DMFT  approaches generalized to handle multipolar moments \cite{Cricchio2009,Cricchio2011} have, however, been employed to treat those systems  \cite{Suzuki2018,Haule2009}. In general, DFT+U-based methods need to be properly constrained by occupation matrix control\cite{Suzuki2018}. The predictive power of these schemes is still limited, because the proper symmetry breaking needs to be introduced from the onset. The space of possible multipolar order parameters is often too vast to explore all possibilities with such total energy calculations.  Further advances in this area will be important to tackle outstanding problems like the true nature of the hidden order in the canonical URu$_2$Si$_2$ compound, for which exotic chiral hexadecapolar orders are proposed (\textbf{Fig.~\ref{fig:f}e}), \cite{Haule2009,Kung2015} as well as a number of alternative multipolar order parameters~\cite{Mydosh2020}, and an interplay between the multipolar order and the Kondo effect \cite{Christovam2024}.

\subsection*{[H2] Magnetoelectric systems}

We conclude the materials section by briefly touching on magnetoelectric systems~\cite{zeldovich1957, gorbatsevich1983, Spaldin2008, Spaldin2013}. 
Magnetoelectric materials exhibit a coupling between magnetic and electric properties, whereby an applied electric field induces a linear magnetization, and conversely, an applied magnetic field generates a linear electric polarization.
In these systems, antisymmetric spin-orbit coupling associated with the simultaneous breaking of spatial inversion and time-reversal symmetries
can induce odd-parity multipoles like second-order (quadrupole) moments of the magnetization density—a departure from the even-parity multipole orders 
hosted by single $d$ or $f$-electron shells
discussed thus far.
Magnetoelectric multipoles, which have been briefly introduced in the section about methods, are central to the fundamental mechanisms of magnetoelectric coupling and have been extensively studied both theoretically~\cite{Hitomi2014, Verbeedk2023,Fu2015, Thole2018,Bhowal2022, Lovesey2023,Hayami2018} and experimentally~\cite{Fiebig1994, Rikken1997,Sannikov01111998,Goulon2000,Goulon2002, Kubota2004, Kimata2021}.
The prototypical magnetoelectric material in which the connection between local magnetoelectric multipolar order and the atomic-scale magnetoelectric response has been studied is Cr$_2$O$_3$\cite{astrov1961}. This compound also exhibits an antiferroic arrangement of magnetoelectric multipoles, representing a distinct form of hidden order\cite{Verbeedk2023}. 
Moreover, fluctuating parity-odd multipoles have been proposed as potential order parameters in hidden-order phases of unconventional superconductors~\cite{Sumita2020}, and may also play a role in the enigmatic hidden order phase observed in cuprates~\cite{Chakravarty2001}.

\vspace{4mm}

\section*{[H1] Conclusions and Future Perspectives}

As emphasized in this Review, a vast space of competing order parameters and interactions in multipolar systems calls for approaches to construct their realistic low-energy effective Hamiltonians in an unbiased way. The progress in modeling insulating multipolar phases stems from recently formulated efficient ab initio methods that enable parameter-free, material-specific descriptions of hidden multipolar orders. Despite these successful applications, several open questions and challenges remain, requiring continued efforts.

Compared to insulating systems, the scope of ab-initio theories for metallic multipolar phases, such as $f$-electron heavy-fermion systems, remains more limited. 
The reason is that the force theorem\cite{Pourovskii2016} and cluster approaches \cite{Iwahara2015,Iwahara2022} developed for high-rank intersite exchange interactions are based on a strong-coupling perturbative picture that becomes generally invalid in the metallic state. 
The conventional magnetic force theorem for symmetry-broken phases can be applied, \cite{Khmelevskyi2007,Szilva2023}  but it lacks a proper generalization to high-rank multipolar couplings.  This limits its applicability not only for heavy-fermion systems, but also for multipolar intermetallics with localized $f$ shells, where the kinetic exchange mechanism 
is enabled by the on-site Hund's rule coupling between localized and conduction electrons.
In this respect, one may notice the promising development of extracting multipolar IEIs from the full DMFT susceptibility\cite{Otsuki2019}, as discussed, and its application to CeB$_6$ \cite{Otsuki2024}. 

Unlike purely spin-dipole moments, multipolar orders such as the
quadrupolar order can linearly couple to phonons, leading to new material
functionalities via strain-tuning or driven phonons. This connects with the emerging area of
``straintronics'' \cite{Roy2011,Miao2021} and ``phononics'' \cite{Mankowsky2016,Ning2023,Hart2024} in bulk oxides and 2D materials, enabling the use of strain or phonons to read and write information into multipolar degrees of freedom.

The coupling of phonons with doping-injected electrons and holes allows for charge localization, the breaking of crystal symmetry~\cite{Voleti2023} and the formation of polarons~\cite{Franchini2021, Celiberti2024}. This can lead to the coexistence of different high-rank multipolar degrees of freedom, as reviewed for the case of $J=3/2$ and $J=2$ states in Ba$_2$Na$_{1x}$Ca$_x$OsO$_6$~\cite{Celiberti2024}. 
The interaction between different types of local high-rank multipoles within the same system remains unexplored. Understanding these interactions as well as the interaction of polarons with the underlying hidden order could lead to the discovery of novel and exciting physics and will provide insights into the evolution of hidden phases upon doping\cite{Cong2023}.

 In the context of dilute impurities, isolated $d$-orbital ions or actinide $f$-orbital ions embedded at high-symmetry sites in cubic crystals may be viewed as single defects that support many sharp low-energy multiplets hosting a variety of multipolar moments. 
Certain
multiplets, for example, have no dipole moments but only possess quadrupole and octupole moments. Such multiplets can thus 
act as sensors of electric field gradients, which couple linearly to quadrupoles, analogous to how an ordinary 
magnetic field couples to spin dipoles. At the same time,
since quadrupole and octupole moments couple relatively weakly to actual magnetic fields, at ${\cal O}(B^2)$ or ${\cal O}(B^3)$ respectively, 
they are relatively immune to stray magnetic field noise. Embedded multipolar ions might also act as strain sensors, because their multiplet
degeneracies can be broken in specific patterns by externally applied strain fields, which can couple linearly to the quadrupole moments.
Multipolar defect centres can thus provide a rich new class of states
for quantum sensing or for application as multipolar quantum bits, potentially extending the capabilities of existing systems such as nitrogen vacancy centers in diamond \cite{Doherty2013,Schirhagl2014,Rovny2024}.

Another intriguing aspect is the evolution of intersite exchange interactions from the bulk to the surface~\cite{Bayer2007, Franchini2005}. Microscopic Hamiltonian approaches have demonstrated that the breaking of crystal symmetry at surfaces in multipolar materials can pin quadrupolar orders \cite{Voleti2021}, potentially giving rise to novel physics at crystal surfaces, interfaces, or within nanoparticles \cite{Weber2023, Bhowal2024b}. This remains a largely unexplored field of research, requiring combined experimental and theoretical efforts. In particular, atomically resolved, surface-sensitive experimental techniques—such as (spin-dependent) scanning tunneling microscopy, atomic force microscopy, and ARPES—paired with first-principles atomistic simulations are crucial to unravel the surface geometry, identify possible reconstructions, and distinguish the electronic and magnetic properties.

Finally, the integration of artificial intelligence algorithms and data-driven approaches into many-body and materials physics~\cite{Schmidt2019} is driving a transformative shift in the field. Machine learning (ML) strategies have recently been employed to accelerate and optimize the study of magnetic phases within first-principles calculations~\cite{Eckhoff2021, Novikov2022, Kostiuchenko_2024, gao2024, Acosta2022, Ponet2024, Baumsteiger2025}, as well as effective spin Hamiltonians and Monte Carlo simulations~\cite{Mills2020, Wang2020, Kwon2020, Fan2023}. A range of ML architectures, including ML-based magnetic interatomic potentials~\cite{Eckhoff2021, Novikov2022, Kostiuchenko_2024, gao2024}, deep neural networks~\cite{Kwon2020, Acosta2022}, reinforcement learning~\cite{Mills2020, Fan2023}, and Bayesian optimization~\cite{Baumsteiger2025}, has been applied to explore collinear and non-collinear magnetic phases, magnetic potential energy landscapes, and spin-spin interaction parameters.
Although still in its early stages, this field is rapidly advancing and is expected to mature sufficiently to address high-rank intersite exchange interactions, (pseudo)spin excited states, temperature effects and spin-lattice unconventional interactions. By mapping ab initio theories onto efficient ML frameworks, these approaches have the potential to greatly enhance the exploration of complex magnetic interactions, such as those reviewed here, ultimately advancing our understanding of quantum magnetism and unconventional hidden orders.

\bibliography{references}

\begin{thebibliography}{100}
\urlstyle{rm}
\expandafter\ifx\csname url\endcsname\relax
  \def\url#1{\texttt{#1}}\fi
\expandafter\ifx\csname urlprefix\endcsname\relax\def\urlprefix{URL }\fi
\expandafter\ifx\csname doiprefix\endcsname\relax\def\doiprefix{DOI: }\fi
\providecommand{\bibinfo}[2]{#2}
\providecommand{\eprint}[2][]{\url{#2}}

\bibitem{landau1965}
\bibinfo{author}{Landau, L.~D.} \& \bibinfo{author}{Lifshitz, E.~M.}
\newblock \emph{\bibinfo{title}{Quantum Mechanics: Non-relativistic Theory}} (\bibinfo{publisher}{Pergamon Press}, \bibinfo{address}{Oxford}, \bibinfo{year}{1965}).

\bibitem{khomskii2014}
\bibinfo{author}{Khomskii, D.~I.}
\newblock \emph{\bibinfo{title}{Transition Metal Compounds}} (\bibinfo{publisher}{Cambridge University Press}, \bibinfo{address}{Cambridge}, \bibinfo{year}{2014}).

\bibitem{Rashba1960}
\bibinfo{author}{Rashba, E.}
\newblock \bibinfo{journal}{\bibinfo{title}{Properties of semiconductors with an extremum loop. i. cyclotron and combinational resonance in a magnetic field perpendicular to the plane of the loop}}.
\newblock {\emph{\JournalTitle{Sov. Phys.-Solid State}}} \textbf{\bibinfo{volume}{2}}, \bibinfo{pages}{1109} (\bibinfo{year}{1960}).

\bibitem{Bihlmayer2022}
\bibinfo{author}{Bihlmayer, G.}, \bibinfo{author}{No{\"e}l, P.}, \bibinfo{author}{Vyalikh, D.~V.}, \bibinfo{author}{Chulkov, E.~V.} \& \bibinfo{author}{Manchon, A.}
\newblock \bibinfo{journal}{\bibinfo{title}{Rashba-like physics in condensed matter}}.
\newblock {\emph{\JournalTitle{Nature Reviews Physics}}} \textbf{\bibinfo{volume}{4}}, \bibinfo{pages}{642--659}, \doiprefix\url{10.1038/s42254-022-00490-y} (\bibinfo{year}{2022}).

\bibitem{hirsch1999}
\bibinfo{author}{Hirsch, J.~E.}
\newblock \bibinfo{journal}{\bibinfo{title}{Spin hall effect}}.
\newblock {\emph{\JournalTitle{Phys. Rev. Lett.}}} \textbf{\bibinfo{volume}{83}}, \bibinfo{pages}{1834} (\bibinfo{year}{1999}).

\bibitem{Maciejko2011}
\bibinfo{author}{Maciejko, J.}, \bibinfo{author}{Hughes, T.~L.} \& \bibinfo{author}{Zhang, S.-C.}
\newblock \bibinfo{journal}{\bibinfo{title}{The quantum spin hall effect}}.
\newblock {\emph{\JournalTitle{Annual Review of Condensed Matter Physics}}} \textbf{\bibinfo{volume}{2}}, \bibinfo{pages}{31--53}, \doiprefix\url{https://doi.org/10.1146/annurev-conmatphys-062910-140538} (\bibinfo{year}{2011}).

\bibitem{fert2017}
\bibinfo{author}{Fert, A.}, \bibinfo{author}{Reyren, N.} \& \bibinfo{author}{Cros, V.}
\newblock \bibinfo{journal}{\bibinfo{title}{Magnetic skyrmions: Advances in physics and potential applications}}.
\newblock {\emph{\JournalTitle{Nat. Rev. Mater.}}} \textbf{\bibinfo{volume}{2}}, \bibinfo{pages}{17031} (\bibinfo{year}{2017}).

\bibitem{Mott1937}
\bibinfo{author}{Mott, N.~F.} \& \bibinfo{author}{Peierls, R.}
\newblock \bibinfo{journal}{\bibinfo{title}{Discussion of the paper by de boer and verwey}}.
\newblock {\emph{\JournalTitle{Proceedings of the Physical Society}}} \textbf{\bibinfo{volume}{49}}, \bibinfo{pages}{72}, \doiprefix\url{10.1088/0959-5309/49/4S/308} (\bibinfo{year}{1937}).

\bibitem{Mott1949}
\bibinfo{author}{Mott, N.~F.}
\newblock \bibinfo{journal}{\bibinfo{title}{The basis of the electron theory of metals, with special reference to the transition metals}}.
\newblock {\emph{\JournalTitle{Proceedings of the Physical Society A}}} \textbf{\bibinfo{volume}{62}}, \bibinfo{pages}{416}, \doiprefix\url{10.1088/0370-1298/62/7/303} (\bibinfo{year}{1949}).

\bibitem{Imada1998}
\bibinfo{author}{Imada, M.}, \bibinfo{author}{Fujimori, A.} \& \bibinfo{author}{Tokura, Y.}
\newblock \bibinfo{journal}{\bibinfo{title}{Metal-insulator transitions}}.
\newblock {\emph{\JournalTitle{Rev. Mod. Phys.}}} \textbf{\bibinfo{volume}{70}}, \bibinfo{pages}{1039--1263}, \doiprefix\url{10.1103/RevModPhys.70.1039} (\bibinfo{year}{1998}).

\bibitem{Morosan2012}
\bibinfo{author}{Morosan, E.}, \bibinfo{author}{Natelson, D.}, \bibinfo{author}{Nevidomskyy, A.~H.} \& \bibinfo{author}{Si, Q.}
\newblock \bibinfo{journal}{\bibinfo{title}{Strongly correlated materials}}.
\newblock {\emph{\JournalTitle{Advanced Materials}}} \textbf{\bibinfo{volume}{24}}, \bibinfo{pages}{4896--4923}, \doiprefix\url{https://doi.org/10.1002/adma.201202018} (\bibinfo{year}{2012}).
\newblock \eprint{https://onlinelibrary.wiley.com/doi/pdf/10.1002/adma.201202018}.

\bibitem{Witczak2014}
\bibinfo{author}{Witczak-Krempa, W.}, \bibinfo{author}{Chen, G.}, \bibinfo{author}{Kim, Y.~B.} \& \bibinfo{author}{Balents, L.}
\newblock \bibinfo{journal}{\bibinfo{title}{Correlated quantum phenomena in the strong spin-orbit regime}}.
\newblock {\emph{\JournalTitle{Annual Review of Condensed Matter Physics}}} \textbf{\bibinfo{volume}{5}}, \bibinfo{pages}{57--82}, \doiprefix\url{https://doi.org/10.1146/annurev-conmatphys-020911-125138} (\bibinfo{year}{2014}).

\bibitem{Khaliullin2005}
\bibinfo{author}{Khaliullin, G.}
\newblock \bibinfo{journal}{\bibinfo{title}{{Orbital Order and Fluctuations in Mott Insulators}}}.
\newblock {\emph{\JournalTitle{Progress of Theoretical Physics Supplement}}} \textbf{\bibinfo{volume}{160}}, \bibinfo{pages}{155--202}, \doiprefix\url{10.1143/PTPS.160.155} (\bibinfo{year}{2005}).
\newblock \eprint{https://academic.oup.com/ptps/article-pdf/doi/10.1143/PTPS.160.155/5162453/160-155.pdf}.

\bibitem{Rau2016}
\bibinfo{author}{Rau, J.~G.}, \bibinfo{author}{Lee, E. K.-H.} \& \bibinfo{author}{Kee, H.-Y.}
\newblock \bibinfo{journal}{\bibinfo{title}{Spin-orbit physics giving rise to novel phases in correlated systems: Iridates and related materials}}.
\newblock {\emph{\JournalTitle{Annual Review of Condensed Matter Physics}}} \textbf{\bibinfo{volume}{7}}, \bibinfo{pages}{195--221}, \doiprefix\url{https://doi.org/10.1146/annurev-conmatphys-031115-011319} (\bibinfo{year}{2016}).

\bibitem{Schaffer2016}
\bibinfo{author}{Schaffer, R.}, \bibinfo{author}{Lee, E. K.-H.}, \bibinfo{author}{Yang, B.-J.} \& \bibinfo{author}{Kim, Y.~B.}
\newblock \bibinfo{journal}{\bibinfo{title}{Recent progress on correlated electron systems with strong spin–orbit coupling}}.
\newblock {\emph{\JournalTitle{Reports on Progress in Physics}}} \textbf{\bibinfo{volume}{79}}, \bibinfo{pages}{094504}, \doiprefix\url{10.1088/0034-4885/79/9/094504} (\bibinfo{year}{2016}).

\bibitem{Takayama2021}
\bibinfo{author}{Takayama, T.}, \bibinfo{author}{Chaloupka, J.}, \bibinfo{author}{Smerald, A.}, \bibinfo{author}{Khaliullin, G.} \& \bibinfo{author}{Takagi, H.}
\newblock \bibinfo{journal}{\bibinfo{title}{Spin–orbit-entangled electronic phases in 4d and 5d transition-metal compounds}}.
\newblock {\emph{\JournalTitle{Journal of the Physical Society of Japan}}} \textbf{\bibinfo{volume}{90}}, \bibinfo{pages}{062001}, \doiprefix\url{10.7566/JPSJ.90.062001} (\bibinfo{year}{2021}).
\newblock \eprint{https://doi.org/10.7566/JPSJ.90.062001}.

\bibitem{Khomskii2021}
\bibinfo{author}{Khomskii, D.~I.} \& \bibinfo{author}{Streltsov, S.~V.}
\newblock \bibinfo{journal}{\bibinfo{title}{Orbital effects in solids: Basics, recent progress, and opportunities}}.
\newblock {\emph{\JournalTitle{Chemical Reviews}}} \textbf{\bibinfo{volume}{121}}, \bibinfo{pages}{2992--3030}, \doiprefix\url{10.1021/acs.chemrev.0c00579} (\bibinfo{year}{2021}).
\newblock \bibinfo{note}{PMID: 33314912}, \eprint{https://doi.org/10.1021/acs.chemrev.0c00579}.

\bibitem{Browne2021}
\bibinfo{author}{Browne, A.~J.}, \bibinfo{author}{Krajewska, A.} \& \bibinfo{author}{Gibbs, A.~S.}
\newblock \bibinfo{journal}{\bibinfo{title}{Quantum materials with strong spin–orbit coupling: challenges and opportunities for materials chemists}}.
\newblock {\emph{\JournalTitle{J. Mater. Chem. C}}} \textbf{\bibinfo{volume}{9}}, \bibinfo{pages}{11640--11654}, \doiprefix\url{10.1039/D1TC02070F} (\bibinfo{year}{2021}).

\bibitem{Jackeli2009}
\bibinfo{author}{Jackeli, G.} \& \bibinfo{author}{Khaliullin, G.}
\newblock \bibinfo{journal}{\bibinfo{title}{Mott insulators in the strong spin-orbit coupling limit: From heisenberg to a quantum compass and kitaev models}}.
\newblock {\emph{\JournalTitle{Phys. Rev. Lett.}}} \textbf{\bibinfo{volume}{102}}, \bibinfo{pages}{017205}, \doiprefix\url{10.1103/PhysRevLett.102.017205} (\bibinfo{year}{2009}).

\bibitem{Calder2012}
\bibinfo{author}{Calder, S.} \emph{et~al.}
\newblock \bibinfo{journal}{\bibinfo{title}{Magnetically driven metal-insulator transition in ${\mathrm{naoso}}_{3}$}}.
\newblock {\emph{\JournalTitle{Phys. Rev. Lett.}}} \textbf{\bibinfo{volume}{108}}, \bibinfo{pages}{257209}, \doiprefix\url{10.1103/PhysRevLett.108.257209} (\bibinfo{year}{2012}).

\bibitem{Kim2016}
\bibinfo{author}{Kim, B.} \emph{et~al.}
\newblock \bibinfo{journal}{\bibinfo{title}{Lifshitz transition driven by spin fluctuations and spin-orbit renormalization in ${\mathrm{naoso}}_{3}$}}.
\newblock {\emph{\JournalTitle{Phys. Rev. B}}} \textbf{\bibinfo{volume}{94}}, \bibinfo{pages}{241113}, \doiprefix\url{10.1103/PhysRevB.94.241113} (\bibinfo{year}{2016}).

\bibitem{Balents2010}
\bibinfo{author}{Balents, L.}
\newblock \bibinfo{journal}{\bibinfo{title}{Spin liquids in frustrated magnets}}.
\newblock {\emph{\JournalTitle{Nature}}} \textbf{\bibinfo{volume}{464}}, \bibinfo{pages}{199--208}, \doiprefix\url{10.1038/nature08917} (\bibinfo{year}{2010}).

\bibitem{deVries2010}
\bibinfo{author}{de~Vries, M.~A.}, \bibinfo{author}{Mclaughlin, A.~C.} \& \bibinfo{author}{Bos, J.-W.~G.}
\newblock \bibinfo{journal}{\bibinfo{title}{Valence bond glass on an fcc lattice in the double perovskite ${\mathrm{ba}}_{2}{\mathrm{ymoo}}_{6}$}}.
\newblock {\emph{\JournalTitle{Phys. Rev. Lett.}}} \textbf{\bibinfo{volume}{104}}, \bibinfo{pages}{177202}, \doiprefix\url{10.1103/PhysRevLett.104.177202} (\bibinfo{year}{2010}).

\bibitem{Sasaki1970}
\bibinfo{author}{Sasaki, K.} \& \bibinfo{author}{Obata, Y.}
\newblock \bibinfo{journal}{\bibinfo{title}{Studies of the dynamical jahn-teller effect on static magnetic susceptibility}}.
\newblock {\emph{\JournalTitle{Journal of the Physical Society of Japan}}} \textbf{\bibinfo{volume}{28}}, \bibinfo{pages}{1157--1167}, \doiprefix\url{10.1143/JPSJ.28.1157} (\bibinfo{year}{1970}).
\newblock \eprint{https://doi.org/10.1143/JPSJ.28.1157}.

\bibitem{Heinze2011}
\bibinfo{author}{Heinze, S.} \emph{et~al.}
\newblock \bibinfo{journal}{\bibinfo{title}{Spontaneous atomic-scale magnetic skyrmion lattice in two dimensions}}.
\newblock {\emph{\JournalTitle{Nature Physics}}} \textbf{\bibinfo{volume}{7}}, \bibinfo{pages}{713--718}, \doiprefix\url{10.1038/nphys2045} (\bibinfo{year}{2011}).

\bibitem{Sitte2018}
\bibinfo{author}{Everschor-Sitte, K.}, \bibinfo{author}{Masell, J.}, \bibinfo{author}{Reeve, R.~M.} \& \bibinfo{author}{Kläui, M.}
\newblock \bibinfo{journal}{\bibinfo{title}{Perspective: Magnetic skyrmions—overview of recent progress in an active research field}}.
\newblock {\emph{\JournalTitle{Journal of Applied Physics}}} \textbf{\bibinfo{volume}{124}}, \bibinfo{pages}{240901}, \doiprefix\url{10.1063/1.5048972} (\bibinfo{year}{2018}).
\newblock \eprint{https://pubs.aip.org/aip/jap/article-pdf/doi/10.1063/1.5048972/19981128/240901\_1\_1.5048972.pdf}.

\bibitem{Dziailoshinskii1957}
\bibinfo{author}{Dzialoshinskii, I.~E.}
\newblock \bibinfo{journal}{\bibinfo{title}{Sov. phys. jetp}}.
\newblock {\emph{\JournalTitle{Soviet Physics JETP}}} \textbf{\bibinfo{volume}{5}}, \bibinfo{pages}{1259} (\bibinfo{year}{1957}).

\bibitem{Moriya1960}
\bibinfo{author}{Moriya, T.}
\newblock \bibinfo{journal}{\bibinfo{title}{New mechanism of anisotropic superexchange interaction}}.
\newblock {\emph{\JournalTitle{Phys. Rev. Lett.}}} \textbf{\bibinfo{volume}{4}}, \bibinfo{pages}{228--230}, \doiprefix\url{10.1103/PhysRevLett.4.228} (\bibinfo{year}{1960}).

\bibitem{Morrish1994}
\bibinfo{author}{Morrish, A.~H.}
\newblock \emph{\bibinfo{title}{Canted Antiferromagnetism: Hematite}} (\bibinfo{publisher}{World Scientific}, \bibinfo{year}{1994}).

\bibitem{Pesin2010}
\bibinfo{author}{Pesin, D.} \& \bibinfo{author}{Balents, L.}
\newblock \bibinfo{journal}{\bibinfo{title}{Mott physics and band topology in materials with strong spin--orbit interaction}}.
\newblock {\emph{\JournalTitle{Nature Physics}}} \textbf{\bibinfo{volume}{6}}, \bibinfo{pages}{376--381}, \doiprefix\url{10.1038/nphys1606} (\bibinfo{year}{2010}).

\bibitem{Celiberti2024}
\bibinfo{author}{Celiberti, L.} \emph{et~al.}
\newblock \bibinfo{journal}{\bibinfo{title}{Spin-orbital jahn-teller bipolarons}}.
\newblock {\emph{\JournalTitle{Nature Communications}}} \textbf{\bibinfo{volume}{15}}, \bibinfo{pages}{2429}, \doiprefix\url{10.1038/s41467-024-46621-0} (\bibinfo{year}{2024}).

\bibitem{Fioremosca2024}
\bibinfo{author}{{Fiore Mosca}, D.}, \bibinfo{author}{Schnait, H.}, \bibinfo{author}{Celiberti, L.}, \bibinfo{author}{Aichhorn, M.} \& \bibinfo{author}{Franchini, C.}
\newblock \bibinfo{journal}{\bibinfo{title}{The mott transition in the 5d1 compound ba2naoso6: A dft+dmft study with paw spinor projectors}}.
\newblock {\emph{\JournalTitle{Computational Materials Science}}} \textbf{\bibinfo{volume}{233}}, \bibinfo{pages}{112764}, \doiprefix\url{https://doi.org/10.1016/j.commatsci.2023.112764} (\bibinfo{year}{2024}).

\bibitem{Hart2024}
\bibinfo{author}{Hart, K.}, \bibinfo{author}{Sutcliffe, R.}, \bibinfo{author}{Refael, G.} \& \bibinfo{author}{Paramekanti, A.}
\newblock \bibinfo{title}{Phonon-driven multipolar dynamics in spin-orbit coupled mott insulators} (\bibinfo{year}{2024}).
\newblock \eprint{2404.17633}.

\bibitem{Khmelevskyi2024}
\bibinfo{author}{Khmelevskyi, S.} \& \bibinfo{author}{Pourovskii, L.~V.}
\newblock \bibinfo{journal}{\bibinfo{title}{Non-collinear magnetism driven by a hidden multipolar order in pro2}}.
\newblock {\emph{\JournalTitle{Communications Physics}}} \textbf{\bibinfo{volume}{7}}, \bibinfo{pages}{12}, \doiprefix\url{10.1038/s42005-023-01503-7} (\bibinfo{year}{2024}).

\bibitem{Soh2024}
\bibinfo{author}{Soh, J.-R.} \emph{et~al.}
\newblock \bibinfo{journal}{\bibinfo{title}{Spectroscopic signatures and origin of hidden order in {Ba$_2$MgReO$_6$}}}.
\newblock {\emph{\JournalTitle{Nature Communications}}} \textbf{\bibinfo{volume}{15}}, \bibinfo{pages}{10383}, \doiprefix\url{10.1038/s41467-024-53893-z} (\bibinfo{year}{2024}).

\bibitem{Yahne2024}
\bibinfo{author}{Yahne, D.~R.} \emph{et~al.}
\newblock \bibinfo{journal}{\bibinfo{title}{Dipolar spin ice regime proximate to an all-in-all-out n\'eel ground state in the dipolar-octupolar pyrochlore ${\mathrm{ce}}_{2}{\mathrm{sn}}_{2}{\mathrm{o}}_{7}$}}.
\newblock {\emph{\JournalTitle{Phys. Rev. X}}} \textbf{\bibinfo{volume}{14}}, \bibinfo{pages}{011005}, \doiprefix\url{10.1103/PhysRevX.14.011005} (\bibinfo{year}{2024}).

\bibitem{Shah2000}
\bibinfo{author}{Shah, N.}, \bibinfo{author}{Chandra, P.}, \bibinfo{author}{Coleman, P.} \& \bibinfo{author}{Mydosh, J.~A.}
\newblock \bibinfo{journal}{\bibinfo{title}{Hidden order in ${\mathrm{uru}}_{2}{\mathrm{si}}_{2}$}}.
\newblock {\emph{\JournalTitle{Phys. Rev. B}}} \textbf{\bibinfo{volume}{61}}, \bibinfo{pages}{564--569}, \doiprefix\url{10.1103/PhysRevB.61.564} (\bibinfo{year}{2000}).

\bibitem{Amitsuka2002}
\bibinfo{author}{Amitsuka, H.} \emph{et~al.}
\newblock \bibinfo{journal}{\bibinfo{title}{Hidden order and weak antiferromagnetism in uru2si2}}.
\newblock {\emph{\JournalTitle{Physica B: Condensed Matter}}} \textbf{\bibinfo{volume}{312-313}}, \bibinfo{pages}{390--396}, \doiprefix\url{https://doi.org/10.1016/S0921-4526(01)01343-6} (\bibinfo{year}{2002}).
\newblock \bibinfo{note}{The International Conference on Strongly Correlated Electron Systems}.

\bibitem{Chandra2002}
\bibinfo{author}{Chandra, P.}, \bibinfo{author}{lian, P.}, \bibinfo{author}{Mydosh, J.~A.} \& \bibinfo{author}{Tripathi, V.}
\newblock \bibinfo{journal}{\bibinfo{title}{Hidden orbital order in the heavy fermion metal uru2si2}}.
\newblock {\emph{\JournalTitle{Nature}}} \textbf{\bibinfo{volume}{417}}, \bibinfo{pages}{831--834}, \doiprefix\url{10.1038/nature00795} (\bibinfo{year}{2002}).

\bibitem{Cricchio2009}
\bibinfo{author}{Cricchio, F.}, \bibinfo{author}{Bultmark, F.}, \bibinfo{author}{Gr\aa{}n\"as, O.} \& \bibinfo{author}{Nordstr\"om, L.}
\newblock \bibinfo{journal}{\bibinfo{title}{Itinerant magnetic multipole moments of rank five as the hidden order in ${\mathrm{uru}}_{2}{\mathrm{si}}_{2}$}}.
\newblock {\emph{\JournalTitle{Phys. Rev. Lett.}}} \textbf{\bibinfo{volume}{103}}, \bibinfo{pages}{107202}, \doiprefix\url{10.1103/PhysRevLett.103.107202} (\bibinfo{year}{2009}).

\bibitem{Jackeli2009b}
\bibinfo{author}{Jackeli, G.} \& \bibinfo{author}{Khaliullin, G.}
\newblock \bibinfo{journal}{\bibinfo{title}{Magnetically hidden order of kramers doublets in ${d}^{1}$ systems: ${\mathrm{sr}}_{2}{\mathrm{vo}}_{4}$}}.
\newblock {\emph{\JournalTitle{Phys. Rev. Lett.}}} \textbf{\bibinfo{volume}{103}}, \bibinfo{pages}{067205}, \doiprefix\url{10.1103/PhysRevLett.103.067205} (\bibinfo{year}{2009}).

\bibitem{Chen2010}
\bibinfo{author}{Chen, G.}, \bibinfo{author}{Pereira, R.} \& \bibinfo{author}{Balents, L.}
\newblock \bibinfo{journal}{\bibinfo{title}{Exotic phases induced by strong spin-orbit coupling in ordered double perovskites}}.
\newblock {\emph{\JournalTitle{Phys. Rev. B}}} \textbf{\bibinfo{volume}{82}}, \bibinfo{pages}{174440}, \doiprefix\url{10.1103/PhysRevB.82.174440} (\bibinfo{year}{2010}).

\bibitem{Gardner2010}
\bibinfo{author}{Gardner, J.~S.}, \bibinfo{author}{Gingras, M. J.~P.} \& \bibinfo{author}{Greedan, J.~E.}
\newblock \bibinfo{journal}{\bibinfo{title}{Magnetic pyrochlore oxides}}.
\newblock {\emph{\JournalTitle{Rev. Mod. Phys.}}} \textbf{\bibinfo{volume}{82}}, \bibinfo{pages}{53--107}, \doiprefix\url{10.1103/RevModPhys.82.53} (\bibinfo{year}{2010}).

\bibitem{Mydosh2011}
\bibinfo{author}{Mydosh, J.~A.} \& \bibinfo{author}{Oppeneer, P.~M.}
\newblock \bibinfo{journal}{\bibinfo{title}{Colloquium: Hidden order, superconductivity, and magnetism: The unsolved case of ${\mathrm{uru}}_{2}{\mathrm{si}}_{2}$}}.
\newblock {\emph{\JournalTitle{Rev. Mod. Phys.}}} \textbf{\bibinfo{volume}{83}}, \bibinfo{pages}{1301--1322}, \doiprefix\url{10.1103/RevModPhys.83.1301} (\bibinfo{year}{2011}).

\bibitem{Tsirlin2012}
\bibinfo{author}{Tsirlin, A.~A.} \emph{et~al.}
\newblock \bibinfo{journal}{\bibinfo{title}{Hidden magnetic order in cuncn}}.
\newblock {\emph{\JournalTitle{Phys. Rev. B}}} \textbf{\bibinfo{volume}{85}}, \bibinfo{pages}{224431}, \doiprefix\url{10.1103/PhysRevB.85.224431} (\bibinfo{year}{2012}).

\bibitem{Kim2012}
\bibinfo{author}{Kim, B.~H.}, \bibinfo{author}{Khaliullin, G.} \& \bibinfo{author}{Min, B.~I.}
\newblock \bibinfo{journal}{\bibinfo{title}{Magnetic couplings, optical spectra, and spin-orbit exciton in $5d$ electron mott insulator ${\mathrm{sr}}_{2}{\mathrm{iro}}_{4}$}}.
\newblock {\emph{\JournalTitle{Phys. Rev. Lett.}}} \textbf{\bibinfo{volume}{109}}, \bibinfo{pages}{167205}, \doiprefix\url{10.1103/PhysRevLett.109.167205} (\bibinfo{year}{2012}).

\bibitem{Sugiyama2014}
\bibinfo{author}{Sugiyama, J.} \emph{et~al.}
\newblock \bibinfo{journal}{\bibinfo{title}{Hidden magnetic order in sr${}_{2}$vo${}_{4}$ clarified with ${\ensuremath{\mu}}^{+}$sr}}.
\newblock {\emph{\JournalTitle{Phys. Rev. B}}} \textbf{\bibinfo{volume}{89}}, \bibinfo{pages}{020402}, \doiprefix\url{10.1103/PhysRevB.89.020402} (\bibinfo{year}{2014}).

\bibitem{Huang2014}
\bibinfo{author}{Huang, Y.-P.}, \bibinfo{author}{Chen, G.} \& \bibinfo{author}{Hermele, M.}
\newblock \bibinfo{journal}{\bibinfo{title}{Quantum spin ices and topological phases from dipolar-octupolar doublets on the pyrochlore lattice}}.
\newblock {\emph{\JournalTitle{Phys. Rev. Lett.}}} \textbf{\bibinfo{volume}{112}}, \bibinfo{pages}{167203}, \doiprefix\url{10.1103/PhysRevLett.112.167203} (\bibinfo{year}{2014}).

\bibitem{Zhao2016}
\bibinfo{author}{Zhao, L.} \emph{et~al.}
\newblock \bibinfo{journal}{\bibinfo{title}{Evidence of an odd-parity hidden order in a spin--orbit coupled correlated iridate}}.
\newblock {\emph{\JournalTitle{Nature Physics}}} \textbf{\bibinfo{volume}{12}}, \bibinfo{pages}{32--36}, \doiprefix\url{10.1038/nphys3517} (\bibinfo{year}{2016}).

\bibitem{Cameron2016}
\bibinfo{author}{Cameron, A.~S.}, \bibinfo{author}{Friemel, G.} \& \bibinfo{author}{Inosov, D.~S.}
\newblock \bibinfo{journal}{\bibinfo{title}{Multipolar phases and magnetically hidden order: review of the heavy-fermion compound $ce_{1-x}la_xb_6$}}.
\newblock {\emph{\JournalTitle{Reports on Progress in Physics}}} \textbf{\bibinfo{volume}{79}}, \bibinfo{pages}{066502}, \doiprefix\url{10.1088/0034-4885/79/6/066502} (\bibinfo{year}{2016}).

\bibitem{Li2016}
\bibinfo{author}{Li, Y.-D.}, \bibinfo{author}{Wang, X.} \& \bibinfo{author}{Chen, G.}
\newblock \bibinfo{journal}{\bibinfo{title}{Hidden multipolar orders of dipole-octupole doublets on a triangular lattice}}.
\newblock {\emph{\JournalTitle{Phys. Rev. B}}} \textbf{\bibinfo{volume}{94}}, \bibinfo{pages}{201114}, \doiprefix\url{10.1103/PhysRevB.94.201114} (\bibinfo{year}{2016}).

\bibitem{Lu2017}
\bibinfo{author}{Lu, L.} \emph{et~al.}
\newblock \bibinfo{journal}{\bibinfo{title}{Magnetism and local symmetry breaking in a mott insulator with strong spin orbit interactions}}.
\newblock {\emph{\JournalTitle{Nature Communications}}} \textbf{\bibinfo{volume}{8}}, \bibinfo{pages}{14407}, \doiprefix\url{10.1038/ncomms14407} (\bibinfo{year}{2017}).

\bibitem{Li2017}
\bibinfo{author}{Li, Y.-D.} \& \bibinfo{author}{Chen, G.}
\newblock \bibinfo{journal}{\bibinfo{title}{Symmetry enriched u(1) topological orders for dipole-octupole doublets on a pyrochlore lattice}}.
\newblock {\emph{\JournalTitle{Phys. Rev. B}}} \textbf{\bibinfo{volume}{95}}, \bibinfo{pages}{041106}, \doiprefix\url{10.1103/PhysRevB.95.041106} (\bibinfo{year}{2017}).

\bibitem{Kim2017}
\bibinfo{author}{Kim, B.}, \bibinfo{author}{Khmelevskyi, S.}, \bibinfo{author}{Mohn, P.} \& \bibinfo{author}{Franchini, C.}
\newblock \bibinfo{journal}{\bibinfo{title}{Competing magnetic interactions in a spin-$\frac{1}{2}$ square lattice: Hidden order in ${\mathrm{sr}}_{2}{\mathrm{vo}}_{4}$}}.
\newblock {\emph{\JournalTitle{Phys. Rev. B}}} \textbf{\bibinfo{volume}{96}}, \bibinfo{pages}{180405}, \doiprefix\url{10.1103/PhysRevB.96.180405} (\bibinfo{year}{2017}).

\bibitem{LiuC2018}
\bibinfo{author}{Liu, C.}, \bibinfo{author}{Li, Y.-D.} \& \bibinfo{author}{Chen, G.}
\newblock \bibinfo{journal}{\bibinfo{title}{Selective measurements of intertwined multipolar orders: Non-kramers doublets on a triangular lattice}}.
\newblock {\emph{\JournalTitle{Phys. Rev. B}}} \textbf{\bibinfo{volume}{98}}, \bibinfo{pages}{045119}, \doiprefix\url{10.1103/PhysRevB.98.045119} (\bibinfo{year}{2018}).

\bibitem{Ishikawa2019}
\bibinfo{author}{Ishikawa, H.} \emph{et~al.}
\newblock \bibinfo{journal}{\bibinfo{title}{Ordering of hidden multipoles in spin-orbit entangled $5{d}^{1}$ ta chlorides}}.
\newblock {\emph{\JournalTitle{Phys. Rev. B}}} \textbf{\bibinfo{volume}{100}}, \bibinfo{pages}{045142}, \doiprefix\url{10.1103/PhysRevB.100.045142} (\bibinfo{year}{2019}).

\bibitem{Gaudet2019}
\bibinfo{author}{Gaudet, J.} \emph{et~al.}
\newblock \bibinfo{journal}{\bibinfo{title}{Quantum spin ice dynamics in the dipole-octupole pyrochlore magnet ${\mathrm{ce}}_{2}{\mathrm{zr}}_{2}{\mathrm{o}}_{7}$}}.
\newblock {\emph{\JournalTitle{Phys. Rev. Lett.}}} \textbf{\bibinfo{volume}{122}}, \bibinfo{pages}{187201}, \doiprefix\url{10.1103/PhysRevLett.122.187201} (\bibinfo{year}{2019}).

\bibitem{Shen2019}
\bibinfo{author}{Shen, Y.} \emph{et~al.}
\newblock \bibinfo{journal}{\bibinfo{title}{Intertwined dipolar and multipolar order in the triangular-lattice magnet tmmggao4}}.
\newblock {\emph{\JournalTitle{Nature Communications}}} \textbf{\bibinfo{volume}{10}}, \bibinfo{pages}{4530}, \doiprefix\url{10.1038/s41467-019-12410-3} (\bibinfo{year}{2019}).

\bibitem{Rau2019}
\bibinfo{author}{Rau, J.~G.} \& \bibinfo{author}{Gingras, M.~J.}
\newblock \bibinfo{journal}{\bibinfo{title}{Frustrated quantum rare-earth pyrochlores}}.
\newblock {\emph{\JournalTitle{Annual Review of Condensed Matter Physics}}} \textbf{\bibinfo{volume}{10}}, \bibinfo{pages}{357--386}, \doiprefix\url{https://doi.org/10.1146/annurev-conmatphys-022317-110520} (\bibinfo{year}{2019}).

\bibitem{Hirai2020}
\bibinfo{author}{Hirai, D.} \emph{et~al.}
\newblock \bibinfo{journal}{\bibinfo{title}{Detection of multipolar orders in the spin-orbit-coupled $5d$ mott insulator $\mathrm{B}{\mathrm{a}}_{2}\mathrm{MgRe}{\mathrm{o}}_{6}$}}.
\newblock {\emph{\JournalTitle{Phys. Rev. Res.}}} \textbf{\bibinfo{volume}{2}}, \bibinfo{pages}{022063}, \doiprefix\url{10.1103/PhysRevResearch.2.022063} (\bibinfo{year}{2020}).

\bibitem{Aeppli2020}
\bibinfo{author}{Aeppli, G.}, \bibinfo{author}{Balatsky, A.~V.}, \bibinfo{author}{R{\o}nnow, H.~M.} \& \bibinfo{author}{Spaldin, N.~A.}
\newblock \bibinfo{journal}{\bibinfo{title}{Hidden, entangled and resonating order}}.
\newblock {\emph{\JournalTitle{Nature Reviews Materials}}} \textbf{\bibinfo{volume}{5}}, \bibinfo{pages}{477--479}, \doiprefix\url{10.1038/s41578-020-0207-z} (\bibinfo{year}{2020}).

\bibitem{Zvereva2020}
\bibinfo{author}{Zvereva, E.~A.} \emph{et~al.}
\newblock \bibinfo{journal}{\bibinfo{title}{Hidden magnetic order in the triangular-lattice magnet ${\mathrm{li}}_{2}{\mathrm{mnteo}}_{6}$}}.
\newblock {\emph{\JournalTitle{Phys. Rev. B}}} \textbf{\bibinfo{volume}{102}}, \bibinfo{pages}{094433}, \doiprefix\url{10.1103/PhysRevB.102.094433} (\bibinfo{year}{2020}).

\bibitem{Maharaj2020}
\bibinfo{author}{Maharaj, D.~D.} \emph{et~al.}
\newblock \bibinfo{journal}{\bibinfo{title}{Octupolar versus n\'eel order in cubic $5{d}^{2}$ double perovskites}}.
\newblock {\emph{\JournalTitle{Phys. Rev. Lett.}}} \textbf{\bibinfo{volume}{124}}, \bibinfo{pages}{087206}, \doiprefix\url{10.1103/PhysRevLett.124.087206} (\bibinfo{year}{2020}).

\bibitem{Sibille2020}
\bibinfo{author}{Sibille, R.} \emph{et~al.}
\newblock \bibinfo{journal}{\bibinfo{title}{A quantum liquid of magnetic octupoles on the pyrochlore lattice}}.
\newblock {\emph{\JournalTitle{Nature Physics}}} \textbf{\bibinfo{volume}{16}}, \bibinfo{pages}{546--552}, \doiprefix\url{10.1038/s41567-020-0827-7} (\bibinfo{year}{2020}).

\bibitem{Fioremosca2021}
\bibinfo{author}{Fiore~Mosca, D.} \emph{et~al.}
\newblock \bibinfo{journal}{\bibinfo{title}{Interplay between multipolar spin interactions, jahn-teller effect, and electronic correlation in a ${J}_{\text{eff}}=\frac{3}{2}$ insulator}}.
\newblock {\emph{\JournalTitle{Phys. Rev. B}}} \textbf{\bibinfo{volume}{103}}, \bibinfo{pages}{104401}, \doiprefix\url{10.1103/PhysRevB.103.104401} (\bibinfo{year}{2021}).

\bibitem{Pourovskii2021}
\bibinfo{author}{Pourovskii, L.~V.}, \bibinfo{author}{Mosca, D.~F.} \& \bibinfo{author}{Franchini, C.}
\newblock \bibinfo{journal}{\bibinfo{title}{Ferro-octupolar order and low-energy excitations in ${\mathrm{d}}^{2}$ double perovskites of osmium}}.
\newblock {\emph{\JournalTitle{Phys. Rev. Lett.}}} \textbf{\bibinfo{volume}{127}}, \bibinfo{pages}{237201}, \doiprefix\url{10.1103/PhysRevLett.127.237201} (\bibinfo{year}{2021}).

\bibitem{Pourovskii2021f}
\bibinfo{author}{Pourovskii, L.~V.} \& \bibinfo{author}{Khmelevskyi, S.}
\newblock \bibinfo{journal}{\bibinfo{title}{Hidden order and multipolar exchange striction in a correlated <i>f</i>-electron system}}.
\newblock {\emph{\JournalTitle{Proceedings of the National Academy of Sciences}}} \textbf{\bibinfo{volume}{118}}, \bibinfo{pages}{e2025317118}, \doiprefix\url{10.1073/pnas.2025317118} (\bibinfo{year}{2021}).
\newblock \eprint{https://www.pnas.org/doi/pdf/10.1073/pnas.2025317118}.

\bibitem{Smith2022}
\bibinfo{author}{Smith, E.~M.} \emph{et~al.}
\newblock \bibinfo{journal}{\bibinfo{title}{Case for a ${\mathrm{u}(1)}_{\ensuremath{\pi}}$ quantum spin liquid ground state in the dipole-octupole pyrochlore ${\mathrm{ce}}_{2}{\mathrm{zr}}_{2}{\mathrm{o}}_{7}$}}.
\newblock {\emph{\JournalTitle{Phys. Rev. X}}} \textbf{\bibinfo{volume}{12}}, \bibinfo{pages}{021015}, \doiprefix\url{10.1103/PhysRevX.12.021015} (\bibinfo{year}{2022}).

\bibitem{Chen2023}
\bibinfo{author}{Chen, G.}
\newblock \bibinfo{journal}{\bibinfo{title}{Distinguishing thermodynamics and spectroscopy for octupolar u(1) spin liquid of ce pyrochlores}}.
\newblock {\emph{\JournalTitle{Phys. Rev. Res.}}} \textbf{\bibinfo{volume}{5}}, \bibinfo{pages}{033169}, \doiprefix\url{10.1103/PhysRevResearch.5.033169} (\bibinfo{year}{2023}).

\bibitem{Voleti2023}
\bibinfo{author}{Voleti, S.}, \bibinfo{author}{Pradhan, K.}, \bibinfo{author}{Bhattacharjee, S.}, \bibinfo{author}{Saha-Dasgupta, T.} \& \bibinfo{author}{Paramekanti, A.}
\newblock \bibinfo{journal}{\bibinfo{title}{Probing octupolar hidden order via janus impurities}}.
\newblock {\emph{\JournalTitle{npj Quantum Materials}}} \textbf{\bibinfo{volume}{8}}, \bibinfo{pages}{42}, \doiprefix\url{10.1038/s41535-023-00575-6} (\bibinfo{year}{2023}).

\bibitem{Verbeedk2023}
\bibinfo{author}{Verbeek, X.~H.}, \bibinfo{author}{Urru, A.} \& \bibinfo{author}{Spaldin, N.~A.}
\newblock \bibinfo{journal}{\bibinfo{title}{Hidden orders and (anti-)magnetoelectric effects in ${\mathrm{cr}}_{2}{\mathrm{o}}_{3}$ and $\ensuremath{\alpha}\text{\ensuremath{-}}{\mathrm{fe}}_{2}{\mathrm{o}}_{3}$}}.
\newblock {\emph{\JournalTitle{Phys. Rev. Res.}}} \textbf{\bibinfo{volume}{5}}, \bibinfo{pages}{L042018}, \doiprefix\url{10.1103/PhysRevResearch.5.L042018} (\bibinfo{year}{2023}).

\bibitem{Szilva2023}
\bibinfo{author}{Szilva, A.} \emph{et~al.}
\newblock \bibinfo{journal}{\bibinfo{title}{Quantitative theory of magnetic interactions in solids}}.
\newblock {\emph{\JournalTitle{Rev. Mod. Phys.}}} \textbf{\bibinfo{volume}{95}}, \bibinfo{pages}{035004}, \doiprefix\url{10.1103/RevModPhys.95.035004} (\bibinfo{year}{2023}).

\bibitem{Santini2009}
\bibinfo{author}{Santini, P.} \emph{et~al.}
\newblock \bibinfo{journal}{\bibinfo{title}{Multipolar interactions in $f$-electron systems: The paradigm of actinide dioxides}}.
\newblock {\emph{\JournalTitle{Rev. Mod. Phys.}}} \textbf{\bibinfo{volume}{81}}, \bibinfo{pages}{807--863}, \doiprefix\url{10.1103/RevModPhys.81.807} (\bibinfo{year}{2009}).

\bibitem{Bultmark2009}
\bibinfo{author}{Bultmark, F.}, \bibinfo{author}{Cricchio, F.}, \bibinfo{author}{Gr\aa{}n\"as, O.} \& \bibinfo{author}{Nordstr\"om, L.}
\newblock \bibinfo{journal}{\bibinfo{title}{Multipole decomposition of $\text{LDA}+u$ energy and its application to actinide compounds}}.
\newblock {\emph{\JournalTitle{Phys. Rev. B}}} \textbf{\bibinfo{volume}{80}}, \bibinfo{pages}{035121}, \doiprefix\url{10.1103/PhysRevB.80.035121} (\bibinfo{year}{2009}).

\bibitem{Pourovskii2016}
\bibinfo{author}{Pourovskii, L.~V.}
\newblock \bibinfo{journal}{\bibinfo{title}{Two-site fluctuations and multipolar intersite exchange interactions in strongly correlated systems}}.
\newblock {\emph{\JournalTitle{Phys. Rev. B}}} \textbf{\bibinfo{volume}{94}}, \bibinfo{pages}{115117}, \doiprefix\url{10.1103/PhysRevB.94.115117} (\bibinfo{year}{2016}).

\bibitem{Pi2014}
\bibinfo{author}{Pi, S.-T.}, \bibinfo{author}{Nanguneri, R.} \& \bibinfo{author}{Savrasov, S.}
\newblock \bibinfo{journal}{\bibinfo{title}{Calculation of multipolar exchange interactions in spin-orbital coupled systems}}.
\newblock {\emph{\JournalTitle{Phys. Rev. Lett.}}} \textbf{\bibinfo{volume}{112}}, \bibinfo{pages}{077203}, \doiprefix\url{10.1103/PhysRevLett.112.077203} (\bibinfo{year}{2014}).

\bibitem{Pi2014b}
\bibinfo{author}{Pi, S.-T.}, \bibinfo{author}{Nanguneri, R.} \& \bibinfo{author}{Savrasov, S.}
\newblock \bibinfo{journal}{\bibinfo{title}{Anisotropic multipolar exchange interactions in systems with strong spin-orbit coupling}}.
\newblock {\emph{\JournalTitle{Phys. Rev. B}}} \textbf{\bibinfo{volume}{90}}, \bibinfo{pages}{045148}, \doiprefix\url{10.1103/PhysRevB.90.045148} (\bibinfo{year}{2014}).

\bibitem{FioreMosca2022}
\bibinfo{author}{Fiore~Mosca, D.}, \bibinfo{author}{Pourovskii, L.~V.} \& \bibinfo{author}{Franchini, C.}
\newblock \bibinfo{journal}{\bibinfo{title}{Modeling magnetic multipolar phases in density functional theory}}.
\newblock {\emph{\JournalTitle{Phys. Rev. B}}} \textbf{\bibinfo{volume}{106}}, \bibinfo{pages}{035127}, \doiprefix\url{10.1103/PhysRevB.106.035127} (\bibinfo{year}{2022}).

\bibitem{Schaufelberger2023}
\bibinfo{author}{Schaufelberger, L.}, \bibinfo{author}{Merkel, M.~E.}, \bibinfo{author}{Tehrani, A.~M.}, \bibinfo{author}{Spaldin, N.~A.} \& \bibinfo{author}{Ederer, C.}
\newblock \bibinfo{journal}{\bibinfo{title}{Exploring energy landscapes of charge multipoles using constrained density functional theory}}.
\newblock {\emph{\JournalTitle{Phys. Rev. Res.}}} \textbf{\bibinfo{volume}{5}}, \bibinfo{pages}{033172}, \doiprefix\url{10.1103/PhysRevResearch.5.033172} (\bibinfo{year}{2023}).

\bibitem{Kimata2021}
\bibinfo{author}{Kimata, M.} \emph{et~al.}
\newblock \bibinfo{journal}{\bibinfo{title}{X-ray study of ferroic octupole order producing anomalous hall effect}}.
\newblock {\emph{\JournalTitle{Nature Communications}}} \textbf{\bibinfo{volume}{12}}, \bibinfo{pages}{5582}, \doiprefix\url{10.1038/s41467-021-25834-7} (\bibinfo{year}{2021}).

\bibitem{McMorrow2001}
\bibinfo{author}{McMorrow, D.~F.}, \bibinfo{author}{McEwen, K.~A.}, \bibinfo{author}{Steigenberger, U.}, \bibinfo{author}{R\o{}nnow, H.~M.} \& \bibinfo{author}{Yakhou, F.}
\newblock \bibinfo{journal}{\bibinfo{title}{X-ray resonant scattering study of the quadrupolar order in ${\mathrm{upd}}_{3}$}}.
\newblock {\emph{\JournalTitle{Phys. Rev. Lett.}}} \textbf{\bibinfo{volume}{87}}, \bibinfo{pages}{057201}, \doiprefix\url{10.1103/PhysRevLett.87.057201} (\bibinfo{year}{2001}).

\bibitem{Santini2006}
\bibinfo{author}{Santini, P.}, \bibinfo{author}{Carretta, S.}, \bibinfo{author}{Magnani, N.}, \bibinfo{author}{Amoretti, G.} \& \bibinfo{author}{Caciuffo, R.}
\newblock \bibinfo{journal}{\bibinfo{title}{Hidden order and low-energy excitations in ${\mathrm{npo}}_{2}$}}.
\newblock {\emph{\JournalTitle{Phys. Rev. Lett.}}} \textbf{\bibinfo{volume}{97}}, \bibinfo{pages}{207203}, \doiprefix\url{10.1103/PhysRevLett.97.207203} (\bibinfo{year}{2006}).

\bibitem{Magnani2008}
\bibinfo{author}{Magnani, N.} \emph{et~al.}
\newblock \bibinfo{journal}{\bibinfo{title}{Inelastic neutron scattering study of the multipolar order parameter in ${\text{npo}}_{2}$}}.
\newblock {\emph{\JournalTitle{Phys. Rev. B}}} \textbf{\bibinfo{volume}{78}}, \bibinfo{pages}{104425}, \doiprefix\url{10.1103/PhysRevB.78.104425} (\bibinfo{year}{2008}).

\bibitem{VASALA2015}
\bibinfo{author}{Vasala, S.} \& \bibinfo{author}{Karppinen, M.}
\newblock \bibinfo{journal}{\bibinfo{title}{A2bbo6 perovskites: A review}}.
\newblock {\emph{\JournalTitle{Progress in Solid State Chemistry}}} \textbf{\bibinfo{volume}{43}}, \bibinfo{pages}{1--36}, \doiprefix\url{https://doi.org/10.1016/j.progsolidstchem.2014.08.001} (\bibinfo{year}{2015}).

\bibitem{Chen2024}
\bibinfo{author}{Chen, J.}, \bibinfo{author}{Feng, H.~L.} \& \bibinfo{author}{Yamaura, K.}
\newblock \bibinfo{journal}{\bibinfo{title}{Review of progress in the materials development of re, os, and ir-based double perovskite oxides}}.
\newblock {\emph{\JournalTitle{Materials Today Physics}}} \textbf{\bibinfo{volume}{40}}, \bibinfo{pages}{101302}, \doiprefix\url{https://doi.org/10.1016/j.mtphys.2023.101302} (\bibinfo{year}{2024}).

\bibitem{Iwahara2022}
\bibinfo{author}{Iwahara, N.}, \bibinfo{author}{Huang, Z.}, \bibinfo{author}{Neefjes, I.} \& \bibinfo{author}{Chibotaru, L.~F.}
\newblock \bibinfo{journal}{\bibinfo{title}{Multipolar exchange interaction and complex order in insulating lanthanides}}.
\newblock {\emph{\JournalTitle{Phys. Rev. B}}} \textbf{\bibinfo{volume}{105}}, \bibinfo{pages}{144401}, \doiprefix\url{10.1103/PhysRevB.105.144401} (\bibinfo{year}{2022}).

\bibitem{Ning2023}
\bibinfo{author}{Ning, H.} \emph{et~al.}
\newblock \bibinfo{journal}{\bibinfo{title}{A coherent phonon-induced hidden quadrupolar ordered state in ca2ruo4}}.
\newblock {\emph{\JournalTitle{Nature Communications}}} \textbf{\bibinfo{volume}{14}}, \bibinfo{pages}{8258}, \doiprefix\url{10.1038/s41467-023-44021-4} (\bibinfo{year}{2023}).

\bibitem{Bersuker2006}
\bibinfo{author}{Bersuker, I.}
\newblock \emph{\bibinfo{title}{The {Jahn}-{Teller} {Effect}}} (\bibinfo{publisher}{Cambridge University Press}, \bibinfo{address}{Cambridge}, \bibinfo{year}{2006}).

\bibitem{Iwahara2023}
\bibinfo{author}{Iwahara, N.} \& \bibinfo{author}{Chibotaru, L.~F.}
\newblock \bibinfo{journal}{\bibinfo{title}{Vibronic order and emergent magnetism in cubic ${d}^{1}$ double perovskites}}.
\newblock {\emph{\JournalTitle{Phys. Rev. B}}} \textbf{\bibinfo{volume}{107}}, \bibinfo{pages}{L220404}, \doiprefix\url{10.1103/PhysRevB.107.L220404} (\bibinfo{year}{2023}).

\bibitem{Fioremosca2024b}
\bibinfo{author}{Fiore~Mosca, D.}, \bibinfo{author}{Franchini, C.} \& \bibinfo{author}{Pourovskii, L.~V.}
\newblock \bibinfo{journal}{\bibinfo{title}{Interplay of superexchange and vibronic effects in the hidden order of ${\mathrm{ba}}_{2}{\mathrm{mgreo}}_{6}$ from first principles}}.
\newblock {\emph{\JournalTitle{Phys. Rev. B}}} \textbf{\bibinfo{volume}{110}}, \bibinfo{pages}{L201101}, \doiprefix\url{10.1103/PhysRevB.110.L201101} (\bibinfo{year}{2024}).

\bibitem{Spaldin2008}
\bibinfo{author}{Spaldin, N.~A.}, \bibinfo{author}{Fiebig, M.} \& \bibinfo{author}{Mostovoy, M.}
\newblock \bibinfo{journal}{\bibinfo{title}{The toroidal moment in condensed-matter physics and its relation to the magnetoelectric effect}}.
\newblock {\emph{\JournalTitle{Journal of Physics: Condensed Matter}}} \textbf{\bibinfo{volume}{20}}, \bibinfo{pages}{434203}, \doiprefix\url{10.1088/0953-8984/20/43/434203} (\bibinfo{year}{2008}).

\bibitem{Smejkal2022}
\bibinfo{author}{\v{S}mejkal, L.}, \bibinfo{author}{Sinova, J.} \& \bibinfo{author}{Jungwirth, T.}
\newblock \bibinfo{journal}{\bibinfo{title}{Emerging research landscape of altermagnetism}}.
\newblock {\emph{\JournalTitle{Phys. Rev. X}}} \textbf{\bibinfo{volume}{12}}, \bibinfo{pages}{040501}, \doiprefix\url{10.1103/PhysRevX.12.040501} (\bibinfo{year}{2022}).

\bibitem{jungwirth2024}
\bibinfo{author}{Jungwirth, T.}, \bibinfo{author}{Fernandes, R.~M.}, \bibinfo{author}{Sinova, J.} \& \bibinfo{author}{Smejkal, L.}
\newblock \bibinfo{title}{Altermagnets and beyond: Nodal magnetically-ordered phases} (\bibinfo{year}{2024}).
\newblock \eprint{2409.10034}.

\bibitem{Ederer2007}
\bibinfo{author}{Ederer, C.} \& \bibinfo{author}{Spaldin, N.~A.}
\newblock \bibinfo{journal}{\bibinfo{title}{Towards a microscopic theory of toroidal moments in bulk periodic crystals}}.
\newblock {\emph{\JournalTitle{Phys. Rev. B}}} \textbf{\bibinfo{volume}{76}}, \bibinfo{pages}{214404}, \doiprefix\url{10.1103/PhysRevB.76.214404} (\bibinfo{year}{2007}).

\bibitem{Hayami2019}
\bibinfo{author}{Hayami, S.}, \bibinfo{author}{Yanagi, Y.}, \bibinfo{author}{Kusunose, H.} \& \bibinfo{author}{Motome, Y.}
\newblock \bibinfo{journal}{\bibinfo{title}{Electric toroidal quadrupoles in the spin-orbit-coupled metal ${\mathrm{cd}}_{2}{\mathrm{re}}_{2}{\mathrm{o}}_{7}$}}.
\newblock {\emph{\JournalTitle{Phys. Rev. Lett.}}} \textbf{\bibinfo{volume}{122}}, \bibinfo{pages}{147602}, \doiprefix\url{10.1103/PhysRevLett.122.147602} (\bibinfo{year}{2019}).

\bibitem{Yatsuhiro2021}
\bibinfo{author}{Yatsushiro, M.}, \bibinfo{author}{Kusunose, H.} \& \bibinfo{author}{Hayami, S.}
\newblock \bibinfo{journal}{\bibinfo{title}{Multipole classification in 122 magnetic point groups for unified understanding of multiferroic responses and transport phenomena}}.
\newblock {\emph{\JournalTitle{Phys. Rev. B}}} \textbf{\bibinfo{volume}{104}}, \bibinfo{pages}{054412}, \doiprefix\url{10.1103/PhysRevB.104.054412} (\bibinfo{year}{2021}).

\bibitem{Bhowal2024a}
\bibinfo{author}{Bhowal, S.} \& \bibinfo{author}{Spaldin, N.~A.}
\newblock \bibinfo{journal}{\bibinfo{title}{Ferroically ordered magnetic octupoles in $d$-wave altermagnets}}.
\newblock {\emph{\JournalTitle{Phys. Rev. X}}} \textbf{\bibinfo{volume}{14}}, \bibinfo{pages}{011019}, \doiprefix\url{10.1103/PhysRevX.14.011019} (\bibinfo{year}{2024}).

\bibitem{DZYALOSHINSKY1958}
\bibinfo{author}{Dzyaloshinsky, I.}
\newblock \bibinfo{journal}{\bibinfo{title}{A thermodynamic theory of “weak” ferromagnetism of antiferromagnetics}}.
\newblock {\emph{\JournalTitle{Journal of Physics and Chemistry of Solids}}} \textbf{\bibinfo{volume}{4}}, \bibinfo{pages}{241--255}, \doiprefix\url{https://doi.org/10.1016/0022-3697(58)90076-3} (\bibinfo{year}{1958}).

\bibitem{Moriyab}
\bibinfo{author}{Moriya, T.}
\newblock \bibinfo{journal}{\bibinfo{title}{Anisotropic superexchange interaction and weak ferromagnetism}}.
\newblock {\emph{\JournalTitle{Phys. Rev.}}} \textbf{\bibinfo{volume}{120}}, \bibinfo{pages}{91--98}, \doiprefix\url{10.1103/PhysRev.120.91} (\bibinfo{year}{1960}).

\bibitem{Takeda1972}
\bibinfo{author}{Takeda, T.}, \bibinfo{author}{Yamaguchi, Y.} \& \bibinfo{author}{Watanabe, H.}
\newblock \bibinfo{journal}{\bibinfo{title}{Magnetic structure of srfeo3}}.
\newblock {\emph{\JournalTitle{Journal of the Physical Society of Japan}}} \textbf{\bibinfo{volume}{33}}, \bibinfo{pages}{967--969}, \doiprefix\url{10.1143/JPSJ.33.967} (\bibinfo{year}{1972}).
\newblock \eprint{https://doi.org/10.1143/JPSJ.33.967}.

\bibitem{Liu2015}
\bibinfo{author}{Liu, P.} \emph{et~al.}
\newblock \bibinfo{journal}{\bibinfo{title}{Anisotropic magnetic couplings and structure-driven canted to collinear transitions in ${\mathrm{sr}}_{2}{\mathrm{iro}}_{4}$ by magnetically constrained noncollinear dft}}.
\newblock {\emph{\JournalTitle{Phys. Rev. B}}} \textbf{\bibinfo{volume}{92}}, \bibinfo{pages}{054428}, \doiprefix\url{10.1103/PhysRevB.92.054428} (\bibinfo{year}{2015}).

\bibitem{Tokura2021}
\bibinfo{author}{Tokura, Y.} \& \bibinfo{author}{Kanazawa, N.}
\newblock \bibinfo{journal}{\bibinfo{title}{Magnetic skyrmion materials}}.
\newblock {\emph{\JournalTitle{Chemical Reviews}}} \textbf{\bibinfo{volume}{121}}, \bibinfo{pages}{2857--2897}, \doiprefix\url{10.1021/acs.chemrev.0c00297} (\bibinfo{year}{2021}).
\newblock \bibinfo{note}{PMID: 33164494}, \eprint{https://doi.org/10.1021/acs.chemrev.0c00297}.

\bibitem{Xu2019}
\bibinfo{author}{Xu, C.}, \bibinfo{author}{Xu, B.}, \bibinfo{author}{Dup\'e, B.} \& \bibinfo{author}{Bellaiche, L.}
\newblock \bibinfo{journal}{\bibinfo{title}{Magnetic interactions in ${\mathrm{bifeo}}_{3}$: A first-principles study}}.
\newblock {\emph{\JournalTitle{Phys. Rev. B}}} \textbf{\bibinfo{volume}{99}}, \bibinfo{pages}{104420}, \doiprefix\url{10.1103/PhysRevB.99.104420} (\bibinfo{year}{2019}).

\bibitem{Yang2023}
\bibinfo{author}{Yang, H.}, \bibinfo{author}{Liang, J.} \& \bibinfo{author}{Cui, Q.}
\newblock \bibinfo{journal}{\bibinfo{title}{First-principles calculations for dzyaloshinskii--moriya interaction}}.
\newblock {\emph{\JournalTitle{Nature Reviews Physics}}} \textbf{\bibinfo{volume}{5}}, \bibinfo{pages}{43--61}, \doiprefix\url{10.1038/s42254-022-00529-0} (\bibinfo{year}{2023}).

\bibitem{Fert2023}
\bibinfo{author}{Fert, A.}, \bibinfo{author}{Chshiev, M.}, \bibinfo{author}{Thiaville, A.} \& \bibinfo{author}{Yang, H.}
\newblock \bibinfo{journal}{\bibinfo{title}{From early theories of dzyaloshinskii–moriya interactions in metallic systems to today’s novel roads}}.
\newblock {\emph{\JournalTitle{Journal of the Physical Society of Japan}}} \textbf{\bibinfo{volume}{92}}, \bibinfo{pages}{081001}, \doiprefix\url{10.7566/JPSJ.92.081001} (\bibinfo{year}{2023}).
\newblock \eprint{https://doi.org/10.7566/JPSJ.92.081001}.

\bibitem{Pan2020}
\bibinfo{author}{Pan, H.}, \bibinfo{author}{Wu, F.} \& \bibinfo{author}{Das~Sarma, S.}
\newblock \bibinfo{journal}{\bibinfo{title}{Band topology, hubbard model, heisenberg model, and dzyaloshinskii-moriya interaction in twisted bilayer ${\mathrm{wse}}_{2}$}}.
\newblock {\emph{\JournalTitle{Phys. Rev. Res.}}} \textbf{\bibinfo{volume}{2}}, \bibinfo{pages}{033087}, \doiprefix\url{10.1103/PhysRevResearch.2.033087} (\bibinfo{year}{2020}).

\bibitem{HEIDE2009}
\bibinfo{author}{Heide, M.}, \bibinfo{author}{Bihlmayer, G.} \& \bibinfo{author}{Blügel, S.}
\newblock \bibinfo{journal}{\bibinfo{title}{Describing dzyaloshinskii–moriya spirals from first principles}}.
\newblock {\emph{\JournalTitle{Physica B: Condensed Matter}}} \textbf{\bibinfo{volume}{404}}, \bibinfo{pages}{2678--2683}, \doiprefix\url{https://doi.org/10.1016/j.physb.2009.06.070} (\bibinfo{year}{2009}).

\bibitem{Moskvin2019}
\bibinfo{author}{Moskvin, A.}
\newblock \bibinfo{journal}{\bibinfo{title}{Dzyaloshinskii–moriya coupling in 3d insulators}}.
\newblock {\emph{\JournalTitle{Condensed Matter}}} \textbf{\bibinfo{volume}{4}}, \doiprefix\url{10.3390/condmat4040084} (\bibinfo{year}{2019}).

\bibitem{Kuepferling2023}
\bibinfo{author}{Kuepferling, M.} \emph{et~al.}
\newblock \bibinfo{journal}{\bibinfo{title}{Measuring interfacial dzyaloshinskii-moriya interaction in ultrathin magnetic films}}.
\newblock {\emph{\JournalTitle{Rev. Mod. Phys.}}} \textbf{\bibinfo{volume}{95}}, \bibinfo{pages}{015003}, \doiprefix\url{10.1103/RevModPhys.95.015003} (\bibinfo{year}{2023}).

\bibitem{CAMLEY2023}
\bibinfo{author}{Camley, R.~E.} \& \bibinfo{author}{Livesey, K.~L.}
\newblock \bibinfo{journal}{\bibinfo{title}{Consequences of the dzyaloshinskii-moriya interaction}}.
\newblock {\emph{\JournalTitle{Surface Science Reports}}} \textbf{\bibinfo{volume}{78}}, \bibinfo{pages}{100605}, \doiprefix\url{https://doi.org/10.1016/j.surfrep.2023.100605} (\bibinfo{year}{2023}).

\bibitem{BUYERS1996}
\bibinfo{author}{Buyers, W.}
\newblock \bibinfo{journal}{\bibinfo{title}{Low moments in heavy-fermion systems}}.
\newblock {\emph{\JournalTitle{Physica B: Condensed Matter}}} \textbf{\bibinfo{volume}{223-224}}, \bibinfo{pages}{9--14}, \doiprefix\url{https://doi.org/10.1016/0921-4526(96)00027-0} (\bibinfo{year}{1996}).
\newblock \bibinfo{note}{Proceedings of the International Conference on Strongly Correlated Electron Systems}.

\bibitem{ERKELENS1987}
\bibinfo{author}{Erkelens, W.} \emph{et~al.}
\newblock \bibinfo{journal}{\bibinfo{title}{Neutron scattering study of the antiferroquadrupolar ordering in ceb6 and ce0.75la0.25b6}}.
\newblock {\emph{\JournalTitle{Journal of Magnetism and Magnetic Materials}}} \textbf{\bibinfo{volume}{63-64}}, \bibinfo{pages}{61--63}, \doiprefix\url{https://doi.org/10.1016/0304-8853(87)90522-1} (\bibinfo{year}{1987}).

\bibitem{Shiina1997}
\bibinfo{author}{Shiina, R.}, \bibinfo{author}{Shiba, H.} \& \bibinfo{author}{Thalmeier, P.}
\newblock \bibinfo{journal}{\bibinfo{title}{Magnetic-field effects on quadrupolar ordering in a $\gamma$ 8-quartet system ceb 6}}.
\newblock {\emph{\JournalTitle{Journal of the Physical Society of Japan}}} \textbf{\bibinfo{volume}{66}}, \bibinfo{pages}{1741--1755}, \doiprefix\url{10.1143/JPSJ.66.1741} (\bibinfo{year}{1997}).
\newblock \eprint{https://doi.org/10.1143/JPSJ.66.1741}.

\bibitem{Kitagawa1996}
\bibinfo{author}{Kitagawa, J.}, \bibinfo{author}{Takeda, N.} \& \bibinfo{author}{Ishikawa, M.}
\newblock \bibinfo{journal}{\bibinfo{title}{Possible quadrupolar ordering in a kondo-lattice compound ${\mathrm{ce}}_{3}$${\mathrm{pd}}_{20}$${\mathrm{ge}}_{6}$}}.
\newblock {\emph{\JournalTitle{Phys. Rev. B}}} \textbf{\bibinfo{volume}{53}}, \bibinfo{pages}{5101--5103}, \doiprefix\url{10.1103/PhysRevB.53.5101} (\bibinfo{year}{1996}).

\bibitem{Tayama2001}
\bibinfo{author}{Tayama, T.} \emph{et~al.}
\newblock \bibinfo{journal}{\bibinfo{title}{Antiferro-quadrupolar ordering and multipole interactions in prpb 3}}.
\newblock {\emph{\JournalTitle{Journal of the Physical Society of Japan}}} \textbf{\bibinfo{volume}{70}}, \bibinfo{pages}{248--258}, \doiprefix\url{10.1143/JPSJ.70.248} (\bibinfo{year}{2001}).
\newblock \eprint{https://doi.org/10.1143/JPSJ.70.248}.

\bibitem{Tanida2006}
\bibinfo{author}{Tanida, H.}, \bibinfo{author}{S.~Suzuki, H.}, \bibinfo{author}{Takagi, S.}, \bibinfo{author}{Onodera, H.} \& \bibinfo{author}{Tanigaki, K.}
\newblock \bibinfo{journal}{\bibinfo{title}{Possible low-temperature strongly correlated electron behavior from multipole fluctuations in prmg3 with cubic non-kramers $\gamma 3$ doublet ground state}}.
\newblock {\emph{\JournalTitle{Journal of the Physical Society of Japan}}} \textbf{\bibinfo{volume}{75}}, \bibinfo{pages}{073705}, \doiprefix\url{10.1143/JPSJ.75.073705} (\bibinfo{year}{2006}).
\newblock \eprint{https://doi.org/10.1143/JPSJ.75.073705}.

\bibitem{Liu2019}
\bibinfo{author}{Liu, H.} \& \bibinfo{author}{Khaliullin, G.}
\newblock \bibinfo{journal}{\bibinfo{title}{Pseudo-jahn-teller effect and magnetoelastic coupling in spin-orbit mott insulators}}.
\newblock {\emph{\JournalTitle{Phys. Rev. Lett.}}} \textbf{\bibinfo{volume}{122}}, \bibinfo{pages}{057203}, \doiprefix\url{10.1103/PhysRevLett.122.057203} (\bibinfo{year}{2019}).

\bibitem{Suzuki2018}
\bibinfo{author}{Suzuki, M.-T.}, \bibinfo{author}{Ikeda, H.} \& \bibinfo{author}{Oppeneer, P.~M.}
\newblock \bibinfo{journal}{\bibinfo{title}{First-principles theory of magnetic multipoles in condensed matter systems}}.
\newblock {\emph{\JournalTitle{Journal of the Physical Society of Japan}}} \textbf{\bibinfo{volume}{87}}, \bibinfo{pages}{041008}, \doiprefix\url{10.7566/JPSJ.87.041008} (\bibinfo{year}{2018}).
\newblock \eprint{https://doi.org/10.7566/JPSJ.87.041008}.

\bibitem{Thole2018}
\bibinfo{author}{Thöle, F.} \& \bibinfo{author}{Spaldin, N.~A.}
\newblock \bibinfo{journal}{\bibinfo{title}{Magnetoelectric multipoles in metals}}.
\newblock {\emph{\JournalTitle{Philosophical Transactions of the Royal Society A: Mathematical, Physical and Engineering Sciences}}} \textbf{\bibinfo{volume}{376}}, \bibinfo{pages}{20170450}, \doiprefix\url{10.1098/rsta.2017.0450} (\bibinfo{year}{2018}).
\newblock \eprint{https://royalsocietypublishing.org/doi/pdf/10.1098/rsta.2017.0450}.

\bibitem{Kuramoto2008}
\bibinfo{author}{Kuramoto, Y.}
\newblock \bibinfo{journal}{\bibinfo{title}{{Electronic Higher Multipoles in Solids}}}.
\newblock {\emph{\JournalTitle{Progress of Theoretical Physics Supplement}}} \textbf{\bibinfo{volume}{176}}, \bibinfo{pages}{77--96}, \doiprefix\url{10.1143/PTPS.176.77} (\bibinfo{year}{2008}).
\newblock \eprint{https://academic.oup.com/ptps/article-pdf/doi/10.1143/PTPS.176.77/5324654/176-77.pdf}.

\bibitem{Kuramoto2009}
\bibinfo{author}{Kuramoto, Y.}, \bibinfo{author}{Kusunose, H.} \& \bibinfo{author}{Kiss, A.}
\newblock \bibinfo{journal}{\bibinfo{title}{Multipole orders and fluctuations in strongly correlated electron systems}}.
\newblock {\emph{\JournalTitle{Journal of the Physical Society of Japan}}} \textbf{\bibinfo{volume}{78}}, \bibinfo{pages}{072001}, \doiprefix\url{10.1143/JPSJ.78.072001} (\bibinfo{year}{2009}).
\newblock \eprint{https://doi.org/10.1143/JPSJ.78.072001}.

\bibitem{Anderson1973}
\bibinfo{author}{Anderson, P.}
\newblock \bibinfo{journal}{\bibinfo{title}{Resonating valence bonds: A new kind of insulator?}}
\newblock {\emph{\JournalTitle{Materials Research Bulletin}}} \textbf{\bibinfo{volume}{8}}, \bibinfo{pages}{153--160}, \doiprefix\url{https://doi.org/10.1016/0025-5408(73)90167-0} (\bibinfo{year}{1973}).

\bibitem{Savary2017}
\bibinfo{author}{Savary, L.} \& \bibinfo{author}{Balents, L.}
\newblock \bibinfo{journal}{\bibinfo{title}{Quantum spin liquids: a review}}.
\newblock {\emph{\JournalTitle{Reports on Progress in Physics}}} \textbf{\bibinfo{volume}{80}}, \bibinfo{pages}{016502}, \doiprefix\url{10.1088/0034-4885/80/1/016502} (\bibinfo{year}{2016}).

\bibitem{Shimizu2003}
\bibinfo{author}{Shimizu, Y.}, \bibinfo{author}{Miyagawa, K.}, \bibinfo{author}{Kanoda, K.}, \bibinfo{author}{Maesato, M.} \& \bibinfo{author}{Saito, G.}
\newblock \bibinfo{journal}{\bibinfo{title}{Spin liquid state in an organic mott insulator with a triangular lattice}}.
\newblock {\emph{\JournalTitle{Phys. Rev. Lett.}}} \textbf{\bibinfo{volume}{91}}, \bibinfo{pages}{107001}, \doiprefix\url{10.1103/PhysRevLett.91.107001} (\bibinfo{year}{2003}).

\bibitem{Powell2011}
\bibinfo{author}{Powell, B.~J.} \& \bibinfo{author}{McKenzie, R.~H.}
\newblock \bibinfo{journal}{\bibinfo{title}{Quantum frustration in organic mott insulators: from spin liquids to unconventional superconductors}}.
\newblock {\emph{\JournalTitle{Reports on Progress in Physics}}} \textbf{\bibinfo{volume}{74}}, \bibinfo{pages}{056501}, \doiprefix\url{10.1088/0034-4885/74/5/056501} (\bibinfo{year}{2011}).

\bibitem{Plumb2014}
\bibinfo{author}{Plumb, K.~W.} \emph{et~al.}
\newblock \bibinfo{journal}{\bibinfo{title}{$\ensuremath{\alpha}\ensuremath{-}{\mathrm{rucl}}_{3}$: A spin-orbit assisted mott insulator on a honeycomb lattice}}.
\newblock {\emph{\JournalTitle{Phys. Rev. B}}} \textbf{\bibinfo{volume}{90}}, \bibinfo{pages}{041112}, \doiprefix\url{10.1103/PhysRevB.90.041112} (\bibinfo{year}{2014}).

\bibitem{Mendels2010}
\bibinfo{author}{Mendels, P.} \& \bibinfo{author}{Bert, F.}
\newblock \bibinfo{journal}{\bibinfo{title}{Quantum kagome antiferromagnet zncu3(oh)6cl2}}.
\newblock {\emph{\JournalTitle{Journal of the Physical Society of Japan}}} \textbf{\bibinfo{volume}{79}}, \bibinfo{pages}{011001}, \doiprefix\url{10.1143/JPSJ.79.011001} (\bibinfo{year}{2010}).
\newblock \eprint{https://doi.org/10.1143/JPSJ.79.011001}.

\bibitem{Winter2017}
\bibinfo{author}{Winter, S.~M.} \emph{et~al.}
\newblock \bibinfo{journal}{\bibinfo{title}{Models and materials for generalized kitaev magnetism}}.
\newblock {\emph{\JournalTitle{Journal of Physics: Condensed Matter}}} \textbf{\bibinfo{volume}{29}}, \bibinfo{pages}{493002}, \doiprefix\url{10.1088/1361-648X/aa8cf5} (\bibinfo{year}{2017}).

\bibitem{Takagi2019}
\bibinfo{author}{Takagi, H.}, \bibinfo{author}{Takayama, T.}, \bibinfo{author}{Jackeli, G.}, \bibinfo{author}{Khaliullin, G.} \& \bibinfo{author}{Nagler, S.~E.}
\newblock \bibinfo{journal}{\bibinfo{title}{Concept and realization of kitaev quantum spin liquids}}.
\newblock {\emph{\JournalTitle{Nature Reviews Physics}}} \textbf{\bibinfo{volume}{1}}, \bibinfo{pages}{264--280}, \doiprefix\url{10.1038/s42254-019-0038-2} (\bibinfo{year}{2019}).

\bibitem{Kim2008}
\bibinfo{author}{Kim, B.~J.} \emph{et~al.}
\newblock \bibinfo{journal}{\bibinfo{title}{Novel ${J}_{\mathrm{eff}}=1/2$ mott state induced by relativistic spin-orbit coupling in ${\mathrm{sr}}_{2}{\mathrm{iro}}_{4}$}}.
\newblock {\emph{\JournalTitle{Phys. Rev. Lett.}}} \textbf{\bibinfo{volume}{101}}, \bibinfo{pages}{076402}, \doiprefix\url{10.1103/PhysRevLett.101.076402} (\bibinfo{year}{2008}).

\bibitem{Comin2012}
\bibinfo{author}{Comin, R.} \emph{et~al.}
\newblock \bibinfo{journal}{\bibinfo{title}{${\mathrm{na}}_{2}{\mathrm{iro}}_{3}$ as a novel relativistic mott insulator with a 340-mev gap}}.
\newblock {\emph{\JournalTitle{Phys. Rev. Lett.}}} \textbf{\bibinfo{volume}{109}}, \bibinfo{pages}{266406}, \doiprefix\url{10.1103/PhysRevLett.109.266406} (\bibinfo{year}{2012}).

\bibitem{Ju2013}
\bibinfo{author}{Ju, W.}, \bibinfo{author}{Liu, G.-Q.} \& \bibinfo{author}{Yang, Z.}
\newblock \bibinfo{journal}{\bibinfo{title}{Exotic spin-orbital mott insulating states in bairo${}_{3}$}}.
\newblock {\emph{\JournalTitle{Phys. Rev. B}}} \textbf{\bibinfo{volume}{87}}, \bibinfo{pages}{075112}, \doiprefix\url{10.1103/PhysRevB.87.075112} (\bibinfo{year}{2013}).

\bibitem{Franchini2014}
\bibinfo{author}{Franchini, C.}
\newblock \bibinfo{journal}{\bibinfo{title}{Hybrid functionals applied to perovskites}}.
\newblock {\emph{\JournalTitle{Journal of Physics: Condensed Matter}}} \textbf{\bibinfo{volume}{26}}, \bibinfo{pages}{253202}, \doiprefix\url{10.1088/0953-8984/26/25/253202} (\bibinfo{year}{2014}).

\bibitem{Martins2017}
\bibinfo{author}{Martins, C.}, \bibinfo{author}{Aichhorn, M.} \& \bibinfo{author}{Biermann, S.}
\newblock \bibinfo{journal}{\bibinfo{title}{Coulomb correlations in 4d and 5d oxides from first principles—or how spin–orbit materials choose their effective orbital degeneracies}}.
\newblock {\emph{\JournalTitle{Journal of Physics: Condensed Matter}}} \textbf{\bibinfo{volume}{29}}, \bibinfo{pages}{263001}, \doiprefix\url{10.1088/1361-648X/aa648f} (\bibinfo{year}{2017}).

\bibitem{Dzero2010}
\bibinfo{author}{Dzero, M.}, \bibinfo{author}{Sun, K.}, \bibinfo{author}{Galitski, V.} \& \bibinfo{author}{Coleman, P.}
\newblock \bibinfo{journal}{\bibinfo{title}{Topological kondo insulators}}.
\newblock {\emph{\JournalTitle{Phys. Rev. Lett.}}} \textbf{\bibinfo{volume}{104}}, \bibinfo{pages}{106408}, \doiprefix\url{10.1103/PhysRevLett.104.106408} (\bibinfo{year}{2010}).

\bibitem{Xiao2011}
\bibinfo{author}{Xiao, D.}, \bibinfo{author}{Zhu, W.}, \bibinfo{author}{Ran, Y.}, \bibinfo{author}{Nagaosa, N.} \& \bibinfo{author}{Okamoto, S.}
\newblock \bibinfo{journal}{\bibinfo{title}{Interface engineering of quantum hall effects in digital transition metal oxide heterostructures}}.
\newblock {\emph{\JournalTitle{Nature Communications}}} \textbf{\bibinfo{volume}{2}}, \bibinfo{pages}{596}, \doiprefix\url{10.1038/ncomms1602} (\bibinfo{year}{2011}).

\bibitem{Dzsaber2017}
\bibinfo{author}{Dzsaber, S.} \emph{et~al.}
\newblock \bibinfo{journal}{\bibinfo{title}{Kondo insulator to semimetal transformation tuned by spin-orbit coupling}}.
\newblock {\emph{\JournalTitle{Phys. Rev. Lett.}}} \textbf{\bibinfo{volume}{118}}, \bibinfo{pages}{246601}, \doiprefix\url{10.1103/PhysRevLett.118.246601} (\bibinfo{year}{2017}).

\bibitem{Chen2022}
\bibinfo{author}{Chen, L.} \emph{et~al.}
\newblock \bibinfo{journal}{\bibinfo{title}{Topological semimetal driven by strong correlations and crystalline symmetry}}.
\newblock {\emph{\JournalTitle{Nature Physics}}} \textbf{\bibinfo{volume}{18}}, \bibinfo{pages}{1341--1346}, \doiprefix\url{10.1038/s41567-022-01743-4} (\bibinfo{year}{2022}).

\bibitem{Caviglia2010}
\bibinfo{author}{Caviglia, A.~D.} \emph{et~al.}
\newblock \bibinfo{journal}{\bibinfo{title}{Tunable rashba spin-orbit interaction at oxide interfaces}}.
\newblock {\emph{\JournalTitle{Phys. Rev. Lett.}}} \textbf{\bibinfo{volume}{104}}, \bibinfo{pages}{126803}, \doiprefix\url{10.1103/PhysRevLett.104.126803} (\bibinfo{year}{2010}).

\bibitem{Wadehra2020}
\bibinfo{author}{Wadehra, N.} \emph{et~al.}
\newblock \bibinfo{journal}{\bibinfo{title}{Planar hall effect and anisotropic magnetoresistance in polar-polar interface of lavo3-ktao3 with strong spin-orbit coupling}}.
\newblock {\emph{\JournalTitle{Nature Communications}}} \textbf{\bibinfo{volume}{11}}, \bibinfo{pages}{874}, \doiprefix\url{10.1038/s41467-020-14689-z} (\bibinfo{year}{2020}).

\bibitem{Okuma2024}
\bibinfo{author}{Okuma, H.}, \bibinfo{author}{Katayama, Y.} \& \bibinfo{author}{Ueno, K.}
\newblock \bibinfo{journal}{\bibinfo{title}{Large rashba parameter for $4d$ strongly correlated perovskite oxide $\mathrm{SrNb}{\mathrm{o}}_{3}$ ultrathin films}}.
\newblock {\emph{\JournalTitle{Phys. Rev. Mater.}}} \textbf{\bibinfo{volume}{8}}, \bibinfo{pages}{015001}, \doiprefix\url{10.1103/PhysRevMaterials.8.015001} (\bibinfo{year}{2024}).

\bibitem{Generalov2017}
\bibinfo{author}{Generalov, A.} \emph{et~al.}
\newblock \bibinfo{journal}{\bibinfo{title}{Spin orientation of two-dimensional electrons driven by temperature-tunable competition of spin–orbit and exchange–magnetic interactions}}.
\newblock {\emph{\JournalTitle{Nano Letters}}} \textbf{\bibinfo{volume}{17}}, \bibinfo{pages}{811--820}, \doiprefix\url{10.1021/acs.nanolett.6b04036} (\bibinfo{year}{2017}).
\newblock \bibinfo{note}{PMID: 28032768}, \eprint{https://doi.org/10.1021/acs.nanolett.6b04036}.

\bibitem{Generalov2018}
\bibinfo{author}{Generalov, A.} \emph{et~al.}
\newblock \bibinfo{journal}{\bibinfo{title}{Strong spin-orbit coupling in the noncentrosymmetric kondo lattice}}.
\newblock {\emph{\JournalTitle{Phys. Rev. B}}} \textbf{\bibinfo{volume}{98}}, \bibinfo{pages}{115157}, \doiprefix\url{10.1103/PhysRevB.98.115157} (\bibinfo{year}{2018}).

\bibitem{Bercioux2015}
\bibinfo{author}{Bercioux, D.} \& \bibinfo{author}{Lucignano, P.}
\newblock \bibinfo{journal}{\bibinfo{title}{Quantum transport in rashba spin–orbit materials: a review}}.
\newblock {\emph{\JournalTitle{Reports on Progress in Physics}}} \textbf{\bibinfo{volume}{78}}, \bibinfo{pages}{106001}, \doiprefix\url{10.1088/0034-4885/78/10/106001} (\bibinfo{year}{2015}).

\bibitem{Manchon2015}
\bibinfo{author}{Manchon, A.}, \bibinfo{author}{Koo, H.~C.}, \bibinfo{author}{Nitta, J.}, \bibinfo{author}{Frolov, S.~M.} \& \bibinfo{author}{Duine, R.~A.}
\newblock \bibinfo{journal}{\bibinfo{title}{New perspectives for rashba spin--orbit coupling}}.
\newblock {\emph{\JournalTitle{Nature Materials}}} \textbf{\bibinfo{volume}{14}}, \bibinfo{pages}{871--882}, \doiprefix\url{10.1038/nmat4360} (\bibinfo{year}{2015}).

\bibitem{DFTmag}
\bibinfo{author}{Bihlmayer, G.}
\newblock \emph{\bibinfo{title}{Density Functional Theory for Magnetism and Magnetic Anisotropy}}, \bibinfo{pages}{1--23} (\bibinfo{publisher}{Springer International Publishing}, \bibinfo{address}{Cham}, \bibinfo{year}{2018}).

\bibitem{Carnall1989}
\bibinfo{author}{Carnall, W.~T.}, \bibinfo{author}{Goodman, G.~L.}, \bibinfo{author}{Rajnak, K.} \& \bibinfo{author}{Rana, R.~S.}
\newblock \bibinfo{journal}{\bibinfo{title}{{A systematic analysis of the spectra of the lanthanides doped into single crystal LaF$_3$}}}.
\newblock {\emph{\JournalTitle{The Journal of Chemical Physics}}} \textbf{\bibinfo{volume}{90}}, \bibinfo{pages}{3443--3457}, \doiprefix\url{10.1063/1.455853} (\bibinfo{year}{1989}).

\bibitem{Moore2009}
\bibinfo{author}{Moore, K.~T.} \& \bibinfo{author}{van~der Laan, G.}
\newblock \bibinfo{journal}{\bibinfo{title}{Nature of the $5f$ states in actinide metals}}.
\newblock {\emph{\JournalTitle{Rev. Mod. Phys.}}} \textbf{\bibinfo{volume}{81}}, \bibinfo{pages}{235--298}, \doiprefix\url{10.1103/RevModPhys.81.235} (\bibinfo{year}{2009}).

\bibitem{Taylor2017}
\bibinfo{author}{Taylor, A.~E.} \emph{et~al.}
\newblock \bibinfo{journal}{\bibinfo{title}{Spin-orbit coupling controlled $j=3/2$ electronic ground state in $5{d}^{3}$ oxides}}.
\newblock {\emph{\JournalTitle{Phys. Rev. Lett.}}} \textbf{\bibinfo{volume}{118}}, \bibinfo{pages}{207202}, \doiprefix\url{10.1103/PhysRevLett.118.207202} (\bibinfo{year}{2017}).

\bibitem{Paramekanti2018}
\bibinfo{author}{Paramekanti, A.} \emph{et~al.}
\newblock \bibinfo{journal}{\bibinfo{title}{Spin-orbit coupled systems in the atomic limit: rhenates, osmates, iridates}}.
\newblock {\emph{\JournalTitle{Phys. Rev. B}}} \textbf{\bibinfo{volume}{97}}, \bibinfo{pages}{235119}, \doiprefix\url{10.1103/PhysRevB.97.235119} (\bibinfo{year}{2018}).

\bibitem{abragam_bleaney_book}
\bibinfo{author}{Abragam, A.} \& \bibinfo{author}{Bleaney, B.}
\newblock \emph{\bibinfo{title}{Electron Paramagnetic Resonance of Transition Ions}}.
\newblock Oxford Classic Texts in the Physical Sciences (\bibinfo{publisher}{OUP Oxford}, \bibinfo{year}{2012}).

\bibitem{Blum_DM}
\bibinfo{author}{Blum, K.}
\newblock \emph{\bibinfo{title}{Density matrix theory and applications}} (\bibinfo{publisher}{Plenum Press, New York}, \bibinfo{year}{1996}).

\bibitem{Sakurai1993Modern}
\bibinfo{author}{Sakurai, J.~J.}
\newblock \emph{\bibinfo{title}{Modern Quantum Mechanics (Revised Edition)}} (\bibinfo{publisher}{Addison Wesley}, \bibinfo{year}{1993}).

\bibitem{Dubovik1990}
\bibinfo{author}{Dubovik, V.} \& \bibinfo{author}{Tugushev, V.}
\newblock \bibinfo{journal}{\bibinfo{title}{Toroid moments in electrodynamics and solid-state physics}}.
\newblock {\emph{\JournalTitle{Physics Reports}}} \textbf{\bibinfo{volume}{187}}, \bibinfo{pages}{145--202}, \doiprefix\url{https://doi.org/10.1016/0370-1573(90)90042-Z} (\bibinfo{year}{1990}).

\bibitem{Spaldin2013}
\bibinfo{author}{Spaldin, N.~A.}, \bibinfo{author}{Fechner, M.}, \bibinfo{author}{Bousquet, E.}, \bibinfo{author}{Balatsky, A.} \& \bibinfo{author}{Nordstr\"om, L.}
\newblock \bibinfo{journal}{\bibinfo{title}{Monopole-based formalism for the diagonal magnetoelectric response}}.
\newblock {\emph{\JournalTitle{Phys. Rev. B}}} \textbf{\bibinfo{volume}{88}}, \bibinfo{pages}{094429}, \doiprefix\url{10.1103/PhysRevB.88.094429} (\bibinfo{year}{2013}).

\bibitem{Hayami2018}
\bibinfo{author}{Hayami, S.} \& \bibinfo{author}{Kusunose, H.}
\newblock \bibinfo{journal}{\bibinfo{title}{Microscopic description of electric and magnetic toroidal multipoles in hybrid orbitals}}.
\newblock {\emph{\JournalTitle{Journal of the Physical Society of Japan}}} \textbf{\bibinfo{volume}{87}}, \bibinfo{pages}{033709}, \doiprefix\url{10.7566/JPSJ.87.033709} (\bibinfo{year}{2018}).

\bibitem{Watanabe2018}
\bibinfo{author}{Watanabe, H.} \& \bibinfo{author}{Yanase, Y.}
\newblock \bibinfo{journal}{\bibinfo{title}{Group-theoretical classification of multipole order: Emergent responses and candidate materials}}.
\newblock {\emph{\JournalTitle{Phys. Rev. B}}} \textbf{\bibinfo{volume}{98}}, \bibinfo{pages}{245129}, \doiprefix\url{10.1103/PhysRevB.98.245129} (\bibinfo{year}{2018}).

\bibitem{Hayami2018_1}
\bibinfo{author}{Hayami, S.}, \bibinfo{author}{Yatsushiro, M.}, \bibinfo{author}{Yanagi, Y.} \& \bibinfo{author}{Kusunose, H.}
\newblock \bibinfo{journal}{\bibinfo{title}{Classification of atomic-scale multipoles under crystallographic point groups and application to linear response tensors}}.
\newblock {\emph{\JournalTitle{Phys. Rev. B}}} \textbf{\bibinfo{volume}{98}}, \bibinfo{pages}{165110}, \doiprefix\url{10.1103/PhysRevB.98.165110} (\bibinfo{year}{2018}).

\bibitem{Suzuki2019}
\bibinfo{author}{Suzuki, M.-T.} \emph{et~al.}
\newblock \bibinfo{journal}{\bibinfo{title}{Multipole expansion for magnetic structures: A generation scheme for a symmetry-adapted orthonormal basis set in the crystallographic point group}}.
\newblock {\emph{\JournalTitle{Phys. Rev. B}}} \textbf{\bibinfo{volume}{99}}, \bibinfo{pages}{174407}, \doiprefix\url{10.1103/PhysRevB.99.174407} (\bibinfo{year}{2019}).

\bibitem{Huebsch2021}
\bibinfo{author}{Huebsch, M.-T.}, \bibinfo{author}{Nomoto, T.}, \bibinfo{author}{Suzuki, M.-T.} \& \bibinfo{author}{Arita, R.}
\newblock \bibinfo{journal}{\bibinfo{title}{Benchmark for ab initio prediction of magnetic structures based on cluster-multipole theory}}.
\newblock {\emph{\JournalTitle{Phys. Rev. X}}} \textbf{\bibinfo{volume}{11}}, \bibinfo{pages}{011031}, \doiprefix\url{10.1103/PhysRevX.11.011031} (\bibinfo{year}{2021}).

\bibitem{Kung2015}
\bibinfo{author}{Kung, H.-H.} \emph{et~al.}
\newblock \bibinfo{journal}{\bibinfo{title}{Chirality density wave of the "hidden order" phase in {URu$_2$Si$_2$}}}.
\newblock {\emph{\JournalTitle{Science}}} \textbf{\bibinfo{volume}{347}}, \bibinfo{pages}{1339--1342}, \doiprefix\url{10.1126/science.1259729} (\bibinfo{year}{2015}).
\newblock \eprint{https://www.science.org/doi/pdf/10.1126/science.1259729}.

\bibitem{Ye2019a}
\bibinfo{author}{Ye, M.}, \bibinfo{author}{Rosenberg, E.~W.}, \bibinfo{author}{Fisher, I.~R.} \& \bibinfo{author}{Blumberg, G.}
\newblock \bibinfo{journal}{\bibinfo{title}{Lattice dynamics, crystal-field excitations, and quadrupolar fluctuations of ${\mathrm{ybru}}_{2}{\mathrm{ge}}_{2}$}}.
\newblock {\emph{\JournalTitle{Phys. Rev. B}}} \textbf{\bibinfo{volume}{99}}, \bibinfo{pages}{235104}, \doiprefix\url{10.1103/PhysRevB.99.235104} (\bibinfo{year}{2019}).

\bibitem{Ye2019b}
\bibinfo{author}{Ye, M.} \emph{et~al.}
\newblock \bibinfo{journal}{\bibinfo{title}{Raman spectroscopy of $f$-electron metals: An example of ${\mathrm{ceb}}_{6}$}}.
\newblock {\emph{\JournalTitle{Phys. Rev. Mater.}}} \textbf{\bibinfo{volume}{3}}, \bibinfo{pages}{065003}, \doiprefix\url{10.1103/PhysRevMaterials.3.065003} (\bibinfo{year}{2019}).

\bibitem{Volkov2021}
\bibinfo{author}{Volkov, P.~A.} \emph{et~al.}
\newblock \bibinfo{journal}{\bibinfo{title}{Critical charge fluctuations and emergent coherence in a strongly correlated excitonic insulator}}.
\newblock {\emph{\JournalTitle{npj Quantum Materials}}} \textbf{\bibinfo{volume}{6}}, \bibinfo{pages}{52}, \doiprefix\url{10.1038/s41535-021-00351-4} (\bibinfo{year}{2021}).

\bibitem{Patri2019}
\bibinfo{author}{Patri, A.~S.} \emph{et~al.}
\newblock \bibinfo{journal}{\bibinfo{title}{Unveiling hidden multipolar orders with magnetostriction}}.
\newblock {\emph{\JournalTitle{Nature Communications}}} \textbf{\bibinfo{volume}{10}}, \bibinfo{pages}{4092}, \doiprefix\url{10.1038/s41467-019-11913-3} (\bibinfo{year}{2019}).

\bibitem{Ye2023}
\bibinfo{author}{Ye, L.}, \bibinfo{author}{Sorensen, M.~E.}, \bibinfo{author}{Bachmann, M.~D.} \& \bibinfo{author}{Fisher, I.~R.}
\newblock \bibinfo{title}{Measurement of the magnetic octupole susceptibility of prv2al20} (\bibinfo{year}{2023}).
\newblock \eprint{2309.04633}.

\bibitem{Lovesey2020}
\bibinfo{author}{Lovesey, S.~W.} \& \bibinfo{author}{Khalyavin, D.~D.}
\newblock \bibinfo{journal}{\bibinfo{title}{Lone octupole and bulk magnetism in osmate $5{d}^{2}$ double perovskites}}.
\newblock {\emph{\JournalTitle{Phys. Rev. B}}} \textbf{\bibinfo{volume}{102}}, \bibinfo{pages}{064407}, \doiprefix\url{10.1103/PhysRevB.102.064407} (\bibinfo{year}{2020}).

\bibitem{Urru2023}
\bibinfo{author}{Urru, A.} \emph{et~al.}
\newblock \bibinfo{journal}{\bibinfo{title}{Neutron scattering from local magnetoelectric multipoles: A combined theoretical, computational, and experimental perspective}}.
\newblock {\emph{\JournalTitle{Phys. Rev. Res.}}} \textbf{\bibinfo{volume}{5}}, \bibinfo{pages}{033147}, \doiprefix\url{10.1103/PhysRevResearch.5.033147} (\bibinfo{year}{2023}).

\bibitem{Urru2022}
\bibinfo{author}{Urru, A.} \& \bibinfo{author}{Spaldin, N.~A.}
\newblock \bibinfo{journal}{\bibinfo{title}{Magnetic octupole tensor decomposition and second-order magnetoelectric effect}}.
\newblock {\emph{\JournalTitle{Annals of Physics}}} \textbf{\bibinfo{volume}{447}}, \bibinfo{pages}{168964}, \doiprefix\url{https://doi.org/10.1016/j.aop.2022.168964} (\bibinfo{year}{2022}).

\bibitem{Harter2017}
\bibinfo{author}{Harter, J.~W.}, \bibinfo{author}{Zhao, Z.~Y.}, \bibinfo{author}{Yan, J.-Q.}, \bibinfo{author}{Mandrus, D.~G.} \& \bibinfo{author}{Hsieh, D.}
\newblock \bibinfo{journal}{\bibinfo{title}{A parity-breaking electronic nematic phase transition in the spin-orbit coupled metal cd<sub>2</sub>re<sub>2</sub>o<sub>7</sub>}}.
\newblock {\emph{\JournalTitle{Science}}} \textbf{\bibinfo{volume}{356}}, \bibinfo{pages}{295--299}, \doiprefix\url{10.1126/science.aad1188} (\bibinfo{year}{2017}).
\newblock \eprint{https://www.science.org/doi/pdf/10.1126/science.aad1188}.

\bibitem{vanderLaan2021}
\bibinfo{author}{van~der Laan, G.} \& \bibinfo{author}{Lovesey, S.~W.}
\newblock \bibinfo{journal}{\bibinfo{title}{Electronic multipoles in second harmonic generation and neutron scattering}}.
\newblock {\emph{\JournalTitle{Phys. Rev. B}}} \textbf{\bibinfo{volume}{103}}, \bibinfo{pages}{125124}, \doiprefix\url{10.1103/PhysRevB.103.125124} (\bibinfo{year}{2021}).

\bibitem{Taniguchi2016}
\bibinfo{author}{Taniguchi, T.} \emph{et~al.}
\newblock \bibinfo{journal}{\bibinfo{title}{Nmr observation of ferro-quadrupole order in prti2al20}}.
\newblock {\emph{\JournalTitle{Journal of the Physical Society of Japan}}} \textbf{\bibinfo{volume}{85}}, \bibinfo{pages}{113703}, \doiprefix\url{10.7566/JPSJ.85.113703} (\bibinfo{year}{2016}).
\newblock \eprint{https://doi.org/10.7566/JPSJ.85.113703}.

\bibitem{Taniguchi2019}
\bibinfo{author}{Taniguchi, T.} \emph{et~al.}
\newblock \bibinfo{journal}{\bibinfo{title}{Field-induced switching of ferro-quadrupole order parameter in prti2al20}}.
\newblock {\emph{\JournalTitle{Journal of the Physical Society of Japan}}} \textbf{\bibinfo{volume}{88}}, \bibinfo{pages}{084707}, \doiprefix\url{10.7566/JPSJ.88.084707} (\bibinfo{year}{2019}).
\newblock \eprint{https://doi.org/10.7566/JPSJ.88.084707}.

\bibitem{Anderson1950}
\bibinfo{author}{Anderson, P.~W.}
\newblock \bibinfo{journal}{\bibinfo{title}{Antiferromagnetism. theory of superexchange interaction}}.
\newblock {\emph{\JournalTitle{Phys. Rev.}}} \textbf{\bibinfo{volume}{79}}, \bibinfo{pages}{350--356}, \doiprefix\url{10.1103/PhysRev.79.350} (\bibinfo{year}{1950}).

\bibitem{Goodenough1955}
\bibinfo{author}{Goodenough, J.~B.}
\newblock \bibinfo{journal}{\bibinfo{title}{Theory of the role of covalence in the perovskite-type manganites $[\mathrm{La}, m(\mathrm{II})]\mathrm{Mn}{\mathrm{o}}_{3}$}}.
\newblock {\emph{\JournalTitle{Phys. Rev.}}} \textbf{\bibinfo{volume}{100}}, \bibinfo{pages}{564--573}, \doiprefix\url{10.1103/PhysRev.100.564} (\bibinfo{year}{1955}).

\bibitem{Kanamori1959}
\bibinfo{author}{Kanamori, J.}
\newblock \bibinfo{journal}{\bibinfo{title}{Superexchange interaction and symmetry properties of electron orbitals}}.
\newblock {\emph{\JournalTitle{Journal of Physics and Chemistry of Solids}}} \textbf{\bibinfo{volume}{10}}, \bibinfo{pages}{87--98}, \doiprefix\url{https://doi.org/10.1016/0022-3697(59)90061-7} (\bibinfo{year}{1959}).

\bibitem{Gehring1975}
\bibinfo{author}{Gehring, G.~A.} \& \bibinfo{author}{Gehring, K.~A.}
\newblock \bibinfo{journal}{\bibinfo{title}{Co-operative jahn-teller effects}}.
\newblock {\emph{\JournalTitle{Reports on Progress in Physics}}} \textbf{\bibinfo{volume}{38}}, \bibinfo{pages}{1}, \doiprefix\url{10.1088/0034-4885/38/1/001} (\bibinfo{year}{1975}).

\bibitem{kugel_khomskii}
\bibinfo{author}{K~I~Kugel', K.~I.} \& \bibinfo{author}{Khomskii, D.~I.}
\newblock \bibinfo{journal}{\bibinfo{title}{The jahn-teller effect and magnetism: transition metal compounds}}.
\newblock {\emph{\JournalTitle{Sov. Phys. Uspekhi}}} \textbf{\bibinfo{volume}{25}}, \bibinfo{pages}{231--256} (\bibinfo{year}{1982}).

\bibitem{Khomskii_book_2014}
\bibinfo{author}{Khomskii, D.~I.}
\newblock \emph{\bibinfo{title}{Transition Metal Compounds}} (\bibinfo{publisher}{Cambridge University Press}, \bibinfo{year}{2014}).

\bibitem{Hirst1978}
\bibinfo{author}{Hirst, L.}
\newblock \bibinfo{journal}{\bibinfo{title}{Theory of the coupling between conduction electrons and moments of 3d and 4f ions in metals}}.
\newblock {\emph{\JournalTitle{Advances in Physics}}} \textbf{\bibinfo{volume}{27}}, \bibinfo{pages}{231--285}, \doiprefix\url{10.1080/00018737800101374} (\bibinfo{year}{1978}).

\bibitem{Fulde1985}
\bibinfo{author}{Fulde, P.} \& \bibinfo{author}{Loewenhaupt, M.}
\newblock \bibinfo{journal}{\bibinfo{title}{Magnetic excitations in crystal-field split 4f systems}}.
\newblock {\emph{\JournalTitle{Advances in Physics}}} \textbf{\bibinfo{volume}{34}}, \bibinfo{pages}{589--661}, \doiprefix\url{10.1080/00018738500101821} (\bibinfo{year}{1985}).

\bibitem{RareEarthMag_book}
\bibinfo{author}{Jensen, J.} \& \bibinfo{author}{Mackintosh, A.~R.}
\newblock \emph{\bibinfo{title}{Rare Earth Magnetism: Structures and Excitations}} (\bibinfo{publisher}{Clarendon Press, Oxford}, \bibinfo{year}{1991}).

\bibitem{Chen2011}
\bibinfo{author}{Chen, G.} \& \bibinfo{author}{Balents, L.}
\newblock \bibinfo{journal}{\bibinfo{title}{Spin-orbit coupling in ${d}^{2}$ ordered double perovskites}}.
\newblock {\emph{\JournalTitle{Phys. Rev. B}}} \textbf{\bibinfo{volume}{84}}, \bibinfo{pages}{094420}, \doiprefix\url{10.1103/PhysRevB.84.094420} (\bibinfo{year}{2011}).

\bibitem{Svoboda2021}
\bibinfo{author}{Svoboda, C.}, \bibinfo{author}{Zhang, W.}, \bibinfo{author}{Randeria, M.} \& \bibinfo{author}{Trivedi, N.}
\newblock \bibinfo{journal}{\bibinfo{title}{Orbital order drives magnetic order in and double perovskite mott insulators}}.
\newblock {\emph{\JournalTitle{Physical Review B}}} \textbf{\bibinfo{volume}{104}}, \doiprefix\url{10.1103/PhysRevB.104.024437} (\bibinfo{year}{2021}).

\bibitem{Jahn_Teller1937}
\bibinfo{author}{Jahn, H.~A.}, \bibinfo{author}{Teller, E.} \& \bibinfo{author}{Donnan, F.~G.}
\newblock \bibinfo{journal}{\bibinfo{title}{Stability of polyatomic molecules in degenerate electronic states - i—orbital degeneracy}}.
\newblock {\emph{\JournalTitle{Proceedings of the Royal Society of London. Series A - Mathematical and Physical Sciences}}} \textbf{\bibinfo{volume}{161}}, \bibinfo{pages}{220--235}, \doiprefix\url{10.1098/rspa.1937.0142} (\bibinfo{year}{1937}).

\bibitem{bersuker_vibronic_1989}
\bibinfo{author}{Bersuker, I.~B.} \& \bibinfo{author}{Polinger, V.~Z.}
\newblock \emph{\bibinfo{title}{Vibronic {Interactions} in {Molecules} and {Crystals}}} (\bibinfo{publisher}{Springer Berlin Heidelberg}, \bibinfo{year}{1989}).

\bibitem{Streltsov2020}
\bibinfo{author}{Streltsov, S.~V.} \& \bibinfo{author}{Khomskii, D.~I.}
\newblock \bibinfo{journal}{\bibinfo{title}{Jahn-teller effect and spin-orbit coupling: Friends or foes?}}
\newblock {\emph{\JournalTitle{Phys. Rev. X}}} \textbf{\bibinfo{volume}{10}}, \bibinfo{pages}{031043}, \doiprefix\url{10.1103/PhysRevX.10.031043} (\bibinfo{year}{2020}).

\bibitem{Streltsov2022}
\bibinfo{author}{Streltsov, S.~V.}, \bibinfo{author}{Temnikov, F.~V.}, \bibinfo{author}{Kugel, K.~I.} \& \bibinfo{author}{Khomskii, D.~I.}
\newblock \bibinfo{journal}{\bibinfo{title}{Interplay of the jahn-teller effect and spin-orbit coupling: The case of trigonal vibrations}}.
\newblock {\emph{\JournalTitle{Phys. Rev. B}}} \textbf{\bibinfo{volume}{105}}, \bibinfo{pages}{205142}, \doiprefix\url{10.1103/PhysRevB.105.205142} (\bibinfo{year}{2022}).

\bibitem{Bersuker2013}
\bibinfo{author}{Bersuker, I.~B.}
\newblock \bibinfo{journal}{\bibinfo{title}{Pseudo-jahn–teller effect—a two-state paradigm in formation, deformation, and transformation of molecular systems and solids}}.
\newblock {\emph{\JournalTitle{Chemical Reviews}}} \textbf{\bibinfo{volume}{113}}, \bibinfo{pages}{1351--1390}, \doiprefix\url{10.1021/cr300279n} (\bibinfo{year}{2013}).

\bibitem{Iwahara2024}
\bibinfo{author}{Iwahara, N.}
\newblock \bibinfo{journal}{\bibinfo{title}{Dynamic jahn–teller phenomena in heavy transition metal compounds}}.
\newblock {\emph{\JournalTitle{Journal of the Physical Society of Japan}}} \textbf{\bibinfo{volume}{93}}, \bibinfo{pages}{121003}, \doiprefix\url{10.7566/JPSJ.93.121003} (\bibinfo{year}{2024}).
\newblock \eprint{https://doi.org/10.7566/JPSJ.93.121003}.

\bibitem{Lejaeghere2016}
\bibinfo{author}{Lejaeghere, K.} \emph{et~al.}
\newblock \bibinfo{journal}{\bibinfo{title}{Reproducibility in density functional theory calculations of solids}}.
\newblock {\emph{\JournalTitle{Science}}} \textbf{\bibinfo{volume}{351}}, \bibinfo{pages}{aad3000}, \doiprefix\url{10.1126/science.aad3000} (\bibinfo{year}{2016}).

\bibitem{Franchini2005}
\bibinfo{author}{Franchini, C.}, \bibinfo{author}{Bayer, V.}, \bibinfo{author}{Podloucky, R.}, \bibinfo{author}{Paier, J.} \& \bibinfo{author}{Kresse, G.}
\newblock \bibinfo{journal}{\bibinfo{title}{Density functional theory study of mno by a hybrid functional approach}}.
\newblock {\emph{\JournalTitle{Phys. Rev. B}}} \textbf{\bibinfo{volume}{72}}, \bibinfo{pages}{045132}, \doiprefix\url{10.1103/PhysRevB.72.045132} (\bibinfo{year}{2005}).

\bibitem{Franchini2011}
\bibinfo{author}{Franchini, C.} \emph{et~al.}
\newblock \bibinfo{journal}{\bibinfo{title}{Exceptionally strong magnetism in the 4$d$ perovskites $r$tco${}_{3}$ ($r=\mathrm{Ca}$, sr, ba)}}.
\newblock {\emph{\JournalTitle{Phys. Rev. B}}} \textbf{\bibinfo{volume}{83}}, \bibinfo{pages}{220402}, \doiprefix\url{10.1103/PhysRevB.83.220402} (\bibinfo{year}{2011}).

\bibitem{Archer2011}
\bibinfo{author}{Archer, T.} \emph{et~al.}
\newblock \bibinfo{journal}{\bibinfo{title}{Exchange interactions and magnetic phases of transition metal oxides: Benchmarking advanced ab initio methods}}.
\newblock {\emph{\JournalTitle{Phys. Rev. B}}} \textbf{\bibinfo{volume}{84}}, \bibinfo{pages}{115114}, \doiprefix\url{10.1103/PhysRevB.84.115114} (\bibinfo{year}{2011}).

\bibitem{Anisimov1991}
\bibinfo{author}{Anisimov, V.~I.}, \bibinfo{author}{Zaanen, J.} \& \bibinfo{author}{Andersen, O.~K.}
\newblock \bibinfo{journal}{\bibinfo{title}{Band theory and mott insulators: Hubbard u instead of stoner i}}.
\newblock {\emph{\JournalTitle{Phys. Rev. B}}} \textbf{\bibinfo{volume}{44}}, \bibinfo{pages}{943--954}, \doiprefix\url{10.1103/PhysRevB.44.943} (\bibinfo{year}{1991}).

\bibitem{Kusunose2008}
\bibinfo{author}{Kusunose, H.}
\newblock \bibinfo{journal}{\bibinfo{title}{Description of multipole in f-electron systems}}.
\newblock {\emph{\JournalTitle{Journal of the Physical Society of Japan}}} \textbf{\bibinfo{volume}{77}}, \bibinfo{pages}{064710}, \doiprefix\url{10.1143/JPSJ.77.064710} (\bibinfo{year}{2008}).
\newblock \eprint{https://doi.org/10.1143/JPSJ.77.064710}.

\bibitem{Lovesey2015}
\bibinfo{author}{Lovesey, S.~W.}
\newblock \bibinfo{journal}{\bibinfo{title}{Theory of neutron scattering by electrons in magnetic materials}}.
\newblock {\emph{\JournalTitle{Physica Scripta}}} \textbf{\bibinfo{volume}{90}}, \bibinfo{pages}{108011}, \doiprefix\url{10.1088/0031-8949/90/10/108011} (\bibinfo{year}{2015}).

\bibitem{Shick2001}
\bibinfo{author}{Shick, A.}, \bibinfo{author}{Pickett, W.} \& \bibinfo{author}{Liechtenstein, A.}
\newblock \bibinfo{journal}{\bibinfo{title}{Ground and metastable states in $\gamma$-ce from correlated band theory}}.
\newblock {\emph{\JournalTitle{Journal of Electron Spectroscopy and Related Phenomena}}} \textbf{\bibinfo{volume}{114-116}}, \bibinfo{pages}{753--758}, \doiprefix\url{https://doi.org/10.1016/S0368-2048(00)00394-7} (\bibinfo{year}{2001}).
\newblock \bibinfo{note}{Proceeding of the Eight International Conference on Electronic Spectroscopy and Structure,}.

\bibitem{Larson2007}
\bibinfo{author}{Larson, P.}, \bibinfo{author}{Lambrecht, W. R.~L.}, \bibinfo{author}{Chantis, A.} \& \bibinfo{author}{van Schilfgaarde, M.}
\newblock \bibinfo{journal}{\bibinfo{title}{Electronic structure of rare-earth nitrides using the $\mathrm{LSDA}+u$ approach: Importance of allowing $4f$ orbitals to break the cubic crystal symmetry}}.
\newblock {\emph{\JournalTitle{Phys. Rev. B}}} \textbf{\bibinfo{volume}{75}}, \bibinfo{pages}{045114}, \doiprefix\url{10.1103/PhysRevB.75.045114} (\bibinfo{year}{2007}).

\bibitem{Gerald2008}
\bibinfo{author}{Jomard, G.}, \bibinfo{author}{Amadon, B.}, \bibinfo{author}{Bottin, F. m.~c.} \& \bibinfo{author}{Torrent, M.}
\newblock \bibinfo{journal}{\bibinfo{title}{Structural, thermodynamic, and electronic properties of plutonium oxides from first principles}}.
\newblock {\emph{\JournalTitle{Phys. Rev. B}}} \textbf{\bibinfo{volume}{78}}, \bibinfo{pages}{075125}, \doiprefix\url{10.1103/PhysRevB.78.075125} (\bibinfo{year}{2008}).

\bibitem{Dorado2009}
\bibinfo{author}{Dorado, B.}, \bibinfo{author}{Amadon, B.}, \bibinfo{author}{Freyss, M.} \& \bibinfo{author}{Bertolus, M.}
\newblock \bibinfo{journal}{\bibinfo{title}{$\text{DFT}+\text{U}$ calculations of the ground state and metastable states of uranium dioxide}}.
\newblock {\emph{\JournalTitle{Phys. Rev. B}}} \textbf{\bibinfo{volume}{79}}, \bibinfo{pages}{235125}, \doiprefix\url{10.1103/PhysRevB.79.235125} (\bibinfo{year}{2009}).

\bibitem{Allen2014}
\bibinfo{author}{Allen, J.~P.} \& \bibinfo{author}{Watson, G.~W.}
\newblock \bibinfo{journal}{\bibinfo{title}{Occupation matrix control of d- and f-electron localisations using dft + u}}.
\newblock {\emph{\JournalTitle{Phys. Chem. Chem. Phys.}}} \textbf{\bibinfo{volume}{16}}, \bibinfo{pages}{21016--21031}, \doiprefix\url{10.1039/C4CP01083C} (\bibinfo{year}{2014}).

\bibitem{Ma2015}
\bibinfo{author}{Ma, P.-W.} \& \bibinfo{author}{Dudarev, S.~L.}
\newblock \bibinfo{journal}{\bibinfo{title}{Constrained density functional for noncollinear magnetism}}.
\newblock {\emph{\JournalTitle{Phys. Rev. B}}} \textbf{\bibinfo{volume}{91}}, \bibinfo{pages}{054420}, \doiprefix\url{10.1103/PhysRevB.91.054420} (\bibinfo{year}{2015}).

\bibitem{Dudarev2019}
\bibinfo{author}{Dudarev, S.~L.} \emph{et~al.}
\newblock \bibinfo{journal}{\bibinfo{title}{Parametrization of $\mathrm{LSDA}+u$ for noncollinear magnetic configurations: Multipolar magnetism in ${\mathrm{uo}}_{2}$}}.
\newblock {\emph{\JournalTitle{Phys. Rev. Mater.}}} \textbf{\bibinfo{volume}{3}}, \bibinfo{pages}{083802}, \doiprefix\url{10.1103/PhysRevMaterials.3.083802} (\bibinfo{year}{2019}).

\bibitem{Dederichs1984}
\bibinfo{author}{Dederichs, P.~H.}, \bibinfo{author}{Bl\"ugel, S.}, \bibinfo{author}{Zeller, R.} \& \bibinfo{author}{Akai, H.}
\newblock \bibinfo{journal}{\bibinfo{title}{Ground states of constrained systems: Application to cerium impurities}}.
\newblock {\emph{\JournalTitle{Phys. Rev. Lett.}}} \textbf{\bibinfo{volume}{53}}, \bibinfo{pages}{2512--2515}, \doiprefix\url{10.1103/PhysRevLett.53.2512} (\bibinfo{year}{1984}).

\bibitem{Tehrani2021}
\bibinfo{author}{Tehrani, A.~M.} \& \bibinfo{author}{Spaldin, N.~A.}
\newblock \bibinfo{journal}{\bibinfo{title}{Untangling the structural, magnetic dipole, and charge multipolar orders in {Ba$_2$MgReO$_6$}}}.
\newblock {\emph{\JournalTitle{Physical Review Materials}}} \textbf{\bibinfo{volume}{5}}, \doiprefix\url{10.1103/PhysRevMaterials.5.104410} (\bibinfo{year}{2021}).

\bibitem{Yoon2019}
\bibinfo{author}{Yoon, S.} \emph{et~al.}
\newblock \bibinfo{journal}{\bibinfo{title}{A “non-dynamical” way of describing room-temperature paramagnetic manganese oxide}}.
\newblock {\emph{\JournalTitle{Phys. Chem. Chem. Phys.}}} \textbf{\bibinfo{volume}{21}}, \bibinfo{pages}{15932--15939}, \doiprefix\url{10.1039/C9CP00280D} (\bibinfo{year}{2019}).

\bibitem{Varignon2019}
\bibinfo{author}{Varignon, J.}, \bibinfo{author}{Bibes, M.} \& \bibinfo{author}{Zunger, A.}
\newblock \bibinfo{journal}{\bibinfo{title}{Mott gapping in $3d\phantom{\rule{0.16em}{0ex}}ab{\mathrm{o}}_{3}$ perovskites without mott-hubbard interelectronic repulsion energy u}}.
\newblock {\emph{\JournalTitle{Phys. Rev. B}}} \textbf{\bibinfo{volume}{100}}, \bibinfo{pages}{035119}, \doiprefix\url{10.1103/PhysRevB.100.035119} (\bibinfo{year}{2019}).

\bibitem{Georges1996}
\bibinfo{author}{Georges, A.}, \bibinfo{author}{Kotliar, G.}, \bibinfo{author}{Krauth, W.} \& \bibinfo{author}{Rozenberg, M.~J.}
\newblock \bibinfo{journal}{\bibinfo{title}{Dynamical mean-field theory of strongly correlated fermion systems and the limit of infinite dimensions}}.
\newblock {\emph{\JournalTitle{Rev. Mod. Phys.}}} \textbf{\bibinfo{volume}{68}}, \bibinfo{pages}{13--125} (\bibinfo{year}{1996}).

\bibitem{Anisimov1997_1}
\bibinfo{author}{Anisimov, V.~I.}, \bibinfo{author}{Poteryaev, A.~I.}, \bibinfo{author}{Korotin, M.~A.}, \bibinfo{author}{Anokhin, A.~O.} \& \bibinfo{author}{Kotliar, G.}
\newblock \bibinfo{journal}{\bibinfo{title}{First-principles calculations of the electronic structure and spectra of strongly correlated systems: dynamical mean-field theory}}.
\newblock {\emph{\JournalTitle{Journal of Physics: Condensed Matter}}} \textbf{\bibinfo{volume}{9}}, \bibinfo{pages}{7359} (\bibinfo{year}{1997}).

\bibitem{Lichtenstein_LDApp}
\bibinfo{author}{Lichtenstein, A.~I.} \& \bibinfo{author}{Katsnelson, M.~I.}
\newblock \bibinfo{journal}{\bibinfo{title}{Ab initio calculations of quasiparticle band structure in correlated systems: Lda++ approach}}.
\newblock {\emph{\JournalTitle{Phys. Rev. B}}} \textbf{\bibinfo{volume}{57}}, \bibinfo{pages}{6884--6895} (\bibinfo{year}{1998}).

\bibitem{Kotliar2006}
\bibinfo{author}{Kotliar, G.} \emph{et~al.}
\newblock \bibinfo{journal}{\bibinfo{title}{Electronic structure calculations with dynamical mean-field theory}}.
\newblock {\emph{\JournalTitle{Rev. Mod. Phys.}}} \textbf{\bibinfo{volume}{78}}, \bibinfo{pages}{865--951}, \doiprefix\url{10.1103/RevModPhys.78.865} (\bibinfo{year}{2006}).

\bibitem{Pavarini2011}
\bibinfo{author}{Pavarini, E.}, \bibinfo{author}{Koch, E.}, \bibinfo{author}{Vollhardt, D.}, \bibinfo{author}{Lichtenstein, A.} \& \bibinfo{author}{with Predictive Power~for Strongly Correlated~Materials, D. F. F. . D. M.-F.~A.}
\newblock \emph{\bibinfo{title}{The LDA+DMFT Approach to Strongly Correlated Materials: Lecture Notes of the Autumn School 2011 Hands-on LDA+DMFT at Forschungszentrum J{\"u}lich, 4 - 7 October 2011, Organized by the DFG Research Unit 1346 Dynamical Mean-Field Approach with Predictive Power for Strongly Correlated Materials}}.
\newblock Schriften des Forschungszentrums J{\"u}lich: Reihe Modeling and Simulation (\bibinfo{publisher}{Universit{\"a}t Augsburg}, \bibinfo{year}{2011}).

\bibitem{Haule2009}
\bibinfo{author}{Haule, K.} \& \bibinfo{author}{Kotliar, G.}
\newblock \bibinfo{journal}{\bibinfo{title}{Arrested kondo effect and hidden order in uru2si2}}.
\newblock {\emph{\JournalTitle{Nature Physics}}} \textbf{\bibinfo{volume}{5}}, \bibinfo{pages}{796--799}, \doiprefix\url{10.1038/nphys1392} (\bibinfo{year}{2009}).

\bibitem{Merkel2024}
\bibinfo{author}{Merkel, M.~E.}, \bibinfo{author}{Tehrani, A.~M.} \& \bibinfo{author}{Ederer, C.}
\newblock \bibinfo{journal}{\bibinfo{title}{Probing the mott insulating behavior of ${\mathrm{ba}}_{2}{\mathrm{mgreo}}_{6}$ with $\mathrm{DFT}+\mathrm{DMFT}$}}.
\newblock {\emph{\JournalTitle{Phys. Rev. Res.}}} \textbf{\bibinfo{volume}{6}}, \bibinfo{pages}{023233}, \doiprefix\url{10.1103/PhysRevResearch.6.023233} (\bibinfo{year}{2024}).

\bibitem{Mackintosh1980}
\bibinfo{author}{Mackintosh, A.~R.} \& \bibinfo{author}{Andersen, O.}
\newblock \bibinfo{title}{Electrons at the fermi surface}.
\newblock chap. \bibinfo{chapter}{The electronic structure of transition metals} (\bibinfo{publisher}{Cambridge University Press}, \bibinfo{year}{1980}).

\bibitem{Liechtenstein1984}
\bibinfo{author}{Liechtenstein, A.~I.}, \bibinfo{author}{Katsnelson, M.~I.} \& \bibinfo{author}{Gubanov, V.~A.}
\newblock \bibinfo{journal}{\bibinfo{title}{Exchange interactions and spin-wave stiffness in ferromagnetic metals}}.
\newblock {\emph{\JournalTitle{Journal of Physics F: Metal Physics}}} \textbf{\bibinfo{volume}{14}}, \bibinfo{pages}{L125}, \doiprefix\url{10.1088/0305-4608/14/7/007} (\bibinfo{year}{1984}).

\bibitem{Liechtenstein1987}
\bibinfo{author}{Liechtenstein, A.}, \bibinfo{author}{Katsnelson, M.}, \bibinfo{author}{Antropov, V.} \& \bibinfo{author}{Gubanov, V.}
\newblock \bibinfo{journal}{\bibinfo{title}{Local spin density functional approach to the theory of exchange interactions in ferromagnetic metals and alloys}}.
\newblock {\emph{\JournalTitle{Journal of Magnetism and Magnetic Materials}}} \textbf{\bibinfo{volume}{67}}, \bibinfo{pages}{65--74}, \doiprefix\url{https://doi.org/10.1016/0304-8853(87)90721-9} (\bibinfo{year}{1987}).

\bibitem{Katsnelson2000}
\bibinfo{author}{Katsnelson, M.~I.} \& \bibinfo{author}{Lichtenstein, A.~I.}
\newblock \bibinfo{journal}{\bibinfo{title}{First-principles calculations of magnetic interactions in correlated systems}}.
\newblock {\emph{\JournalTitle{Phys. Rev. B}}} \textbf{\bibinfo{volume}{61}}, \bibinfo{pages}{8906--8912}, \doiprefix\url{10.1103/PhysRevB.61.8906} (\bibinfo{year}{2000}).

\bibitem{Bruno2003}
\bibinfo{author}{Bruno, P.}
\newblock \bibinfo{journal}{\bibinfo{title}{Exchange interaction parameters and adiabatic spin-wave spectra of ferromagnets: A ``renormalized magnetic force theorem''}}.
\newblock {\emph{\JournalTitle{Phys. Rev. Lett.}}} \textbf{\bibinfo{volume}{90}}, \bibinfo{pages}{087205}, \doiprefix\url{10.1103/PhysRevLett.90.087205} (\bibinfo{year}{2003}).

\bibitem{Kvashnin2015}
\bibinfo{author}{Kvashnin, Y.~O.} \emph{et~al.}
\newblock \bibinfo{journal}{\bibinfo{title}{Exchange parameters of strongly correlated materials: Extraction from spin-polarized density functional theory plus dynamical mean-field theory}}.
\newblock {\emph{\JournalTitle{Phys. Rev. B}}} \textbf{\bibinfo{volume}{91}}, \bibinfo{pages}{125133}, \doiprefix\url{10.1103/PhysRevB.91.125133} (\bibinfo{year}{2015}).

\bibitem{Solovyev1996}
\bibinfo{author}{Solovyev, I.}, \bibinfo{author}{Hamada, N.} \& \bibinfo{author}{Terakura, K.}
\newblock \bibinfo{journal}{\bibinfo{title}{Crucial role of the lattice distortion in the magnetism of ${\mathrm{lamno}}_{3}$}}.
\newblock {\emph{\JournalTitle{Phys. Rev. Lett.}}} \textbf{\bibinfo{volume}{76}}, \bibinfo{pages}{4825--4828}, \doiprefix\url{10.1103/PhysRevLett.76.4825} (\bibinfo{year}{1996}).

\bibitem{Hubbard1963}
\bibinfo{author}{Hubbard, J.}
\newblock \bibinfo{journal}{\bibinfo{title}{Electron correlations in narrow energy bands}}.
\newblock {\emph{\JournalTitle{Proc. Roy. Soc. (London)}}} \textbf{\bibinfo{volume}{A 276}}, \bibinfo{pages}{238}, \doiprefix\url{10.1098/rspa.1963.0204} (\bibinfo{year}{1963}).

\bibitem{Lebeque2005}
\bibinfo{author}{Leb\`egue, S.} \emph{et~al.}
\newblock \bibinfo{journal}{\bibinfo{title}{Electronic structure and spectroscopic properties of thulium monochalcogenides}}.
\newblock {\emph{\JournalTitle{Phys. Rev. B}}} \textbf{\bibinfo{volume}{72}}, \bibinfo{pages}{245102}, \doiprefix\url{10.1103/PhysRevB.72.245102} (\bibinfo{year}{2005}).

\bibitem{Pourovskii2009}
\bibinfo{author}{Pourovskii, L.~V.}, \bibinfo{author}{Delaney, K.~T.}, \bibinfo{author}{Van~de Walle, C.~G.}, \bibinfo{author}{Spaldin, N.~A.} \& \bibinfo{author}{Georges, A.}
\newblock \bibinfo{journal}{\bibinfo{title}{Role of atomic multiplets in the electronic structure of rare-earth semiconductors and semimetals}}.
\newblock {\emph{\JournalTitle{Phys. Rev. Lett.}}} \textbf{\bibinfo{volume}{102}}, \bibinfo{pages}{096401}, \doiprefix\url{10.1103/PhysRevLett.102.096401} (\bibinfo{year}{2009}).

\bibitem{Shick2009}
\bibinfo{author}{Shick, A.~B.}, \bibinfo{author}{Koloren\ifmmode~\check{c}\else \v{c}\fi{}, J.}, \bibinfo{author}{Lichtenstein, A.~I.} \& \bibinfo{author}{Havela, L.}
\newblock \bibinfo{journal}{\bibinfo{title}{Electronic structure and spectral properties of am, cm, and bk: Charge-density self-consistent $\text{LDA}+\text{HIA}$ calculations in the fp-lapw basis}}.
\newblock {\emph{\JournalTitle{Phys. Rev. B}}} \textbf{\bibinfo{volume}{80}}, \bibinfo{pages}{085106}, \doiprefix\url{10.1103/PhysRevB.80.085106} (\bibinfo{year}{2009}).

\bibitem{Locht2016}
\bibinfo{author}{Locht, I. L.~M.} \emph{et~al.}
\newblock \bibinfo{journal}{\bibinfo{title}{Standard model of the rare earths analyzed from the hubbard i approximation}}.
\newblock {\emph{\JournalTitle{Phys. Rev. B}}} \textbf{\bibinfo{volume}{94}}, \bibinfo{pages}{085137}, \doiprefix\url{10.1103/PhysRevB.94.085137} (\bibinfo{year}{2016}).

\bibitem{Georges2004}
\bibinfo{author}{Georges, A.}
\newblock \bibinfo{journal}{\bibinfo{title}{{Strongly Correlated Electron Materials: Dynamical Mean‐Field Theory and Electronic Structure}}}.
\newblock {\emph{\JournalTitle{AIP Conference Proceedings}}} \textbf{\bibinfo{volume}{715}}, \bibinfo{pages}{3--74}, \doiprefix\url{10.1063/1.1800733} (\bibinfo{year}{2004}).

\bibitem{Iwahara2018b}
\bibinfo{author}{Iwahara, N.}, \bibinfo{author}{Ungur, L.} \& \bibinfo{author}{Chibotaru, L.~F.}
\newblock \bibinfo{journal}{\bibinfo{title}{$\stackrel{\ifmmode \tilde{}\else \~{}\fi{}}{J}$-pseudospin states and the crystal field of cubic systems}}.
\newblock {\emph{\JournalTitle{Phys. Rev. B}}} \textbf{\bibinfo{volume}{98}}, \bibinfo{pages}{054436}, \doiprefix\url{10.1103/PhysRevB.98.054436} (\bibinfo{year}{2018}).

\bibitem{Polinger2009}
\bibinfo{author}{Polinger, V.}
\newblock \emph{\bibinfo{title}{Orbital Ordering Versus the Traditional Approach in the Cooperative Jahn--Teller Effect: A Comparative Study}}, \bibinfo{pages}{685--725} (\bibinfo{publisher}{Springer Berlin Heidelberg}, \bibinfo{address}{Berlin, Heidelberg}, \bibinfo{year}{2009}).

\bibitem{Pourovskii2023}
\bibinfo{author}{Pourovskii, L.~V.}
\newblock \bibinfo{journal}{\bibinfo{title}{Multipolar interactions and magnetic excitation gap in ${d}^{3}$ spin-orbit mott insulators}}.
\newblock {\emph{\JournalTitle{Phys. Rev. B}}} \textbf{\bibinfo{volume}{108}}, \bibinfo{pages}{054436}, \doiprefix\url{10.1103/PhysRevB.108.054436} (\bibinfo{year}{2023}).

\bibitem{Jarell1992}
\bibinfo{author}{Jarrell, M.}
\newblock \bibinfo{journal}{\bibinfo{title}{Hubbard model in infinite dimensions: A quantum monte carlo study}}.
\newblock {\emph{\JournalTitle{Phys. Rev. Lett.}}} \textbf{\bibinfo{volume}{69}}, \bibinfo{pages}{168--171}, \doiprefix\url{10.1103/PhysRevLett.69.168} (\bibinfo{year}{1992}).

\bibitem{Gull2011}
\bibinfo{author}{Gull, E.} \emph{et~al.}
\newblock \bibinfo{journal}{\bibinfo{title}{Continuous-time monte carlo methods for quantum impurity models}}.
\newblock {\emph{\JournalTitle{Rev. Mod. Phys.}}} \textbf{\bibinfo{volume}{83}}, \bibinfo{pages}{349--404}, \doiprefix\url{10.1103/RevModPhys.83.349} (\bibinfo{year}{2011}).

\bibitem{Otsuki2019}
\bibinfo{author}{Otsuki, J.}, \bibinfo{author}{Yoshimi, K.}, \bibinfo{author}{Shinaoka, H.} \& \bibinfo{author}{Nomura, Y.}
\newblock \bibinfo{journal}{\bibinfo{title}{Strong-coupling formula for momentum-dependent susceptibilities in dynamical mean-field theory}}.
\newblock {\emph{\JournalTitle{Phys. Rev. B}}} \textbf{\bibinfo{volume}{99}}, \bibinfo{pages}{165134}, \doiprefix\url{10.1103/PhysRevB.99.165134} (\bibinfo{year}{2019}).

\bibitem{Otsuki2024}
\bibinfo{author}{Otsuki, J.}, \bibinfo{author}{Yoshimi, K.}, \bibinfo{author}{Shinaoka, H.} \& \bibinfo{author}{Jeschke, H.~O.}
\newblock \bibinfo{journal}{\bibinfo{title}{Multipolar ordering from dynamical mean field theory with application to ${\mathrm{ceb}}_{6}$}}.
\newblock {\emph{\JournalTitle{Phys. Rev. B}}} \textbf{\bibinfo{volume}{110}}, \bibinfo{pages}{035104}, \doiprefix\url{10.1103/PhysRevB.110.035104} (\bibinfo{year}{2024}).

\bibitem{Anderson1959}
\bibinfo{author}{Anderson, P.~W.}
\newblock \bibinfo{journal}{\bibinfo{title}{New approach to the theory of superexchange interactions}}.
\newblock {\emph{\JournalTitle{Phys. Rev.}}} \textbf{\bibinfo{volume}{115}}, \bibinfo{pages}{2--13}, \doiprefix\url{10.1103/PhysRev.115.2} (\bibinfo{year}{1959}).

\bibitem{Anderson1963}
\bibinfo{author}{Anderson, P.~W.}
\newblock \bibinfo{title}{Theory of magnetic exchange interactions:exchange in insulators and semiconductors}.
\newblock vol.~\bibinfo{volume}{14} of \emph{\bibinfo{series}{Solid State Physics}}, \bibinfo{pages}{99--214}, \doiprefix\url{https://doi.org/10.1016/S0081-1947(08)60260-X} (\bibinfo{publisher}{Academic Press}, \bibinfo{year}{1963}).

\bibitem{Mironov2003}
\bibinfo{author}{Mironov, V.}, \bibinfo{author}{Chibotaru, L.} \& \bibinfo{author}{Ceulemans, A.}
\newblock \bibinfo{title}{First-order phase transition in uo2: The interplay of the 5f2–5f2 superexchange interaction and jahn–teller effect}.
\newblock In \emph{\bibinfo{booktitle}{Advances in Quantum Chemistry}}, vol.~\bibinfo{volume}{44} of \emph{\bibinfo{series}{Advances in Quantum Chemistry}}, \bibinfo{pages}{599--616}, \doiprefix\url{https://doi.org/10.1016/S0065-3276(03)44040-9} (\bibinfo{publisher}{Academic Press}, \bibinfo{year}{2003}).

\bibitem{Iwahara2015}
\bibinfo{author}{Iwahara, N.} \& \bibinfo{author}{Chibotaru, L.~F.}
\newblock \bibinfo{journal}{\bibinfo{title}{Exchange interaction between $j$ multiplets}}.
\newblock {\emph{\JournalTitle{Phys. Rev. B}}} \textbf{\bibinfo{volume}{91}}, \bibinfo{pages}{174438}, \doiprefix\url{10.1103/PhysRevB.91.174438} (\bibinfo{year}{2015}).

\bibitem{Voleti2021}
\bibinfo{author}{Voleti, S.}, \bibinfo{author}{Haldar, A.} \& \bibinfo{author}{Paramekanti, A.}
\newblock \bibinfo{journal}{\bibinfo{title}{Octupolar order and ising quantum criticality tuned by strain and dimensionality: Application to $d$-orbital mott insulators}}.
\newblock {\emph{\JournalTitle{Phys. Rev. B}}} \textbf{\bibinfo{volume}{104}}, \bibinfo{pages}{174431}, \doiprefix\url{10.1103/PhysRevB.104.174431} (\bibinfo{year}{2021}).

\bibitem{Schrieffer1966}
\bibinfo{author}{Schrieffer, J.~R.} \& \bibinfo{author}{Wolff, P.~A.}
\newblock \bibinfo{journal}{\bibinfo{title}{Relation between the anderson and kondo hamiltonians}}.
\newblock {\emph{\JournalTitle{Phys. Rev.}}} \textbf{\bibinfo{volume}{149}}, \bibinfo{pages}{491--492}, \doiprefix\url{10.1103/PhysRev.149.491} (\bibinfo{year}{1966}).

\bibitem{Beom2012}
\bibinfo{author}{Kim, B.~H.}, \bibinfo{author}{Khaliullin, G.} \& \bibinfo{author}{Min, B.~I.}
\newblock \bibinfo{journal}{\bibinfo{title}{Magnetic couplings, optical spectra, and spin-orbit exciton in $5d$ electron mott insulator ${\mathrm{sr}}_{2}{\mathrm{iro}}_{4}$}}.
\newblock {\emph{\JournalTitle{Phys. Rev. Lett.}}} \textbf{\bibinfo{volume}{109}}, \bibinfo{pages}{167205}, \doiprefix\url{10.1103/PhysRevLett.109.167205} (\bibinfo{year}{2012}).

\bibitem{Winter2016}
\bibinfo{author}{Winter, S.~M.}, \bibinfo{author}{Li, Y.}, \bibinfo{author}{Jeschke, H.~O.} \& \bibinfo{author}{Valent\'{\i}, R.}
\newblock \bibinfo{journal}{\bibinfo{title}{Challenges in design of kitaev materials: Magnetic interactions from competing energy scales}}.
\newblock {\emph{\JournalTitle{Phys. Rev. B}}} \textbf{\bibinfo{volume}{93}}, \bibinfo{pages}{214431}, \doiprefix\url{10.1103/PhysRevB.93.214431} (\bibinfo{year}{2016}).

\bibitem{Beom2019}
\bibinfo{author}{Kim, B.~H.}, \bibinfo{author}{Efremov, D.~V.} \& \bibinfo{author}{van~den Brink, J.}
\newblock \bibinfo{journal}{\bibinfo{title}{Spin-orbital excitons and their potential condensation in pentavalent iridates}}.
\newblock {\emph{\JournalTitle{Phys. Rev. Mater.}}} \textbf{\bibinfo{volume}{3}}, \bibinfo{pages}{014414}, \doiprefix\url{10.1103/PhysRevMaterials.3.014414} (\bibinfo{year}{2019}).

\bibitem{Iwahara2018}
\bibinfo{author}{Iwahara, N.}, \bibinfo{author}{Vieru, V.} \& \bibinfo{author}{Chibotaru, L.~F.}
\newblock \bibinfo{journal}{\bibinfo{title}{Spin-orbital-lattice entangled states in cubic ${d}^{1}$ double perovskites}}.
\newblock {\emph{\JournalTitle{Phys. Rev. B}}} \textbf{\bibinfo{volume}{98}}, \bibinfo{pages}{075138}, \doiprefix\url{10.1103/PhysRevB.98.075138} (\bibinfo{year}{2018}).

\bibitem{Peil2019}
\bibinfo{author}{Peil, O.~E.}, \bibinfo{author}{Hampel, A.}, \bibinfo{author}{Ederer, C.} \& \bibinfo{author}{Georges, A.}
\newblock \bibinfo{journal}{\bibinfo{title}{Mechanism and control parameters of the coupled structural and metal-insulator transition in nickelates}}.
\newblock {\emph{\JournalTitle{Phys. Rev. B}}} \textbf{\bibinfo{volume}{99}}, \bibinfo{pages}{245127}, \doiprefix\url{10.1103/PhysRevB.99.245127} (\bibinfo{year}{2019}).

\bibitem{Georgescu2019}
\bibinfo{author}{Georgescu, A.~B.}, \bibinfo{author}{Peil, O.~E.}, \bibinfo{author}{Disa, A.~S.}, \bibinfo{author}{Georges, A.} \& \bibinfo{author}{Millis, A.~J.}
\newblock \bibinfo{journal}{\bibinfo{title}{Disentangling lattice and electronic contributions to the metal–insulator transition from bulk vs. layer confined rnio 3}}.
\newblock {\emph{\JournalTitle{Proceedings of the National Academy of Sciences}}} \textbf{\bibinfo{volume}{116}}, \bibinfo{pages}{14434–14439}, \doiprefix\url{10.1073/pnas.1818728116} (\bibinfo{year}{2019}).

\bibitem{Delange2017}
\bibinfo{author}{Delange, P.}, \bibinfo{author}{Biermann, S.}, \bibinfo{author}{Miyake, T.} \& \bibinfo{author}{Pourovskii, L.}
\newblock \bibinfo{journal}{\bibinfo{title}{Crystal-field splittings in rare-earth-based hard magnets: An ab initio approach}}.
\newblock {\emph{\JournalTitle{Phys. Rev. B}}} \textbf{\bibinfo{volume}{96}}, \bibinfo{pages}{155132}, \doiprefix\url{10.1103/PhysRevB.96.155132} (\bibinfo{year}{2017}).

\bibitem{Baroni2001}
\bibinfo{author}{Baroni, S.}, \bibinfo{author}{de~Gironcoli, S.}, \bibinfo{author}{Dal~Corso, A.} \& \bibinfo{author}{Giannozzi, P.}
\newblock \bibinfo{journal}{\bibinfo{title}{Phonons and related crystal properties from density-functional perturbation theory}}.
\newblock {\emph{\JournalTitle{Rev. Mod. Phys.}}} \textbf{\bibinfo{volume}{73}}, \bibinfo{pages}{515--562}, \doiprefix\url{10.1103/RevModPhys.73.515} (\bibinfo{year}{2001}).

\bibitem{Pourovskii2019}
\bibinfo{author}{Pourovskii, L.~V.} \& \bibinfo{author}{Khmelevskyi, S.}
\newblock \bibinfo{journal}{\bibinfo{title}{Quadrupolar superexchange interactions, multipolar order, and magnetic phase transition in ${\mathrm{uo}}_{2}$}}.
\newblock {\emph{\JournalTitle{Phys. Rev. B}}} \textbf{\bibinfo{volume}{99}}, \bibinfo{pages}{094439}, \doiprefix\url{10.1103/PhysRevB.99.094439} (\bibinfo{year}{2019}).

\bibitem{Trimarchi2018}
\bibinfo{author}{Trimarchi, G.}, \bibinfo{author}{Wang, Z.} \& \bibinfo{author}{Zunger, A.}
\newblock \bibinfo{journal}{\bibinfo{title}{Polymorphous band structure model of gapping in the antiferromagnetic and paramagnetic phases of the mott insulators mno, feo, coo, and nio}}.
\newblock {\emph{\JournalTitle{Phys. Rev. B}}} \textbf{\bibinfo{volume}{97}}, \bibinfo{pages}{035107}, \doiprefix\url{10.1103/PhysRevB.97.035107} (\bibinfo{year}{2018}).

\bibitem{Rotter2012}
\bibinfo{author}{Rotter, M.}, \bibinfo{author}{Le, M.~D.}, \bibinfo{author}{Boothroyd, A.~T.} \& \bibinfo{author}{Blanco, J.~A.}
\newblock \bibinfo{journal}{\bibinfo{title}{Dynamical matrix diagonalization for the calculation of dispersive excitations}}.
\newblock {\emph{\JournalTitle{Journal of Physics: Condensed Matter}}} \textbf{\bibinfo{volume}{24}}, \bibinfo{pages}{213201}, \doiprefix\url{10.1088/0953-8984/24/21/213201} (\bibinfo{year}{2012}).

\bibitem{Shiina2007}
\bibinfo{author}{Shiina, R.}, \bibinfo{author}{Sakai, O.} \& \bibinfo{author}{Shiba, H.}
\newblock \bibinfo{journal}{\bibinfo{title}{Magnetic form factor of elastic neutron scattering expected for octupolar phases in {Ce$_{1-x}$La$_x$B$_6$} and {NpO$_2$}}}.
\newblock {\emph{\JournalTitle{Journal of the Physical Society of Japan}}} \textbf{\bibinfo{volume}{76}}, \bibinfo{pages}{094702}, \doiprefix\url{10.1143/JPSJ.76.094702} (\bibinfo{year}{2007}).

\bibitem{quantum_magnetism_2004}
\bibinfo{editor}{Schollw\"ock, U.}, \bibinfo{editor}{Richter, J.}, \bibinfo{editor}{Farnell, D. J.~J.} \& \bibinfo{editor}{Bishop, R.~F.} (eds.) \emph{\bibinfo{title}{Quantum Magnetism}} (\bibinfo{publisher}{Springer Berlin Heidelberg}, \bibinfo{year}{2004}).

\bibitem{Sandvik2010}
\bibinfo{author}{Sandvik, A.~W.}
\newblock \bibinfo{journal}{\bibinfo{title}{{Computational Studies of Quantum Spin Systems}}}.
\newblock {\emph{\JournalTitle{AIP Conference Proceedings}}} \textbf{\bibinfo{volume}{1297}}, \bibinfo{pages}{135--338}, \doiprefix\url{10.1063/1.3518900} (\bibinfo{year}{2010}).

\bibitem{Schollwock2011}
\bibinfo{author}{Schollw\"ock, U.}
\newblock \bibinfo{journal}{\bibinfo{title}{The density-matrix renormalization group in the age of matrix product states}}.
\newblock {\emph{\JournalTitle{Annals of Physics}}} \textbf{\bibinfo{volume}{326}}, \bibinfo{pages}{96--192}, \doiprefix\url{https://doi.org/10.1016/j.aop.2010.09.012} (\bibinfo{year}{2011}).
\newblock \bibinfo{note}{January 2011 Special Issue}.

\bibitem{}
\bibinfo{author}{Thalmeier, P.}, \bibinfo{author}{Shiina, R.}, \bibinfo{author}{Shiba, H.} \& \bibinfo{author}{Sakai, O.}
\newblock \bibinfo{journal}{\bibinfo{title}{Theory of multipolar excitations in ceb 6}}.
\newblock {\emph{\JournalTitle{Journal of the Physical Society of Japan}}} \textbf{\bibinfo{volume}{67}}, \bibinfo{pages}{2363--2371}, \doiprefix\url{10.1143/JPSJ.67.2363} (\bibinfo{year}{1998}).
\newblock \eprint{https://doi.org/10.1143/JPSJ.67.2363}.

\bibitem{Englman1970}
\bibinfo{author}{Englman, R.} \& \bibinfo{author}{Halperin, B.}
\newblock \bibinfo{journal}{\bibinfo{title}{Cooperative dynamic jahn-teller effect. i. molecular field treatment of spinels}}.
\newblock {\emph{\JournalTitle{Phys. Rev. B}}} \textbf{\bibinfo{volume}{2}}, \bibinfo{pages}{75--94}, \doiprefix\url{10.1103/PhysRevB.2.75} (\bibinfo{year}{1970}).

\bibitem{Khaliullin2021}
\bibinfo{author}{Khaliullin, G.}, \bibinfo{author}{Churchill, D.}, \bibinfo{author}{Stavropoulos, P.~P.} \& \bibinfo{author}{Kee, H.-Y.}
\newblock \bibinfo{journal}{\bibinfo{title}{Exchange interactions, jahn-teller coupling, and multipole orders in pseudospin one-half $5{d}^{2}$ mott insulators}}.
\newblock {\emph{\JournalTitle{Phys. Rev. Res.}}} \textbf{\bibinfo{volume}{3}}, \bibinfo{pages}{033163}, \doiprefix\url{10.1103/PhysRevResearch.3.033163} (\bibinfo{year}{2021}).

\bibitem{Devereaux2007}
\bibinfo{author}{Devereaux, T.~P.} \& \bibinfo{author}{Hackl, R.}
\newblock \bibinfo{journal}{\bibinfo{title}{Inelastic light scattering from correlated electrons}}.
\newblock {\emph{\JournalTitle{Rev. Mod. Phys.}}} \textbf{\bibinfo{volume}{79}}, \bibinfo{pages}{175--233}, \doiprefix\url{10.1103/RevModPhys.79.175} (\bibinfo{year}{2007}).

\bibitem{Kim2017rixs}
\bibinfo{author}{Kim, B.~J.} \& \bibinfo{author}{Khaliullin, G.}
\newblock \bibinfo{journal}{\bibinfo{title}{Resonant inelastic x-ray scattering operators for ${t}_{2g}$ orbital systems}}.
\newblock {\emph{\JournalTitle{Phys. Rev. B}}} \textbf{\bibinfo{volume}{96}}, \bibinfo{pages}{085108}, \doiprefix\url{10.1103/PhysRevB.96.085108} (\bibinfo{year}{2017}).

\bibitem{Lovesey_book}
\bibinfo{author}{Lovesey, S.~W.}
\newblock \emph{\bibinfo{title}{Theory of Neutron Scattering from Condensed Matter}} (\bibinfo{publisher}{Clarendon Press, Oxford}, \bibinfo{year}{1984}).

\bibitem{Paramekanti2020}
\bibinfo{author}{Paramekanti, A.}, \bibinfo{author}{Maharaj, D.~D.} \& \bibinfo{author}{Gaulin, B.~D.}
\newblock \bibinfo{journal}{\bibinfo{title}{Octupolar order in $d$-orbital mott insulators}}.
\newblock {\emph{\JournalTitle{Phys. Rev. B}}} \textbf{\bibinfo{volume}{101}}, \bibinfo{pages}{054439}, \doiprefix\url{10.1103/PhysRevB.101.054439} (\bibinfo{year}{2020}).

\bibitem{Yuan2017}
\bibinfo{author}{Yuan, B.} \emph{et~al.}
\newblock \bibinfo{journal}{\bibinfo{title}{Determination of hund's coupling in $5d$ oxides using resonant inelastic x-ray scattering}}.
\newblock {\emph{\JournalTitle{Phys. Rev. B}}} \textbf{\bibinfo{volume}{95}}, \bibinfo{pages}{235114}, \doiprefix\url{10.1103/PhysRevB.95.235114} (\bibinfo{year}{2017}).

\bibitem{Nag2018}
\bibinfo{author}{Nag, A.} \emph{et~al.}
\newblock \bibinfo{journal}{\bibinfo{title}{Origin of magnetic moments and presence of spin-orbit singlets in ${\mathrm{ba}}_{2}{\mathrm{yiro}}_{6}$}}.
\newblock {\emph{\JournalTitle{Phys. Rev. B}}} \textbf{\bibinfo{volume}{98}}, \bibinfo{pages}{014431}, \doiprefix\url{10.1103/PhysRevB.98.014431} (\bibinfo{year}{2018}).

\bibitem{Cong2023}
\bibinfo{author}{Cong, R.} \emph{et~al.}
\newblock \bibinfo{journal}{\bibinfo{title}{Effects of charge doping on mott insulator with strong spin-orbit coupling, ${\mathrm{ba}}_{2}{\mathrm{na}}_{1\ensuremath{-}x}{\mathrm{ca}}_{x}{\mathrm{oso}}_{6}$}}.
\newblock {\emph{\JournalTitle{Phys. Rev. Mater.}}} \textbf{\bibinfo{volume}{7}}, \bibinfo{pages}{084409}, \doiprefix\url{10.1103/PhysRevMaterials.7.084409} (\bibinfo{year}{2023}).

\bibitem{Stitzer2002}
\bibinfo{author}{Stitzer, K.~E.}, \bibinfo{author}{Smith, M.~D.} \& \bibinfo{author}{{zur Loye}, H.-C.}
\newblock \bibinfo{journal}{\bibinfo{title}{Crystal growth of ba2moso6 (m=li, na) from reactive hydroxide fluxes}}.
\newblock {\emph{\JournalTitle{Solid State Sciences}}} \textbf{\bibinfo{volume}{4}}, \bibinfo{pages}{311--316}, \doiprefix\url{https://doi.org/10.1016/S1293-2558(01)01257-2} (\bibinfo{year}{2002}).

\bibitem{Erickson2007}
\bibinfo{author}{Erickson, A.~S.} \emph{et~al.}
\newblock \bibinfo{journal}{\bibinfo{title}{Ferromagnetism in the mott insulator ${\mathrm{ba}}_{2}{\mathrm{naoso}}_{6}$}}.
\newblock {\emph{\JournalTitle{Phys. Rev. Lett.}}} \textbf{\bibinfo{volume}{99}}, \bibinfo{pages}{016404}, \doiprefix\url{10.1103/PhysRevLett.99.016404} (\bibinfo{year}{2007}).

\bibitem{Steele2011}
\bibinfo{author}{Steele, A.~J.} \emph{et~al.}
\newblock \bibinfo{journal}{\bibinfo{title}{Low-moment magnetism in the double perovskites ba${}_{2}$$m$oso${}_{6}$ ($m=\text{Li},\text{Na}$)}}.
\newblock {\emph{\JournalTitle{Phys. Rev. B}}} \textbf{\bibinfo{volume}{84}}, \bibinfo{pages}{144416}, \doiprefix\url{10.1103/PhysRevB.84.144416} (\bibinfo{year}{2011}).

\bibitem{Carlo2011}
\bibinfo{author}{Carlo, J.~P.} \emph{et~al.}
\newblock \bibinfo{journal}{\bibinfo{title}{Triplet and in-gap magnetic states in the ground state of the quantum frustrated fcc antiferromagnet ba${}_{2}$ymoo${}_{6}$}}.
\newblock {\emph{\JournalTitle{Phys. Rev. B}}} \textbf{\bibinfo{volume}{84}}, \bibinfo{pages}{100404}, \doiprefix\url{10.1103/PhysRevB.84.100404} (\bibinfo{year}{2011}).

\bibitem{Yamamura2006}
\bibinfo{author}{Yamamura, K.}, \bibinfo{author}{Wakeshima, M.} \& \bibinfo{author}{Hinatsu, Y.}
\newblock \bibinfo{journal}{\bibinfo{title}{Structural phase transition and magnetic properties of double perovskites ba2camo6 (m=w, re, os)}}.
\newblock {\emph{\JournalTitle{Journal of Solid State Chemistry}}} \textbf{\bibinfo{volume}{179}}, \bibinfo{pages}{605--612}, \doiprefix\url{https://doi.org/10.1016/j.jssc.2005.10.003} (\bibinfo{year}{2006}).

\bibitem{Ishikawa2021a}
\bibinfo{author}{Ishikawa, H.} \emph{et~al.}
\newblock \bibinfo{journal}{\bibinfo{title}{Phase transition in the $5{d}^{1}$ double perovskite ${\mathrm{ba}}_{2}{\mathrm{careo}}_{6}$ induced by high magnetic field}}.
\newblock {\emph{\JournalTitle{Phys. Rev. B}}} \textbf{\bibinfo{volume}{104}}, \bibinfo{pages}{174422}, \doiprefix\url{10.1103/PhysRevB.104.174422} (\bibinfo{year}{2021}).

\bibitem{Marjerrison2016}
\bibinfo{author}{Marjerrison, C.~A.} \emph{et~al.}
\newblock \bibinfo{journal}{\bibinfo{title}{Cubic re6+ (5d1) double perovskites, ba2mgreo6, ba2znreo6, and ba2y2/3reo6: Magnetism, heat capacity, $\mu$sr, and neutron scattering studies and comparison with theory}}.
\newblock {\emph{\JournalTitle{Inorganic Chemistry}}} \textbf{\bibinfo{volume}{55}}, \bibinfo{pages}{10701--10713}, \doiprefix\url{10.1021/acs.inorgchem.6b01933} (\bibinfo{year}{2016}).
\newblock \eprint{https://doi.org/10.1021/acs.inorgchem.6b01933}.

\bibitem{Hirai2019}
\bibinfo{author}{Hirai, D.} \& \bibinfo{author}{Hiroi, Z.}
\newblock \bibinfo{journal}{\bibinfo{title}{Successive symmetry breaking in a jeff = 3/2 quartet in the spin–orbit coupled insulator ba2mgreo6}}.
\newblock {\emph{\JournalTitle{Journal of the Physical Society of Japan}}} \textbf{\bibinfo{volume}{88}}, \bibinfo{pages}{064712}, \doiprefix\url{10.7566/JPSJ.88.064712} (\bibinfo{year}{2019}).
\newblock \eprint{https://doi.org/10.7566/JPSJ.88.064712}.

\bibitem{LiuW2018}
\bibinfo{author}{Liu, W.} \emph{et~al.}
\newblock \bibinfo{journal}{\bibinfo{title}{Phase diagram of ba2naoso6, a mott insulator with strong spin orbit interactions}}.
\newblock {\emph{\JournalTitle{Physica B: Condensed Matter}}} \textbf{\bibinfo{volume}{536}}, \bibinfo{pages}{863--866}, \doiprefix\url{https://doi.org/10.1016/j.physb.2017.08.062} (\bibinfo{year}{2018}).

\bibitem{Willa2019}
\bibinfo{author}{Willa, K.} \emph{et~al.}
\newblock \bibinfo{journal}{\bibinfo{title}{Phase transition preceding magnetic long-range order in the double perovskite ${\mathrm{ba}}_{2}{\mathrm{naoso}}_{6}$}}.
\newblock {\emph{\JournalTitle{Phys. Rev. B}}} \textbf{\bibinfo{volume}{100}}, \bibinfo{pages}{041108}, \doiprefix\url{10.1103/PhysRevB.100.041108} (\bibinfo{year}{2019}).

\bibitem{Kesavan2020}
\bibinfo{author}{Kesavan, J.~K.} \emph{et~al.}
\newblock \bibinfo{journal}{\bibinfo{title}{Doping evolution of the local electronic and structural properties of the double perovskite ba2na1–xcaxoso6}}.
\newblock {\emph{\JournalTitle{The Journal of Physical Chemistry C}}} \textbf{\bibinfo{volume}{124}}, \bibinfo{pages}{16577--16585}, \doiprefix\url{10.1021/acs.jpcc.0c04807} (\bibinfo{year}{2020}).
\newblock \bibinfo{note}{PMID: 33643515}, \eprint{https://doi.org/10.1021/acs.jpcc.0c04807}.

\bibitem{Barbosa2022}
\bibinfo{author}{da~Cruz Pinha~Barbosa, V.} \emph{et~al.}
\newblock \bibinfo{journal}{\bibinfo{title}{The impact of structural distortions on the magnetism of double perovskites containing 5d1 transition-metal ions}}.
\newblock {\emph{\JournalTitle{Chemistry of Materials}}} \textbf{\bibinfo{volume}{34}}, \bibinfo{pages}{1098--1109}, \doiprefix\url{10.1021/acs.chemmater.1c03456} (\bibinfo{year}{2022}).
\newblock \eprint{https://doi.org/10.1021/acs.chemmater.1c03456}.

\bibitem{Arima2022}
\bibinfo{author}{Arima, H.}, \bibinfo{author}{Oshita, Y.}, \bibinfo{author}{Hirai, D.}, \bibinfo{author}{Hiroi, Z.} \& \bibinfo{author}{Matsubayashi, K.}
\newblock \bibinfo{journal}{\bibinfo{title}{Interplay between quadrupolar and magnetic interactions in 5d1 double perovskite ba2mgreo6 under pressure}}.
\newblock {\emph{\JournalTitle{Journal of the Physical Society of Japan}}} \textbf{\bibinfo{volume}{91}}, \bibinfo{pages}{013702}, \doiprefix\url{10.7566/JPSJ.91.013702} (\bibinfo{year}{2022}).
\newblock \eprint{https://doi.org/10.7566/JPSJ.91.013702}.

\bibitem{Frontini2023}
\bibinfo{author}{Frontini, F.~I.} \emph{et~al.}
\newblock \bibinfo{title}{Spin-orbit-lattice entangled state in {A$_2$MgReO$_6$ (A = Ca, Sr, Ba)} revealed by resonant inelastic x-ray scattering} (\bibinfo{year}{2023}).
\newblock \eprint{2311.01621}.

\bibitem{Zivkovic2024}
\bibinfo{author}{{\v{Z}}ivkovi{\'{c}}, I.} \emph{et~al.}
\newblock \bibinfo{journal}{\bibinfo{title}{Dynamic jahn-teller effect in the strong spin-orbit coupling regime}}.
\newblock {\emph{\JournalTitle{Nature Communications}}} \textbf{\bibinfo{volume}{15}}, \bibinfo{pages}{8587}, \doiprefix\url{10.1038/s41467-024-52935-w} (\bibinfo{year}{2024}).

\bibitem{Aharen2010}
\bibinfo{author}{Aharen, T.} \emph{et~al.}
\newblock \bibinfo{journal}{\bibinfo{title}{Magnetic properties of the geometrically frustrated $s=\frac{1}{2}$ antiferromagnets, ${\text{la}}_{2}{\text{limoo}}_{6}$ and ${\text{ba}}_{2}{\text{ymoo}}_{6}$, with the b-site ordered double perovskite structure: Evidence for a collective spin-singlet ground state}}.
\newblock {\emph{\JournalTitle{Phys. Rev. B}}} \textbf{\bibinfo{volume}{81}}, \bibinfo{pages}{224409}, \doiprefix\url{10.1103/PhysRevB.81.224409} (\bibinfo{year}{2010}).

\bibitem{Pasztorova2023}
\bibinfo{author}{P\'asztorov\'a', J.}, \bibinfo{author}{Tehrani, A.~M.}, \bibinfo{author}{\u{Z}ivković, I.}, \bibinfo{author}{Spaldin, N.~A.} \& \bibinfo{author}{R{\o}nnow, H.~M.}
\newblock \bibinfo{journal}{\bibinfo{title}{Experimental and theoretical thermodynamic studies in ba2mgreo6—the ground state in the context of jahn-teller effect}}.
\newblock {\emph{\JournalTitle{Journal of Physics: Condensed Matter}}} \textbf{\bibinfo{volume}{35}}, \bibinfo{pages}{245603}, \doiprefix\url{10.1088/1361-648X/acc62a} (\bibinfo{year}{2023}).

\bibitem{Voleti2020}
\bibinfo{author}{Voleti, S.}, \bibinfo{author}{Maharaj, D.~D.}, \bibinfo{author}{Gaulin, B.~D.}, \bibinfo{author}{Luke, G.} \& \bibinfo{author}{Paramekanti, A.}
\newblock \bibinfo{journal}{\bibinfo{title}{Multipolar magnetism in $d$-orbital systems: Crystal field levels, octupolar order, and orbital loop currents}}.
\newblock {\emph{\JournalTitle{Phys. Rev. B}}} \textbf{\bibinfo{volume}{101}}, \bibinfo{pages}{155118}, \doiprefix\url{10.1103/PhysRevB.101.155118} (\bibinfo{year}{2020}).

\bibitem{Churchill2022}
\bibinfo{author}{Churchill, D.} \& \bibinfo{author}{Kee, H.-Y.}
\newblock \bibinfo{journal}{\bibinfo{title}{Competing multipolar orders in a face-centered cubic lattice: Application to the osmium double perovskites}}.
\newblock {\emph{\JournalTitle{Phys. Rev. B}}} \textbf{\bibinfo{volume}{105}}, \bibinfo{pages}{014438}, \doiprefix\url{10.1103/PhysRevB.105.014438} (\bibinfo{year}{2022}).

\bibitem{Rayyan2023}
\bibinfo{author}{Rayyan, A.}, \bibinfo{author}{Liu, X.} \& \bibinfo{author}{Kee, H.-Y.}
\newblock \bibinfo{journal}{\bibinfo{title}{Fate of multipolar physics in $5{d}^{2}$ double perovskites}}.
\newblock {\emph{\JournalTitle{Phys. Rev. B}}} \textbf{\bibinfo{volume}{108}}, \bibinfo{pages}{045149}, \doiprefix\url{10.1103/PhysRevB.108.045149} (\bibinfo{year}{2023}).

\bibitem{Aharen2009}
\bibinfo{author}{Aharen, T.} \emph{et~al.}
\newblock \bibinfo{journal}{\bibinfo{title}{Magnetic properties of the $s=\frac{3}{2}$ geometrically frustrated double perovskites ${\text{la}}_{2}{\text{liruo}}_{6}$ and ${\text{ba}}_{2}{\text{yruo}}_{6}$}}.
\newblock {\emph{\JournalTitle{Phys. Rev. B}}} \textbf{\bibinfo{volume}{80}}, \bibinfo{pages}{134423}, \doiprefix\url{10.1103/PhysRevB.80.134423} (\bibinfo{year}{2009}).

\bibitem{Carlo2013}
\bibinfo{author}{Carlo, J.~P.} \emph{et~al.}
\newblock \bibinfo{journal}{\bibinfo{title}{Spin gap and the nature of the 4${d}^{3}$ magnetic ground state in the frustrated fcc antiferromagnet ba${}_{2}$yruo${}_{6}$}}.
\newblock {\emph{\JournalTitle{Phys. Rev. B}}} \textbf{\bibinfo{volume}{88}}, \bibinfo{pages}{024418}, \doiprefix\url{10.1103/PhysRevB.88.024418} (\bibinfo{year}{2013}).

\bibitem{paddison2023}
\bibinfo{author}{Paddison, J. A.~M.} \emph{et~al.}
\newblock \bibinfo{title}{Cubic double perovskites host noncoplanar spin textures} (\bibinfo{year}{2023}).
\newblock \eprint{2301.11395}.

\bibitem{Kermarrec2015}
\bibinfo{author}{Kermarrec, E.} \emph{et~al.}
\newblock \bibinfo{journal}{\bibinfo{title}{Frustrated fcc antiferromagnet ${\mathrm{ba}}_{2}{\mathrm{yoso}}_{6}$: Structural characterization, magnetic properties, and neutron scattering studies}}.
\newblock {\emph{\JournalTitle{Phys. Rev. B}}} \textbf{\bibinfo{volume}{91}}, \bibinfo{pages}{075133}, \doiprefix\url{10.1103/PhysRevB.91.075133} (\bibinfo{year}{2015}).

\bibitem{Ishikawa2021b}
\bibinfo{author}{Ishikawa, H.}, \bibinfo{author}{Yajima, T.}, \bibinfo{author}{Matsuo, A.} \& \bibinfo{author}{Kindo, K.}
\newblock \bibinfo{journal}{\bibinfo{title}{Ligand dependent magnetism of the {${J}_{\text{eff}} = 3/2$} {Mott insulator Cs$_2$MX$_6$ (M = Ta, Nb, X = Br, Cl)}}}.
\newblock {\emph{\JournalTitle{Journal of Physics: Condensed Matter}}} \textbf{\bibinfo{volume}{33}}, \bibinfo{pages}{125802}, \doiprefix\url{10.1088/1361-648X/abd7b5} (\bibinfo{year}{2021}).

\bibitem{Mansouri2023}
\bibinfo{author}{Mansouri~Tehrani, A.} \emph{et~al.}
\newblock \bibinfo{journal}{\bibinfo{title}{Charge multipole correlations and order in {${\mathrm{Cs}}_{2}\mathrm{Ta}{\mathrm{Cl}}_{6}$}}}.
\newblock {\emph{\JournalTitle{Phys. Rev. Res.}}} \textbf{\bibinfo{volume}{5}}, \bibinfo{pages}{L012010}, \doiprefix\url{10.1103/PhysRevResearch.5.L012010} (\bibinfo{year}{2023}).

\bibitem{Morgan2023}
\bibinfo{author}{Morgan, E.~E.} \emph{et~al.}
\newblock \bibinfo{journal}{\bibinfo{title}{Hybrid and inorganic vacancy-ordered double perovskites a2wcl6}}.
\newblock {\emph{\JournalTitle{Chemistry of Materials}}} \textbf{\bibinfo{volume}{35}}, \bibinfo{pages}{7032--7038}, \doiprefix\url{10.1021/acs.chemmater.3c01300} (\bibinfo{year}{2023}).

\bibitem{Pradhan2024}
\bibinfo{author}{Pradhan, K.}, \bibinfo{author}{Paramekanti, A.} \& \bibinfo{author}{Saha-Dasgupta, T.}
\newblock \bibinfo{journal}{\bibinfo{title}{Multipolar magnetism in {$5{d}^{2}$} vacancy-ordered halide double perovskites}}.
\newblock {\emph{\JournalTitle{Phys. Rev. B}}} \textbf{\bibinfo{volume}{109}}, \bibinfo{pages}{184416}, \doiprefix\url{10.1103/PhysRevB.109.184416} (\bibinfo{year}{2024}).

\bibitem{Li2024}
\bibinfo{author}{Li, Y.}, \bibinfo{author}{Seshadri, R.}, \bibinfo{author}{Wilson, S.~D.}, \bibinfo{author}{Cheetham, A.~K.} \& \bibinfo{author}{Valenti, R.}
\newblock \bibinfo{title}{Microscopic origin of temperature-dependent magnetism in spin-orbit-coupled transition metal compounds} (\bibinfo{year}{2024}).
\newblock \eprint{2402.14064}.

\bibitem{Warzanowski2024}
\bibinfo{author}{Warzanowski, P.} \emph{et~al.}
\newblock \bibinfo{title}{Spin-orbital-lattice entanglement in the ideal j=1/2 compound k$_2$ircl$_6$} (\bibinfo{year}{2024}).
\newblock \eprint{2407.12133}.

\bibitem{Osborn1953}
\bibinfo{author}{Osborne, D.~W.} \& \bibinfo{author}{Westrum, E.~F.}
\newblock \bibinfo{journal}{\bibinfo{title}{The heat capacity of thorium dioxide from 10 to 305~k. the heat capacity anomalies in uranium dioxide and neptunium dioxide}}.
\newblock {\emph{\JournalTitle{The Journal of Chemical Physics}}} \textbf{\bibinfo{volume}{21}}, \bibinfo{pages}{1884--1887}, \doiprefix\url{10.1063/1.1698683} (\bibinfo{year}{1953}).

\bibitem{Cox1967}
\bibinfo{author}{Cox, D.} \& \bibinfo{author}{Frazer, B.}
\newblock \bibinfo{journal}{\bibinfo{title}{A neutron diffraction study of {NpO$_2$}}}.
\newblock {\emph{\JournalTitle{Journal of Physics and Chemistry of Solids}}} \textbf{\bibinfo{volume}{28}}, \bibinfo{pages}{1649 -- 1650}, \doiprefix\url{https://doi.org/10.1016/0022-3697(67)90137-0} (\bibinfo{year}{1967}).

\bibitem{Dunlap1968}
\bibinfo{author}{Dunlap, B.}, \bibinfo{author}{Kalvius, G.}, \bibinfo{author}{Lam, D.} \& \bibinfo{author}{Brodsky, M.}
\newblock \bibinfo{journal}{\bibinfo{title}{Hyperfine field of $^{237}${Np} in {NpO$_2$}}}.
\newblock {\emph{\JournalTitle{Journal of Physics and Chemistry of Solids}}} \textbf{\bibinfo{volume}{29}}, \bibinfo{pages}{1365 -- 1367}, \doiprefix\url{10.1016/0022-3697(68)90188-1} (\bibinfo{year}{1968}).

\bibitem{Tsujimoto2014}
\bibinfo{author}{Tsujimoto, M.}, \bibinfo{author}{Matsumoto, Y.}, \bibinfo{author}{Tomita, T.}, \bibinfo{author}{Sakai, A.} \& \bibinfo{author}{Nakatsuji, S.}
\newblock \bibinfo{journal}{\bibinfo{title}{Heavy-fermion superconductivity in the quadrupole ordered state of ${\mathrm{prv}}_{2}{\mathrm{al}}_{20}$}}.
\newblock {\emph{\JournalTitle{Phys. Rev. Lett.}}} \textbf{\bibinfo{volume}{113}}, \bibinfo{pages}{267001}, \doiprefix\url{10.1103/PhysRevLett.113.267001} (\bibinfo{year}{2014}).

\bibitem{Sumita2020}
\bibinfo{author}{Sumita, S.} \& \bibinfo{author}{Yanase, Y.}
\newblock \bibinfo{journal}{\bibinfo{title}{Superconductivity induced by fluctuations of momentum-based multipoles}}.
\newblock {\emph{\JournalTitle{Phys. Rev. Res.}}} \textbf{\bibinfo{volume}{2}}, \bibinfo{pages}{033225}, \doiprefix\url{10.1103/PhysRevResearch.2.033225} (\bibinfo{year}{2020}).

\bibitem{Yamauchi1999}
\bibinfo{author}{Yamauchi, H.} \emph{et~al.}
\newblock \bibinfo{journal}{\bibinfo{title}{Antiferroquadrupolar ordering and magnetic properties of the tetragonal dyb$_2$c$_2$ compound}}.
\newblock {\emph{\JournalTitle{Journal of the Physical Society of Japan}}} \textbf{\bibinfo{volume}{68}}, \bibinfo{pages}{2057--2066}, \doiprefix\url{10.1143/jpsj.68.2057} (\bibinfo{year}{1999}).

\bibitem{Staub2005}
\bibinfo{author}{Staub, U.} \emph{et~al.}
\newblock \bibinfo{journal}{\bibinfo{title}{Orbital dynamics of the $4f$ shell in ${\mathrm{d}\mathrm{y}\mathrm{b}}_{2}{\mathrm{c}}_{2}$}}.
\newblock {\emph{\JournalTitle{Phys. Rev. Lett.}}} \textbf{\bibinfo{volume}{94}}, \bibinfo{pages}{036408}, \doiprefix\url{10.1103/PhysRevLett.94.036408} (\bibinfo{year}{2005}).

\bibitem{Morin1988}
\bibinfo{author}{Morin, P.}, \bibinfo{author}{Rouchy, J.} \& \bibinfo{author}{Schmitt, D.}
\newblock \bibinfo{journal}{\bibinfo{title}{Susceptibility formalism for magnetic and quadrupolar interactions in hexagonal and tetragonal rare-earth compounds}}.
\newblock {\emph{\JournalTitle{Phys. Rev. B}}} \textbf{\bibinfo{volume}{37}}, \bibinfo{pages}{5401--5413}, \doiprefix\url{10.1103/PhysRevB.37.5401} (\bibinfo{year}{1988}).

\bibitem{Aleonard1979}
\bibinfo{author}{Al\'eonard, R.} \& \bibinfo{author}{Morin, P.}
\newblock \bibinfo{journal}{\bibinfo{title}{Tmcd quadrupolar ordering and magnetic interactions}}.
\newblock {\emph{\JournalTitle{Phys. Rev. B}}} \textbf{\bibinfo{volume}{19}}, \bibinfo{pages}{3868--3872}, \doiprefix\url{10.1103/PhysRevB.19.3868} (\bibinfo{year}{1979}).

\bibitem{Giraud1986}
\bibinfo{author}{Giraud, M.}, \bibinfo{author}{Morin, P.}, \bibinfo{author}{Rouchy, J.} \& \bibinfo{author}{Schmitt, D.}
\newblock \bibinfo{journal}{\bibinfo{title}{Magnetic and quadropolar properties of the tmmg compound}}.
\newblock {\emph{\JournalTitle{Journal of Magnetism and Magnetic Materials}}} \textbf{\bibinfo{volume}{59}}, \bibinfo{pages}{255--265}, \doiprefix\url{https://doi.org/10.1016/0304-8853(86)90421-X} (\bibinfo{year}{1986}).

\bibitem{Morin1990}
\bibinfo{author}{Morin, P.} \& \bibinfo{author}{Schmitt, D.}
\newblock \emph{\bibinfo{title}{Quadrupolar interactions and magneto-elastic effects in rare earth intermetallic compounds}} (\bibinfo{publisher}{North-Holland}, \bibinfo{address}{Netherlands}, \bibinfo{year}{1990}).

\bibitem{Morin1987}
\bibinfo{author}{Morin, P.}, \bibinfo{author}{Giraud, M.}, \bibinfo{author}{Burlet, P.} \& \bibinfo{author}{Czopnik, A.}
\newblock \bibinfo{journal}{\bibinfo{title}{Antiferroquadrupolar and antiferromagnetic structures in tmga3}}.
\newblock {\emph{\JournalTitle{Journal of Magnetism and Magnetic Materials}}} \textbf{\bibinfo{volume}{68}}, \bibinfo{pages}{107--114}, \doiprefix\url{https://doi.org/10.1016/0304-8853(87)90103-X} (\bibinfo{year}{1987}).

\bibitem{Morin1982}
\bibinfo{author}{Morin, P.}, \bibinfo{author}{Schmitt, D.} \& \bibinfo{author}{{du Tremolet de Lacheisserie}, E.}
\newblock \bibinfo{journal}{\bibinfo{title}{Magnetic and quadrupolar properties of prpb3}}.
\newblock {\emph{\JournalTitle{Journal of Magnetism and Magnetic Materials}}} \textbf{\bibinfo{volume}{30}}, \bibinfo{pages}{257--264}, \doiprefix\url{https://doi.org/10.1016/0304-8853(82)90206-2} (\bibinfo{year}{1982}).

\bibitem{Paddison2015}
\bibinfo{author}{Paddison, J. A.~M.} \emph{et~al.}
\newblock \bibinfo{journal}{\bibinfo{title}{Hidden order in spin-liquid gd$_3$ga$_5$o$_{12}$.}}
\newblock {\emph{\JournalTitle{Science (New York, N.Y.)}}} \textbf{\bibinfo{volume}{350}}, \bibinfo{pages}{179--81} (\bibinfo{year}{2015}).

\bibitem{Amoretti1989}
\bibinfo{author}{Amoretti, G.} \emph{et~al.}
\newblock \bibinfo{journal}{\bibinfo{title}{5f-electron states in uranium dioxide investigated using high-resolution neutron spectroscopy}}.
\newblock {\emph{\JournalTitle{Phys. Rev. B}}} \textbf{\bibinfo{volume}{40}}, \bibinfo{pages}{1856--1870}, \doiprefix\url{10.1103/PhysRevB.40.1856} (\bibinfo{year}{1989}).

\bibitem{Giannozzi1987}
\bibinfo{author}{Giannozzi, P.} \& \bibinfo{author}{Erdös, P.}
\newblock \bibinfo{journal}{\bibinfo{title}{Theoretical analysis of the 3-k magnetic structure and distortion of uranium dioxide}}.
\newblock {\emph{\JournalTitle{Journal of Magnetism and Magnetic Materials}}} \textbf{\bibinfo{volume}{67}}, \bibinfo{pages}{75--87}, \doiprefix\url{https://doi.org/10.1016/0304-8853(87)90722-0} (\bibinfo{year}{1987}).

\bibitem{Shiina2022}
\bibinfo{author}{Shiina, R.}
\newblock \bibinfo{journal}{\bibinfo{title}{Mechanism of non-coplanar magnetic ordering assisted by the jahn–teller effect in uo2}}.
\newblock {\emph{\JournalTitle{Journal of the Physical Society of Japan}}} \textbf{\bibinfo{volume}{91}}, \bibinfo{pages}{023704}, \doiprefix\url{10.7566/JPSJ.91.023704} (\bibinfo{year}{2022}).
\newblock \eprint{https://doi.org/10.7566/JPSJ.91.023704}.

\bibitem{Allen1968}
\bibinfo{author}{Allen, S.~J.}
\newblock \bibinfo{journal}{\bibinfo{title}{Spin-lattice interaction in u${\mathrm{o}}_{2}$. ii. theory of the first-order phase transition}}.
\newblock {\emph{\JournalTitle{Phys. Rev.}}} \textbf{\bibinfo{volume}{167}}, \bibinfo{pages}{492--496}, \doiprefix\url{10.1103/PhysRev.167.492} (\bibinfo{year}{1968}).

\bibitem{Gardiner2004}
\bibinfo{author}{Gardiner, C.~H.}, \bibinfo{author}{Boothroyd, A.~T.}, \bibinfo{author}{McKelvy, M.~J.}, \bibinfo{author}{McIntyre, G.~J.} \& \bibinfo{author}{Proke\ifmmode~\check{s}\else \v{s}\fi{}, K.}
\newblock \bibinfo{journal}{\bibinfo{title}{Field-induced magnetic and structural domain alignment in ${\text{pro}}_{2}$}}.
\newblock {\emph{\JournalTitle{Phys. Rev. B}}} \textbf{\bibinfo{volume}{70}}, \bibinfo{pages}{024416}, \doiprefix\url{10.1103/PhysRevB.70.024416} (\bibinfo{year}{2004}).

\bibitem{Jensen2007}
\bibinfo{author}{Jensen, J.}
\newblock \bibinfo{journal}{\bibinfo{title}{Static and dynamic jahn-teller effects and antiferromagnetic order in $\mathrm{Pr}{\mathrm{o}}_{2}$: A mean-field analysis}}.
\newblock {\emph{\JournalTitle{Phys. Rev. B}}} \textbf{\bibinfo{volume}{76}}, \bibinfo{pages}{144428}, \doiprefix\url{10.1103/PhysRevB.76.144428} (\bibinfo{year}{2007}).

\bibitem{Osborne1953}
\bibinfo{author}{Osborne, D.~W.} \& \bibinfo{author}{Westrum, E. F.~J.}
\newblock \bibinfo{journal}{\bibinfo{title}{The heat capacity of thorium dioxide from 10 to 305°k. the heat capacity anomalies in uranium dioxide and neptunium dioxide}}.
\newblock {\emph{\JournalTitle{Journal of Chemical Physics}}} \textbf{\bibinfo{volume}{21}}, \bibinfo{pages}{1884--1887} (\bibinfo{year}{1953}).

\bibitem{Tokunaga2005}
\bibinfo{author}{Tokunaga, Y.} \emph{et~al.}
\newblock \bibinfo{journal}{\bibinfo{title}{Nmr evidence for triple-$\stackrel{\ensuremath{\rightarrow}}{q}$ multipole structure in ${\mathrm{npo}}_{2}$}}.
\newblock {\emph{\JournalTitle{Phys. Rev. Lett.}}} \textbf{\bibinfo{volume}{94}}, \bibinfo{pages}{137209}, \doiprefix\url{10.1103/PhysRevLett.94.137209} (\bibinfo{year}{2005}).

\bibitem{Paixao2002}
\bibinfo{author}{Paix\~ao, J.~A.} \emph{et~al.}
\newblock \bibinfo{journal}{\bibinfo{title}{Triple-$\stackrel{\ensuremath{\rightarrow}}{q}$ octupolar ordering in ${\mathrm{n}\mathrm{p}\mathrm{o}}_{2}$}}.
\newblock {\emph{\JournalTitle{Phys. Rev. Lett.}}} \textbf{\bibinfo{volume}{89}}, \bibinfo{pages}{187202}, \doiprefix\url{10.1103/PhysRevLett.89.187202} (\bibinfo{year}{2002}).

\bibitem{Amoretti1992}
\bibinfo{author}{Amoretti, G.} \emph{et~al.}
\newblock \bibinfo{journal}{\bibinfo{title}{Neutron-scattering investigation of the electronic ground state of neptunium dioxide}}.
\newblock {\emph{\JournalTitle{Journal of Physics: Condensed Matter}}} \textbf{\bibinfo{volume}{4}}, \bibinfo{pages}{3459}, \doiprefix\url{10.1088/0953-8984/4/13/010} (\bibinfo{year}{1992}).

\bibitem{Duan2007}
\bibinfo{author}{Duan, C.-G.} \emph{et~al.}
\newblock \bibinfo{journal}{\bibinfo{title}{Electronic, magnetic and transport properties of rare-earth monopnictides}}.
\newblock {\emph{\JournalTitle{Journal of Physics: Condensed Matter}}} \textbf{\bibinfo{volume}{19}}, \bibinfo{pages}{315220}, \doiprefix\url{10.1088/0953-8984/19/31/315220} (\bibinfo{year}{2007}).

\bibitem{Cricchio2011}
\bibinfo{author}{{Cricchio, F.}}, \bibinfo{author}{{Grånäs, O.}} \& \bibinfo{author}{{Nordström, L.}}
\newblock \bibinfo{journal}{\bibinfo{title}{Polarization of an open shell in the presence of spin-orbit coupling}}.
\newblock {\emph{\JournalTitle{EPL}}} \textbf{\bibinfo{volume}{94}}, \bibinfo{pages}{57009}, \doiprefix\url{10.1209/0295-5075/94/57009} (\bibinfo{year}{2011}).

\bibitem{Mydosh2020}
\bibinfo{author}{Mydosh, J.~A.}, \bibinfo{author}{Oppeneer, P.~M.} \& \bibinfo{author}{Riseborough, P.~S.}
\newblock \bibinfo{journal}{\bibinfo{title}{Hidden order and beyond: an experimental—theoretical overview of the multifaceted behavior of uru2si2}}.
\newblock {\emph{\JournalTitle{Journal of Physics: Condensed Matter}}} \textbf{\bibinfo{volume}{32}}, \bibinfo{pages}{143002}, \doiprefix\url{10.1088/1361-648X/ab5eba} (\bibinfo{year}{2020}).

\bibitem{Christovam2024}
\bibinfo{author}{Christovam, D.~S.} \emph{et~al.}
\newblock \bibinfo{journal}{\bibinfo{title}{Spectroscopic evidence of kondo-induced quasiquartet in ${\mathrm{cerh}}_{2}{\mathrm{as}}_{2}$}}.
\newblock {\emph{\JournalTitle{Phys. Rev. Lett.}}} \textbf{\bibinfo{volume}{132}}, \bibinfo{pages}{046401}, \doiprefix\url{10.1103/PhysRevLett.132.046401} (\bibinfo{year}{2024}).

\bibitem{zeldovich1957}
\bibinfo{author}{Zel'dovich, Y.~B.}
\newblock \bibinfo{journal}{\bibinfo{title}{Electromagnetic interaction with parity violation}}.
\newblock {\emph{\JournalTitle{Zh. Eksp. Teor. Fiz.}}} \textbf{\bibinfo{volume}{33}}, \bibinfo{pages}{1531} (\bibinfo{year}{1957}).
\newblock \bibinfo{note}{[Sov. Phys. JETP 6, 1184 (1958)]}.

\bibitem{gorbatsevich1983}
\bibinfo{author}{Gorbatsevich, A.~A.}, \bibinfo{author}{Kopaev, Y.~V.} \& \bibinfo{author}{Tugushev, V.~V.}
\newblock \bibinfo{journal}{\bibinfo{title}{Anomalous nonlinear effects at phase transitions to ferroelectric and magnetoelectric states}}.
\newblock {\emph{\JournalTitle{Zh. Eksp. Teor. Fiz.}}} \textbf{\bibinfo{volume}{85}}, \bibinfo{pages}{1107} (\bibinfo{year}{1983}).
\newblock \bibinfo{note}{[Sov. Phys. JETP 58, 643 (1983)]}.

\bibitem{Hitomi2014}
\bibinfo{author}{Hitomi, T.} \& \bibinfo{author}{Yanase, Y.}
\newblock \bibinfo{journal}{\bibinfo{title}{Electric octupole order in bilayer ruthenate sr3ru2o7}}.
\newblock {\emph{\JournalTitle{Journal of the Physical Society of Japan}}} \textbf{\bibinfo{volume}{83}}, \bibinfo{pages}{114704}, \doiprefix\url{10.7566/JPSJ.83.114704} (\bibinfo{year}{2014}).
\newblock \eprint{https://doi.org/10.7566/JPSJ.83.114704}.

\bibitem{Fu2015}
\bibinfo{author}{Fu, L.}
\newblock \bibinfo{journal}{\bibinfo{title}{Parity-breaking phases of spin-orbit-coupled metals with gyrotropic, ferroelectric, and multipolar orders}}.
\newblock {\emph{\JournalTitle{Phys. Rev. Lett.}}} \textbf{\bibinfo{volume}{115}}, \bibinfo{pages}{026401}, \doiprefix\url{10.1103/PhysRevLett.115.026401} (\bibinfo{year}{2015}).

\bibitem{Bhowal2022}
\bibinfo{author}{Bhowal, S.}, \bibinfo{author}{Collins, S.~P.} \& \bibinfo{author}{Spaldin, N.~A.}
\newblock \bibinfo{journal}{\bibinfo{title}{Hidden $k$-space magnetoelectric multipoles in nonmagnetic ferroelectrics}}.
\newblock {\emph{\JournalTitle{Phys. Rev. Lett.}}} \textbf{\bibinfo{volume}{128}}, \bibinfo{pages}{116402}, \doiprefix\url{10.1103/PhysRevLett.128.116402} (\bibinfo{year}{2022}).

\bibitem{Lovesey2023}
\bibinfo{author}{Lovesey, S.~W.}
\newblock \bibinfo{journal}{\bibinfo{title}{Antiferromagnetic iron-based magnetoelectric compounds}}.
\newblock {\emph{\JournalTitle{Phys. Rev. B}}} \textbf{\bibinfo{volume}{107}}, \bibinfo{pages}{144432}, \doiprefix\url{10.1103/PhysRevB.107.144432} (\bibinfo{year}{2023}).

\bibitem{Fiebig1994}
\bibinfo{author}{Fiebig, M.}, \bibinfo{author}{Fr\"ohlich, D.}, \bibinfo{author}{Krichevtsov, B.~B.} \& \bibinfo{author}{Pisarev, R.~V.}
\newblock \bibinfo{journal}{\bibinfo{title}{Second harmonic generation and magnetic-dipole-electric-dipole interference in antiferromagnetic ${\mathrm{cr}}_{2}$${\mathrm{o}}_{3}$}}.
\newblock {\emph{\JournalTitle{Phys. Rev. Lett.}}} \textbf{\bibinfo{volume}{73}}, \bibinfo{pages}{2127--2130}, \doiprefix\url{10.1103/PhysRevLett.73.2127} (\bibinfo{year}{1994}).

\bibitem{Rikken1997}
\bibinfo{author}{Rikken, G. L. J.~A.} \& \bibinfo{author}{Raupach, E.}
\newblock \bibinfo{journal}{\bibinfo{title}{Observation of magneto-chiral dichroism}}.
\newblock {\emph{\JournalTitle{Nature}}} \textbf{\bibinfo{volume}{390}}, \bibinfo{pages}{493--494}, \doiprefix\url{10.1038/37323} (\bibinfo{year}{1997}).

\bibitem{Sannikov01111998}
\bibinfo{author}{and, D. G.~S.}
\newblock \bibinfo{journal}{\bibinfo{title}{Ferrotoroic phase transition in boracites}}.
\newblock {\emph{\JournalTitle{Ferroelectrics}}} \textbf{\bibinfo{volume}{219}}, \bibinfo{pages}{177--181}, \doiprefix\url{10.1080/00150199808213514} (\bibinfo{year}{1998}).
\newblock \eprint{https://doi.org/10.1080/00150199808213514}.

\bibitem{Goulon2000}
\bibinfo{author}{Goulon, J.} \emph{et~al.}
\newblock \bibinfo{journal}{\bibinfo{title}{First observation of nonreciprocal x-ray gyrotropy}}.
\newblock {\emph{\JournalTitle{Phys. Rev. Lett.}}} \textbf{\bibinfo{volume}{85}}, \bibinfo{pages}{4385--4388}, \doiprefix\url{10.1103/PhysRevLett.85.4385} (\bibinfo{year}{2000}).

\bibitem{Goulon2002}
\bibinfo{author}{Goulon, J.} \emph{et~al.}
\newblock \bibinfo{journal}{\bibinfo{title}{X-ray magnetochiral dichroism: A new spectroscopic probe of parity nonconserving magnetic solids}}.
\newblock {\emph{\JournalTitle{Phys. Rev. Lett.}}} \textbf{\bibinfo{volume}{88}}, \bibinfo{pages}{237401}, \doiprefix\url{10.1103/PhysRevLett.88.237401} (\bibinfo{year}{2002}).

\bibitem{Kubota2004}
\bibinfo{author}{Kubota, M.} \emph{et~al.}
\newblock \bibinfo{journal}{\bibinfo{title}{X-ray directional dichroism of a polar ferrimagnet}}.
\newblock {\emph{\JournalTitle{Phys. Rev. Lett.}}} \textbf{\bibinfo{volume}{92}}, \bibinfo{pages}{137401}, \doiprefix\url{10.1103/PhysRevLett.92.137401} (\bibinfo{year}{2004}).

\bibitem{astrov1961}
\bibinfo{author}{Astrov, D.~N.}
\newblock \bibinfo{journal}{\bibinfo{title}{Magnetoelectric effect in chromium oxide}}.
\newblock {\emph{\JournalTitle{Zh. Eksp. Teor. Fiz.}}} \textbf{\bibinfo{volume}{40}}, \bibinfo{pages}{1035} (\bibinfo{year}{1961}).
\newblock \bibinfo{note}{[Sov. Phys. JETP 13, 729 (1961)]}.

\bibitem{Chakravarty2001}
\bibinfo{author}{Chakravarty, S.}, \bibinfo{author}{Laughlin, R.~B.}, \bibinfo{author}{Morr, D.~K.} \& \bibinfo{author}{Nayak, C.}
\newblock \bibinfo{journal}{\bibinfo{title}{Hidden order in the cuprates}}.
\newblock {\emph{\JournalTitle{Phys. Rev. B}}} \textbf{\bibinfo{volume}{63}}, \bibinfo{pages}{094503}, \doiprefix\url{10.1103/PhysRevB.63.094503} (\bibinfo{year}{2001}).

\bibitem{Khmelevskyi2007}
\bibinfo{author}{Khmelevskyi, S.}, \bibinfo{author}{Khmelevska, T.}, \bibinfo{author}{Ruban, A.~V.} \& \bibinfo{author}{Mohn, P.}
\newblock \bibinfo{journal}{\bibinfo{title}{Magnetic exchange interactions in the paramagnetic state of hcp gd}}.
\newblock {\emph{\JournalTitle{Journal of Physics: Condensed Matter}}} \textbf{\bibinfo{volume}{19}}, \bibinfo{pages}{326218}, \doiprefix\url{10.1088/0953-8984/19/32/326218} (\bibinfo{year}{2007}).

\bibitem{Roy2011}
\bibinfo{author}{Roy, K.}, \bibinfo{author}{Bandyopadhyay, S.} \& \bibinfo{author}{Atulasimha, J.}
\newblock \bibinfo{journal}{\bibinfo{title}{Hybrid spintronics and straintronics: A magnetic technology for ultra low energy computing and signal processing}}.
\newblock {\emph{\JournalTitle{Applied Physics Letters}}} \textbf{\bibinfo{volume}{99}}, \bibinfo{pages}{063108}, \doiprefix\url{10.1063/1.3624900} (\bibinfo{year}{2011}).
\newblock \eprint{https://pubs.aip.org/aip/apl/article-pdf/doi/10.1063/1.3624900/14456471/063108\_1\_online.pdf}.

\bibitem{Miao2021}
\bibinfo{author}{Miao, F.}, \bibinfo{author}{Liang, S.-J.} \& \bibinfo{author}{Cheng, B.}
\newblock \bibinfo{journal}{\bibinfo{title}{Straintronics with van der waals materials}}.
\newblock {\emph{\JournalTitle{npj Quantum Materials}}} \textbf{\bibinfo{volume}{6}}, \bibinfo{pages}{59}, \doiprefix\url{10.1038/s41535-021-00360-3} (\bibinfo{year}{2021}).

\bibitem{Mankowsky2016}
\bibinfo{author}{Mankowsky, R.}, \bibinfo{author}{Först, M.} \& \bibinfo{author}{Cavalleri, A.}
\newblock \bibinfo{journal}{\bibinfo{title}{Non-equilibrium control of complex solids by nonlinear phononics}}.
\newblock {\emph{\JournalTitle{Reports on Progress in Physics}}} \textbf{\bibinfo{volume}{79}}, \bibinfo{pages}{064503}, \doiprefix\url{10.1088/0034-4885/79/6/064503} (\bibinfo{year}{2016}).

\bibitem{Franchini2021}
\bibinfo{author}{Franchini, C.}, \bibinfo{author}{Reticcioli, M.}, \bibinfo{author}{Setvin, M.} \& \bibinfo{author}{Diebold, U.}
\newblock \bibinfo{journal}{\bibinfo{title}{Polarons in materials}}.
\newblock {\emph{\JournalTitle{Nature Reviews Materials}}} \textbf{\bibinfo{volume}{6}}, \bibinfo{pages}{560--586}, \doiprefix\url{10.1038/s41578-021-00289-w} (\bibinfo{year}{2021}).

\bibitem{Doherty2013}
\bibinfo{author}{Doherty, M.~W.} \emph{et~al.}
\newblock \bibinfo{journal}{\bibinfo{title}{The nitrogen-vacancy colour centre in diamond}}.
\newblock {\emph{\JournalTitle{Physics Reports}}} \textbf{\bibinfo{volume}{528}}, \bibinfo{pages}{1--45}, \doiprefix\url{https://doi.org/10.1016/j.physrep.2013.02.001} (\bibinfo{year}{2013}).
\newblock \bibinfo{note}{The nitrogen-vacancy colour centre in diamond}.

\bibitem{Schirhagl2014}
\bibinfo{author}{Schirhagl, R.}, \bibinfo{author}{Chang, K.}, \bibinfo{author}{Loretz, M.} \& \bibinfo{author}{Degen, C.~L.}
\newblock \bibinfo{journal}{\bibinfo{title}{Nitrogen-vacancy centers in diamond: Nanoscale sensors for physics and biology}}.
\newblock {\emph{\JournalTitle{Annual Review of Physical Chemistry}}} \textbf{\bibinfo{volume}{65}}, \bibinfo{pages}{83--105}, \doiprefix\url{https://doi.org/10.1146/annurev-physchem-040513-103659} (\bibinfo{year}{2014}).

\bibitem{Rovny2024}
\bibinfo{author}{Rovny, J.} \emph{et~al.}
\newblock \bibinfo{journal}{\bibinfo{title}{Nanoscale diamond quantum sensors for many-body physics}}.
\newblock {\emph{\JournalTitle{Nature Reviews Physics}}} \textbf{\bibinfo{volume}{6}}, \bibinfo{pages}{753--768}, \doiprefix\url{10.1038/s42254-024-00775-4} (\bibinfo{year}{2024}).

\bibitem{Bayer2007}
\bibinfo{author}{Bayer, V.}, \bibinfo{author}{Franchini, C.} \& \bibinfo{author}{Podloucky, R.}
\newblock \bibinfo{journal}{\bibinfo{title}{Ab initio study of the structural, electronic, and magnetic properties of $\mathrm{MnO}(100)$ and $\mathrm{MnO}(110)$}}.
\newblock {\emph{\JournalTitle{Phys. Rev. B}}} \textbf{\bibinfo{volume}{75}}, \bibinfo{pages}{035404}, \doiprefix\url{10.1103/PhysRevB.75.035404} (\bibinfo{year}{2007}).

\bibitem{Weber2023}
\bibinfo{author}{Weber, S.~F.}, \bibinfo{author}{Urru, A.}, \bibinfo{author}{Bhowal, S.}, \bibinfo{author}{Ederer, C.} \& \bibinfo{author}{Spaldin, N.~A.}
\newblock \bibinfo{title}{Surface magnetization in antiferromagnets: Classification, example materials, and relation to magnetoelectric responses} (\bibinfo{year}{2023}).
\newblock \eprint{2306.06631}.

\bibitem{Bhowal2024b}
\bibinfo{author}{Bhowal, S.}, \bibinfo{author}{Urru, A.}, \bibinfo{author}{Weber, S.~F.} \& \bibinfo{author}{Spaldin, N.~A.}
\newblock \bibinfo{title}{Emergent surface multiferroicity} (\bibinfo{year}{2024}).
\newblock \eprint{2411.12434}.

\bibitem{Schmidt2019}
\bibinfo{author}{Schmidt, J.}, \bibinfo{author}{Marques, M. R.~G.}, \bibinfo{author}{Botti, S.} \& \bibinfo{author}{Marques, M. A.~L.}
\newblock \bibinfo{journal}{\bibinfo{title}{Recent advances and applications of machine learning in solid-state materials science}}.
\newblock {\emph{\JournalTitle{npj Computational Materials}}} \textbf{\bibinfo{volume}{5}}, \bibinfo{pages}{83}, \doiprefix\url{10.1038/s41524-019-0221-0} (\bibinfo{year}{2019}).

\bibitem{Eckhoff2021}
\bibinfo{author}{Eckhoff, M.} \& \bibinfo{author}{Behler, J.}
\newblock \bibinfo{journal}{\bibinfo{title}{High-dimensional neural network potentials for magnetic systems using spin-dependent atom-centered symmetry functions}}.
\newblock {\emph{\JournalTitle{npj Computational Materials}}} \textbf{\bibinfo{volume}{7}}, \bibinfo{pages}{170}, \doiprefix\url{10.1038/s41524-021-00636-z} (\bibinfo{year}{2021}).

\bibitem{Novikov2022}
\bibinfo{author}{Novikov, I.}, \bibinfo{author}{Grabowski, B.}, \bibinfo{author}{K{\"o}rmann, F.} \& \bibinfo{author}{Shapeev, A.}
\newblock \bibinfo{journal}{\bibinfo{title}{Magnetic moment tensor potentials for collinear spin-polarized materials reproduce different magnetic states of bcc fe}}.
\newblock {\emph{\JournalTitle{npj Computational Materials}}} \textbf{\bibinfo{volume}{8}}, \bibinfo{pages}{13}, \doiprefix\url{10.1038/s41524-022-00696-9} (\bibinfo{year}{2022}).

\bibitem{Kostiuchenko_2024}
\bibinfo{author}{Kostiuchenko, T.~S.}, \bibinfo{author}{Shapeev, A.~V.} \& \bibinfo{author}{Novikov, I.~S.}
\newblock \bibinfo{journal}{\bibinfo{title}{Interatomic interaction models for magnetic materials: Recent advances}}.
\newblock {\emph{\JournalTitle{Chinese Physics Letters}}} \textbf{\bibinfo{volume}{41}}, \bibinfo{pages}{066101}, \doiprefix\url{10.1088/0256-307X/41/6/066101} (\bibinfo{year}{2024}).

\bibitem{gao2024}
\bibinfo{author}{Gao, Y.}, \bibinfo{author}{Bokdam, M.} \& \bibinfo{author}{Kelly, P.~J.}
\newblock \bibinfo{title}{Machine learning exchange fields for ab-initio spin dynamics} (\bibinfo{year}{2024}).
\newblock \eprint{2403.10769}.

\bibitem{Acosta2022}
\bibinfo{author}{Acosta, C.~M.}, \bibinfo{author}{Ogoshi, E.}, \bibinfo{author}{Souza, J.~A.} \& \bibinfo{author}{Dalpian, G.~M.}
\newblock \bibinfo{journal}{\bibinfo{title}{Machine learning study of the magnetic ordering in 2d materials}}.
\newblock {\emph{\JournalTitle{ACS Applied Materials \& Interfaces}}} \textbf{\bibinfo{volume}{14}}, \bibinfo{pages}{9418--9432}, \doiprefix\url{10.1021/acsami.1c21558} (\bibinfo{year}{2022}).
\newblock \bibinfo{note}{PMID: 35133125}, \eprint{https://doi.org/10.1021/acsami.1c21558}.

\bibitem{Ponet2024}
\bibinfo{author}{Ponet, L.}, \bibinfo{author}{Lucente, E.~D.} \& \bibinfo{author}{Marzari, N.}
\newblock \bibinfo{journal}{\bibinfo{title}{The energy landscape of magnetic materials}}.
\newblock {\emph{\JournalTitle{npj Computational Materials}}} \textbf{\bibinfo{volume}{10}}, \bibinfo{pages}{151}, \doiprefix\url{10.1038/s41524-024-01310-w} (\bibinfo{year}{2024}).

\bibitem{Baumsteiger2025}
\bibinfo{author}{Baumsteiger, J.}, \bibinfo{author}{Celiberti, L.}, \bibinfo{author}{Rinke, P.}, \bibinfo{author}{Todorović, M.} \& \bibinfo{author}{Franchini, C.}
\newblock \bibinfo{title}{Exploring noncollinear magnetic energy landscapes with bayesian optimization} (\bibinfo{year}{2024}).
\newblock \eprint{2412.16433}.

\bibitem{Mills2020}
\bibinfo{author}{Mills, K.}, \bibinfo{author}{Ronagh, P.} \& \bibinfo{author}{Tamblyn, I.}
\newblock \bibinfo{journal}{\bibinfo{title}{Finding the ground state of spin hamiltonians with reinforcement learning}}.
\newblock {\emph{\JournalTitle{Nature Machine Intelligence}}} \textbf{\bibinfo{volume}{2}}, \bibinfo{pages}{509--517}, \doiprefix\url{10.1038/s42256-020-0226-x} (\bibinfo{year}{2020}).

\bibitem{Wang2020}
\bibinfo{author}{Wang, D.} \emph{et~al.}
\newblock \bibinfo{journal}{\bibinfo{title}{Machine learning magnetic parameters from spin configurations}}.
\newblock {\emph{\JournalTitle{Advanced Science}}} \textbf{\bibinfo{volume}{7}}, \bibinfo{pages}{2000566}, \doiprefix\url{https://doi.org/10.1002/advs.202000566} (\bibinfo{year}{2020}).
\newblock \eprint{https://onlinelibrary.wiley.com/doi/pdf/10.1002/advs.202000566}.

\bibitem{Kwon2020}
\bibinfo{author}{Kwon, H.~Y.} \emph{et~al.}
\newblock \bibinfo{journal}{\bibinfo{title}{Magnetic hamiltonian parameter estimation using deep learning techniques}}.
\newblock {\emph{\JournalTitle{Science Advances}}} \textbf{\bibinfo{volume}{6}}, \bibinfo{pages}{eabb0872}, \doiprefix\url{10.1126/sciadv.abb0872} (\bibinfo{year}{2020}).
\newblock \eprint{https://www.science.org/doi/pdf/10.1126/sciadv.abb0872}.

\bibitem{Fan2023}
\bibinfo{author}{Fan, C.} \emph{et~al.}
\newblock \bibinfo{journal}{\bibinfo{title}{Searching for spin glass ground states through deep reinforcement learning}}.
\newblock {\emph{\JournalTitle{Nature Communications}}} \textbf{\bibinfo{volume}{14}}, \bibinfo{pages}{725}, \doiprefix\url{10.1038/s41467-023-36363-w} (\bibinfo{year}{2023}).

\end{thebibliography}

\noindent\textbf{Acknowledgements}\\
We thank the Erwin Schrödinger Institute (ESI) for hosting the
ESI-PsiK workshop “Spin-Orbit Entangled Quantum Magnetism” and all
participants for the many enlightening discussions.
This research was funded in part by the Austrian Science Fund (FWF) projects I4506, I1490-N19, and J4698. For Open Access purposes, the authors have applied a CC BY public copyright
license to any author accepted manuscript version arising from this submission 
The work here presented is partly funded by the European
Union—Next Generation EU—“PNRR—M4C2, investi-
mento 1.1—Fondo PRIN 2022”—“Superlattices of rela-
tivistic oxides” (ID No. 2022L28H97, CUP
D53D23002260006). AP acknowledges support from
the Natural Sciences Engineering Council of
Canada.
The authors sincerely thank Francesca Perpetuini for her professional assistance in the design and preparation of Figure 1.
\\

\noindent\textbf{Author contributions}\\
C.F. conceived the initial idea for the review and coordinated its writing.
L.V.P., C.F., and A.P. outlined the content.
The initial draft was written by C.F. (Introduction), L.V.P. (Theory and Methods), A.P. ($d$-electron materials), and S.K. ($f$-electron materials).
L.C. and D.F.M. contributed to drafting, compiled the tables, and designed the major figures. All authors participated in the final editing of the text and figures.
\\

\noindent\textbf{Competing interests}\\
The authors declare no competing interests. \\

\end{document}